\documentclass[12pt,a4paper]{article}            
 \usepackage[skins,theorems]{tcolorbox}
\tcbset{highlight math style={enhanced,
  colframe=red,colback=white,arc=0pt,boxrule=1pt}}
  \usepackage[bookmarksopen, bookmarksnumbered, bookmarksopenlevel=2]{hyperref}
  \usepackage{tikz}
  \usepackage{tikz-3dplot}
 \usetikzlibrary{calc}

 \usetikzlibrary{decorations}
 \usepackage[UKenglish]{babel}
 \usepackage[toc,page]{appendix}
 \usepackage{amsmath}
 \usepackage{amssymb}
 \usepackage{float}
 \usepackage{amsthm}
 \usepackage{graphicx}
 \usepackage{hhline}
 \usepackage[bf]{caption}
\usepackage{cite}
\usepackage[vcentermath]{youngtab}
\usepackage{geometry}
\usepackage{slashed}
\usepackage{color}
\usepackage{stackrel}
\usepackage{tikz-cd} 
\usepackage{mathtools}
\usepackage{cancel} 
\usepackage{multirow}
\usepackage[margin=0pt,font=small,labelfont=normalfont,skip=22pt]{subcaption}
\usepackage{empheq}
\usepackage{arydshln}
\usepackage{fancyvrb}
\usepackage{color,soul}

\newcommand{\bqa}{\begin{eqnarray}}
\newcommand{\eqa}{\end{eqnarray}}

\newenvironment{eqn*}{\begin{equation*}\begin{aligned}}{\end{aligned}\end{equation*}\noindent}
\hypersetup{
    pdftitle={},
    pdfauthor={},
    pdfsubject={}
}

\numberwithin{equation}{section}
\numberwithin{table}{section}

\makeatletter

\newcommand{\be}{\begin{equation}}
\newcommand{\ee}{\end{equation}}
\newcommand{\beq}{\begin{equation}}
\newcommand{\eeq}{\end{equation}}
\newcommand{\ba}{\begin{aligned}}
\newcommand{\ea}{\end{aligned}}

\newcommand{\bea}{\begin{eqnarray}}
\newcommand{\eea}{\end{eqnarray}}

\newcommand{\cO}{\mathcal{O}}

\newcommand{\cD}{\mathcal{D}}

\newcommand{\cR}{\mathcal{R}}

\newcommand\bi{\begin{itemize}}
\newcommand\ei{\end{itemize}}

\def\unit{{1\kern-.65ex {\rm l}}}
\def\1{{1\kern-.65ex {\rm l}}}

\def\pl{{\mathrm{pl}}}

\def\dd{{\mathrm{d}}}

\newcount\hour \newcount\minute
\hour=\time \divide \hour by 60
\minute=\time
\def\now{%
\ifnum \hour<13
  \ifnum \hour=0 \advance \hour by 12 \number\hour:\else \number\hour:\fi%
     \ifnum \minute<10 0\fi%
     \number\minute%
\ A.M.%
\else \advance \hour by -12 \number\hour:%
  \ifnum \minute<10 0\fi%
  \number\minute%
  \ P.M.%
\fi%
}

\makeatother

\begin{document}

\begin{titlepage}
\begin{center}
\rightline{\small }

\vskip 15 mm

{\large \bf
String stars in $d\geq 7$
} 
\vskip 11 mm

Alek Bedroya$^{1,2}$ and David H. Wu$^{2}$
\vskip 11 mm
\small ${}^{1}$ 
{\it Princeton Gravity Initiative, Princeton University, Princeton, NJ 08544, USA} \\[3 mm]
\small ${}^{2}$
{\it Jefferson Physical Laboratory, Harvard University, Cambridge, MA 02138, USA}

\end{center}
\vskip 17mm

\begin{abstract}
We raise a thermodynamic puzzle for Horowitz--Polchinski (HP) solutions in the presence of extra compact dimensions and show that it can be resolved by the existence of higher-dimensional string stars. We provide non-trivial evidence for the existence of such string stars in spacetime dimensions $d\geq 7$ as higher-dimensional counterparts of HP solutions.  In particular, we explicitly construct string star solutions in $d=7$ that are under perturbative control. In $d>7$, at the Hagedorn temperature, we identify these string stars as a specific normalizable representative of a new one-parameter bounded family of Euclidean solutions which can be under perturbative control. The higher-order $\alpha'$ corrections play a crucial role in our arguments and, as pointed by other works, nullify the previous arguments against the existence of string stars in $d\geq 7$. The higher-dimensional string stars have non-zero free energy at Hagedorn temperature and their mass and free energy are of the same order as those of a string-sized black hole. In $d>7$, these solutions are string sized, but in $d=7$, the size of these solutions diverges as $\sim (T_{\rm H}-T)^{-1/4}$ near the Hagedorn temperature.
\end{abstract}

\vfill
\end{titlepage}

\newpage

\tableofcontents

\setcounter{page}{1}

\section{Introduction}

Quantum gravitational theories in $d>3$ spacetime dimensions have two defining features: 1) there is a massless spin-2 particle and 2) heavy states are described by black holes. The first feature renders the field theory non-renormalizable and suggests the emergence of new physics and new states at high energies. In string theory, the new physics typically involves a tower of weakly coupled states \cite{Ooguri:2006in}. Moreover, it was observed that the states in the lightest tower are always either Kaluza--Klein (KK) states or excitations of a light string \cite{Lee:2019wij}. This pattern was conjectured to be a universal feature of quantum gravity which is known as the Emergent String Conjecture (ESC). For recent bottom-up arguments for ESC see \cite{Bedroya:2024ubj,Basile:2023blg,Herraez:2024kux}. As we keep climbing the mass ladder, we eventually cross the black hole formation threshold and sufficiently heavy states are described by black holes. Interestingly, there is a nice connection between the towers of light states and the much heavier black hole states that was observed in \cite{Bedroya:2024uva} and we review below.

Consider a large extra dimension of size $R$ which leads to a tower of light KK states with masses proportional to $m_{\text{KK}}\sim 1/R$. The large dimension also gives rise to higher-dimensional black holes that are localized in the extra dimension. These higher-dimensional black holes are more thermodynamically stable when the temperature of the black hole exceeds $\sim m_{\text{KK}}$. This transition is known as the Gregory--Laflamme (GL) transition which we review in section~\ref{sec:GL instabilities for black holes}. The other type of known weakly coupled tower is a tower of string excitations with masses proportional to string mass $M_s$. Interestingly, the existence of this tower is also correlated with a thermodynamic phase transition of black holes. Specifically, in spacetime dimensions $d<7$, there are known Euclidean saddles that become more stable than black holes at a temperature proportional to $M_s$ \cite{Horowitz:1997jc}. These saddles are called the Horowitz--Polchinski (HP) solutions (or also string stars) which we review in section~\ref{sec:HP review}. These observations led the authors of this paper to conjecture that black holes always undergo a thermodynamic phase transition to a more stable saddle \cite{Bedroya:2024uva}. However, as was observed in \cite{Horowitz:1997jc,Chen:2021dsw}, such saddles are missing in weakly coupled string theories in spacetime dimensions $d\geq 7$. Recent progress in studying symmetries of the worldsheet CFT has shed some light on the mysterious behavior of these higher-dimensional counter parts to HP solutions \cite{Balthazar:2022hno,Balthazar:2022szl}, particularly in $d=7$. However, the precise profile, or even existence, of a local bounded solution of self-gravitating winding strings in a weakly coupled string theory in $d>7$ remains mysterious. These missing saddles, and their thermodynamic properties, are the subject of this work. We provide non-trivial evidence for the existence of a new class of worldsheet CFT corresponding to string stars in spacetime dimensions $d> 7$. Note that we refer to these backgrounds as string stars rather than HP solutions in $d<7$ due to their qualitatively different behaviors. 

We raise a puzzle in section~\ref{sec:3} which hints at the existence of string stars in $d\geq 7$. We consider spacetimes with $d<7$ non-compact dimensions and $D\geq7$ total dimensions. If higher-dimensional string stars exist, we can have such solutions that are localized in the compact dimensions in addition to the $d$-dimensional HP solutions. This is reminiscent of different types of black hole solutions in the presence of large extra dimension which was also observed and studied in \cite{Urbach:2022xzw,Chu:2024ggi,Emparan:2024mbp}. Similar to the GL instability of the lower-dimensional black holes, we show that the HP solution is unstable which points to a more stable higher-dimensional solution. We later give independent evidence for the existence of such higher-dimensional string stars in section~\ref{sec:worldsheet description}. We first find a new class of bounded Euclidean backgrounds at the Hagedorn temperature in $d>7$. We can parametrize these solutions by the core value of the winding condensate. For small values of this core value, these backgrounds are described by a weakly coupled worldsheet CFT. These backgrounds are bounded, but due to the asymptotic behavior $\chi(r)\sim r^{-2}$ they generically have diverging free energy. However, as we argue, there exist a specific representative of this family corresponding to the largest value of the core winding condensate that has a different asymptotic behavior $\chi(r) \sim r^{3-d}$ and a finite free energy. The higher-order $\alpha'$ corrections play a crucial role in the existence of the new string star solutions. After formally extending our equations to non-integer dimensions, in the limit of $d\rightarrow 7$, we see that the string stars at the Hagedorn temperature converge to $\chi=0$ while the free energy converges to a non-zero value $F=\frac{1152\pi^3\alpha'^2}{10\kappa^2}M_{\pl}^5$. This is similar to the results of \cite{Balthazar:2022hno}, where the limit of $d\rightarrow 7$ is accompanied by a co-scaling of $T_{\rm H}-T\sim d-7$. Motivated by these findings we found perturbative string star solutions at spacetime dimensions $d=7$ at temperatures $T<T_{\rm H}$ and verified that in the limit of $T\rightarrow T_{\rm H}$ the free energy of the solution converges $F=\frac{1152\pi^3\alpha'^2}{10\kappa^2}M_{\pl}^5$. The fact that we recovered the free energy of these perturbative solutions by taking the limit of $d\rightarrow 7$ lends more confidence to our prescription in $d>7$. 

\section{Lightning review}

As explained in the introduction, in the latter sections of this work, we will work out the details that relate the higher-dimensional string stars to the lower-dimensional HP solutions via a similar thermodynamic transition that was observed for black holes in the presence of extra compact dimensions, namely the GL transitions. Therefore, prior to delving into this renewed perspective on higher-dimensional string stars as a thermodynamically more stable phase of lower-dimensional HP solutions, we will dedicate this section to reviewing the construction of HP solutions in $3<d<7$ spacetime dimensions along with the story of GL transitions for black holes.

\subsection{Horowitz--Polchinski (HP) solutions and their missing cousins in $d>7$}
\label{sec:HP review}

The Horowitz--Polchinski solutions are localized self-gravitating backgrounds in string theory that were constructed in \cite{Horowitz:1997jc} to describe self-gravitating strings at temperatures close to the Hagedorn temperature \cite{Hagedorn:1965st}. The setup of interest is a Euclidean theory with a thermal circle of circumference $\beta$. Since we are studying a thermodynamic system, we impose anti-periodic boundary conditions on the fermions along the thermal circle. Accordingly, the contributions to the zero-point energy of a winding string from bosons and fermions no longer cancel each other out and we are left with a negative contribution to the square of the mass of the winding string. At sufficiently small $\beta$, this negative contribution dominates over the positive contribution from the string tension, and the lightest winding state becomes tachyonic. This transition happens at a temperature that is referred to as the Hagedorn temperature $T_{\rm H}$. 

The mass of these winding modes become light in the regime $\beta\sim \beta_{\rm H}=T_{\rm H}^{-1}$. Therefore, when the temperature of the system is near the Hagedorn temperature $T_{\rm H}$, one can study the EFT describing the strings wrapping the thermal circle. In this regime, Horowitz and Polchinski argued that the coupling between the dilaton and the winding mode is negligible while the dominating terms are those involving couplings between the radion and winding mode. Thus, the leading-order $(d-1)$-dimensional action, in terms of the radion and the winding mode, can be written down explicitly as \cite{Chen:2021dsw}
\begin{equation}
\label{eq:HP initial action}
    S=\pi M_{\pl}^{d-2}R\int \dd^{d-1}x\sqrt{g}e^{-2\phi}\left[-\cR-4(\nabla\phi)^2+(\nabla\varphi)^2+|\nabla\chi|^2+m^2(\varphi)|\chi|^2\right]\,,
\end{equation}
where $\phi$ denotes the $(d-1)$-dimensional dilaton, $Re^{\varphi}$ denotes the radius of the thermal circle which asymptotes to $R=\beta/2\pi$, and $\chi$ is a complex field denoting the two lightest winding modes with winding numbers $\pm1$. The mass coupling can be expanded in $\varphi$ 
such that we have 
\begin{equation*}
    m^2(\varphi)=m_{\infty}^2+\frac{\kappa}{\alpha'}\varphi\,,
 \end{equation*}
 where

 \vspace{.1cm}\noindent \makebox[\textwidth]{\parbox{1.5\textwidth}{
 \begin{equation*}
 \text{Bosonic: }
\begin{cases}
    m_{\infty}^2= \frac{R^2-R_{\rm H}^2}{(\alpha')^2}\,,\\
    R_{\rm H}=2M_s^{-1}\,,\\
    \kappa=8\,,
\end{cases}
\quad
\text{Type II: }
\begin{cases}
    m_{\infty}^2= \frac{R^2-R_{\rm H}^2}{(\alpha')^2}\,,\\
    R_{\rm H}=\sqrt{2}M_s^{-1}\,,\\
    \kappa=4\,,
\end{cases}
\quad \text{Heterotic: }
\begin{cases}
    m_{\infty}^2= \frac{R^2}{\alpha'^2}+\frac{1}{4R^2}-\frac{R_{\rm H}^2}{\alpha'^2}-\frac{1}{4R_{\rm H}^2}\,,\\
    R_{\rm H}=(1+1/\sqrt{2})M_s^{-1}\,,\\
    \kappa=4\sqrt{2}\,,
\end{cases}
\end{equation*}
}}
in which $m_{\infty}$ is the mass of the winding state at spatial infinity, and higher-order terms in $\varphi$ are omitted. In particular, 
as the above expansion is done in $R-R_{\rm H}$, to ensure the validity of any solution found using the above EFT, we must work in the regime where
\begin{equation}
\label{eq:HP regime of good control}
   0<R-R_{\rm H}\ll M_s^{-1}\,.
\end{equation}
Thus, the HP topology is $\mathbb{R}^{d-1}\times S^1_{\beta}$ where $S^1_{\beta}$ is nowhere vanishing in spacetime. Notice that this is a distinct topological configuration from that of black holes where the thermal circle shrinks to zero size. With this setup, the equations of motion (EoM) for $\varphi$ and $\chi$ are 
\begin{subequations}
\begin{align}
\label{eq:winding mode EoM}
    -\nabla^2\chi+\left(m_{\infty}^2+\frac{\kappa}{\alpha'}\varphi\right)\chi&=0\,,\\
\label{eq:radion EoM}
    -2\nabla^2\varphi+\frac{\kappa}{\alpha'}|\chi|^2&=0\,.
\end{align}
\end{subequations}
where $\nabla^2$ is the Laplacian with respect to the flat $(d-1)$-dimensional spatial metric. For simplicity, we will be looking for spherically symmetric solutions which the Laplacian reduces to 
\begin{equation*}
\nabla^2:=\partial_r^2+\frac{d-2}{r}\partial_r\,.
\end{equation*}

\begin{figure}
    \centering
    \includegraphics[width=.7\linewidth]{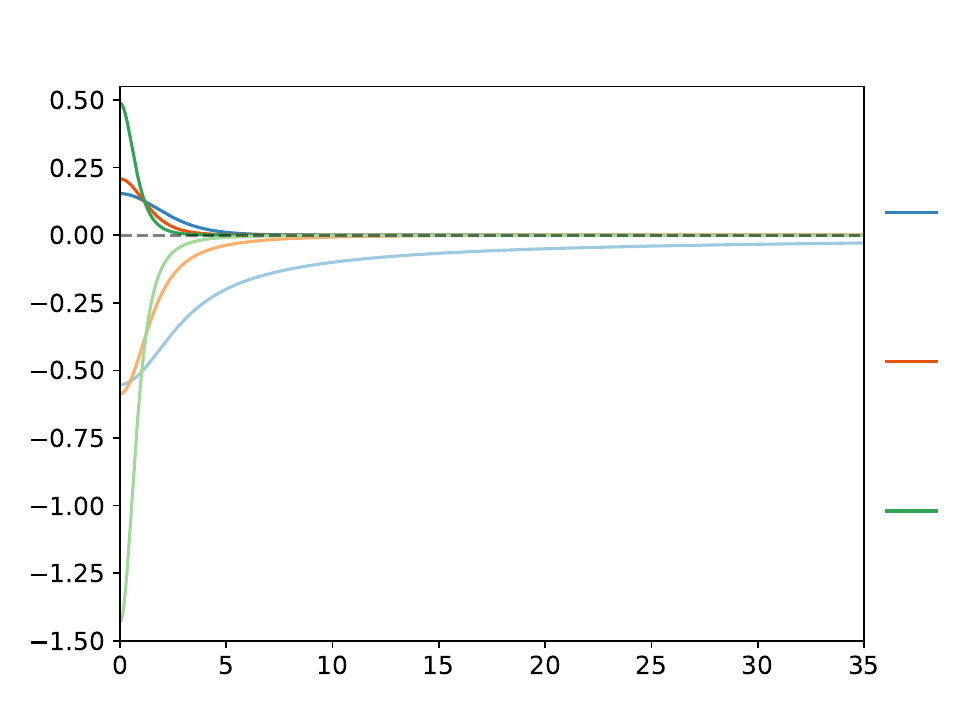}
        \begin{picture}(0,0)\vspace*{-1.2cm}
        \put(-15,210){$d$}
        \put(0,177){$4$}
        \put(0,124){$5$}
        \put(0,70){$6$}
        \put(-180,0){$\hat{r}$}
        \put(-350,215){$\hat{\chi}$}
        \put(-350,25){$\hat{\varphi}$}
        \end{picture}\vspace*{0cm}
    \caption{The spatial profile of the dimensionless normalized winding mode (dark) and radion (light) of the HP solutions \eqref{eq:normalized HP values} in spacetime dimensions $4,5,6$.}
    \label{fig:HP solutions}
\end{figure}
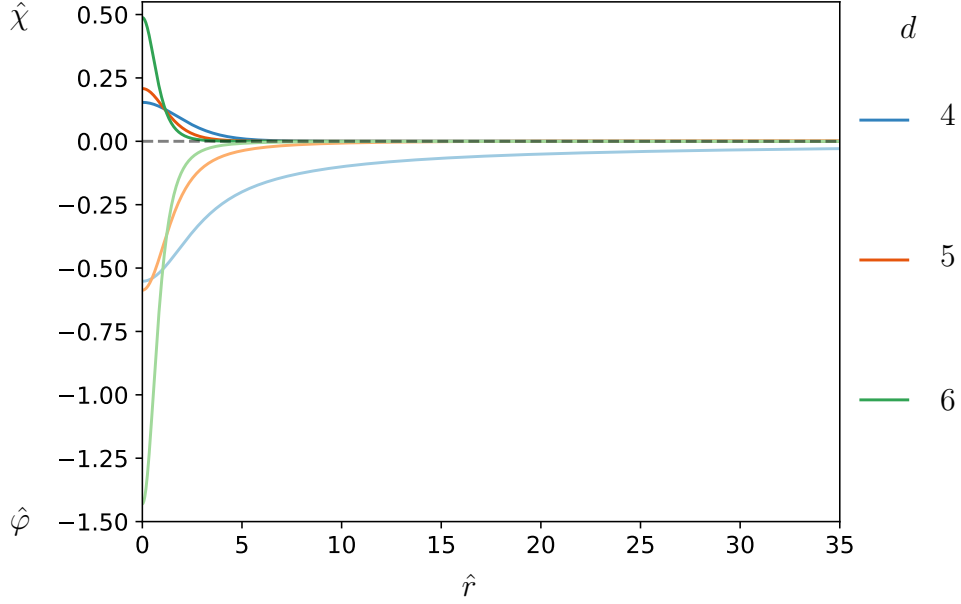

As was argued in \cite{Chen:2021dsw}, the above coupled ODEs have a scaling symmetries that allows for the following reparametrization
\begin{align}
\label{eq:normalized HP values}
\begin{split}
x&=m_\infty \hat x\,,\\
\chi&=\frac{\alpha'\sqrt{2(d-3)\omega_{d-2}}m_{\infty}^2}{\kappa\zeta}\hat\chi\,,\\
\varphi&=\frac{m_{\infty}^2\alpha'}{\zeta^2\kappa}\hat\varphi\,.
\end{split}
\end{align}
The hatted variables are dimenionless, $\omega_{n-1}=\frac{2\pi^{n/2}}{\Gamma[n/2]}$ is the area of a unit $n$-sphere, and $\zeta$ is a constant chosen such that $\int d^{d-1}\hat x|\hat\chi^2|=1$ and admits discrete values.\footnote{The authors in \cite{Chen:2021dsw} use this normalization only in $4+1$ spacetime dimensions and use different normalization in other dimensions which cannot implement the normalization condition on $\hat \chi$ in $4+1$ dimensions. However, we use this normalization in all dimensions given that it does not have that disadvantage.} The equations of motions for the normalized fields takes the following form
\begin{subequations}
\begin{align}
    -\hat\nabla^2\hat\chi+\left(1+\frac{1}{\zeta^2}\hat\varphi\right)\hat\chi&=0\,,\\
    -2\hat\nabla^2\hat\varphi+2(d-3)\omega_{d-2}|\hat\chi|^2&=0\,,\\
    \int d^{d-1}\hat x|\hat\chi^2|&=1\,,
\end{align}
\end{subequations}
where the last equation is the normalization condition. Therefore, the normalization of $\chi$ is given by 
\begin{align}
    \int d^{d-1} x|\chi^2|=\frac{2(d-3)\omega_{d-2}}{\kappa^2\zeta^2}M_s^{-4}m_\infty^{5-d}\,,
\end{align}
where we have made use of $\alpha'=M_s^{-2}$. The profiles of these solutions for spacetime dimensions $d=4,5,6$ are given in fig.~\ref{fig:HP solutions}. As we can construct normalizable solutions, the free energy of HP solutions can be computed from the action \eqref{eq:HP initial action} and is indeed finite for these solutions. Furthermore, for a solution to the equations of motions, one can show that the total action which measures the free energy is given by
\begin{equation}
\label{eq:free energy of HP}
    F_{\rm HP}^{(d)}=\frac{M_{\pl,d}^{d-2}m_\infty^2}{7-d}\int d^{d-1}x |\chi|^2=\frac{2(d-3)\omega_{d-2}}{(7-d)\kappa^2\zeta^2} M_{\pl,d}^{d-2}M_s^{-4}m_{\infty}^{7-d}\,.
\end{equation}
As we will review shortly, such normalizable solutions only exist in $d<7$. The HP solutions, while do have similarities with black holes, have some distinct features in addition to the aforementioned topological differences. Namely, these solutions are under perturbative control near Hagedorn temperature and, particularly, have vanishing free energy in the limit $T\to T_{\rm H}$. Furthermore, the mass and size of these solutions scale as $M\sim m_{\infty}^{7-d}$ and $l\sim m_{\infty}^{-1}$ which implies that as we increase the temperature $T\sim R^{-1}$, the mass decreases while the size increases, opposite to that of black holes. At last, numerical search for the solutions \cite{Chen:2021dsw} shows that the profile of HP solutions is very sensitive to the boundary conditions for \eqref{eq:winding mode EoM} and \eqref{eq:radion EoM} due to the non-linearity in the equations. The spectrum of bounded solutions to these non-linear equations is discrete.

As we mentioned earlier, another important feature of HP solutions is that they only exist in $3<d<7$ spacetime dimensions. One can see that the free energy equation~\eqref{eq:free energy of HP} associated to HP solutions would not vanish at the Hagedorn temperature for $d\geq 7$. The nonexistence of HP solutions for $d\geq 7$ can be further concretely illustrated via a scaling argument, as was done in \cite{Chen:2021dsw} and extended in \cite{Balthazar:2022hno} which we will briefly review below. The starting point is by assuming we have an HP solution $\chi_*,\varphi_*$. Then, by the rescaling 
\begin{equation}
\label{eq:rescaling parameters}
    \chi=\iota\chi_*(x/\gamma)\,,\qquad \varphi=\rho\varphi_*(x/\gamma)\,,
\end{equation}
in which $\iota,\rho,\gamma>0$, we have the relevant components in the action become
\begin{equation*}
    S=\rho^2\gamma^{d-3}S_{1,*}+\iota^2\gamma^{d-3}S_{2,*}+\iota^2\gamma^{d-1}S_{3,*}+\iota^2\rho\gamma^{d-1}S_{4,*}\,,
\end{equation*}
where $S_{1,*},S_{2,*},S_{3,*},S_{4,*}$ are terms in the on-shell action proportional to $(\nabla\varphi)^2$, $|\nabla\chi|^2$, $|\chi|^2$, and $\varphi|\chi|^2$ respectively. We choose $\rho=\gamma^{2}\iota^2$ which is a symmetry of the equation of the motion for $\phi$ and ensures $S_1$ and $S_4$ scale in the same way. With this choice of $\rho$ we have
\begin{equation}\label{RSE}
    S=\iota^2\gamma^{d-1}S_{2,*}+\iota^2\gamma^{d-1}S_{3,*}+\iota^4\gamma^{d+1}(S_{1,*}+S_{4,*})\,.
\end{equation}
Now, since $\chi_*$ is chosen to be a solution to the equations of motion, $\gamma,\iota=1$ is a saddle point for the action. Therefore, $\partial_\iota S\vert_{\iota,\gamma=1}=\partial_\gamma S\vert_{\iota,\gamma=1}=0$. Using these equations, we arrive at
\begin{align}
\label{eq:scaling conclusion for HP}
    S_{3,*}&=\frac{7-d}{d-3}S_{2,*}\,,\nonumber\\
    S_{1,*}+S_{4,*}&=\frac{2}{3-d}S_{2,*}\,,\nonumber\\
    S&=\frac{2}{d-3}S_{2,*}\,.
\end{align}
which cannot be correct when $d>7$ because both $S_{2,*}$ and $S_{3,*}$ are positive. Therefore, from the scaling argument of the EFT describing leading-order couplings between the radion and the winding mode, it seems that HP solutions do not exist in $d>7$ spacetime dimensions. 

\subsection{Gregory--Laflamme (GL) instabilities for black holes}
\label{sec:GL instabilities for black holes}

The GL instabilities for black holes arise in the presence of extra (compact) dimensions. The starting point is to consider the lower-dimensional black holes in the non-compact dimensions. These black holes can be uplifted to black solutions that are uniform in the extra compact dimensions. In other words, these lower-dimensional black holes can be thought of as higher-dimensional black branes that are uniformly wrapped around the internal compact dimensions. However, there are also new black hole solutions that correspond to higher-dimensional black holes that are localized in the internal dimensions. The Bekenstein--Hawking area law gives different free energy/mass relations for these two families of black hole solutions and the crossover between their different free energies leads to a thermodynamic phase transition between the two. Namely, at temperatures much below the KK scale (i.e. inverse of the size of the extra dimensions), the higher-dimensional black holes that are localized in the internal dimensions have lower free energy. Thus, at such temperatures, the higher-dimensional black holes become more thermodynamically favorable in the canonical ensemble. Gregory and Laflamme studied the decay process of the lower-dimensional black holes and showed that they become perturbatively unstable at temperatures of the order of KK mass scale. Note that the temperature at which lower-dimensional black holes become unstable could be different from when they become non-perturbatively unstable (i.e., when the free energies of two black holes cross, as a function of temperature). Whether the two temperatures match depends on the order of the phase transition, but nonetheless, they are expected to be of the same order. 

There is another closely related instability which is that of Euclidean black holes. In the following, we first review the perturbative instability of Euclidean black holes and then relate it to the GL perturbative instability of Lorentzian black holes in the presence of large compact dimensions. This relation between Euclidean instability of 
Schwarzschild black holes and their GL instabilities in Lorentzian signature has been remarked in \cite{Marsano:2006kr}.

Consider a Euclidean Schwarzschild black hole. Suppose we perturb the metric as
\begin{align}
    \delta g_{\mu\nu}=h_{\mu\nu}(r)\,.
\end{align}
The change in the Euclidean action is given by 
\begin{equation}
\label{eq:perturbed EQG action}
    \delta S_{E}=\int \delta\left[\sqrt{g}\mathcal{R}\right]=-\int \sqrt{g} h_{\mu\nu} \Delta_L h^{\mu\nu}\,,
\end{equation}
where $\Delta_L$ is the Lichnerowicz operator associated to the original metric, $g_{\mu\nu}$. In principle, the Euclidean and Lorentzian Lichnerowicz operators are different. However, since our mode does not depend on time, the operators will be the same. To identify a Euclidean instability of the above action~\eqref{eq:perturbed EQG action}, notice that the eigenfunction of the Lichnerowicz operator with a positive eigenvalue will lead to a quadratic perturbation that decreases the overall Euclidean action. This is precisely the mode that corresponds to the Euclidean instability observed in \cite{Gross:1982cv}
\begin{equation}
    \Delta_L h_E^{\mu\nu}=k_{\text{instability}}^2h_E^{\mu\nu}\,,
\end{equation}
with $k_{\rm instability}>0$.
Now, if we go back to the Lorentzian signature, we can use this eigenfunction to construct a perturbation to the metric that solves the equation of motion. The equation of motion for the perturbations to the metric must satisfy 
\begin{align}
    \Delta_L h^{\mu\nu}=0\,,
\end{align}
which is clearly not satisfied by $h^{\mu\nu}_E$. However, if we assume an extra dimension parametrized by $y$ such that the metric is 
\begin{align*}
    ds^2=-f(r)dt^2+ f(r)^{-1}dr^2+r^2d\Omega_{d-2}^2+dy^2\,,
\end{align*}
we can assume a perturbation given by $h^{\mu\nu}=h^{\mu\nu}_E(r)e^{iky}$ where $\mu,\nu=1,\dots,d$. The extra dimension adds a second-order derivative $\partial_y^2$ to the lower-dimensional Lichnerowicz operator. Therefore, the higher-dimensional Lichnerowicz operator acting on the above choice of perturbation becomes
\begin{equation}
    \Delta_L^{(d+1)}h^{\mu\nu}=e^{iky}\left(-k^2+\Delta_L^{(d)}\right)h^{\mu\nu}_E\,,
\end{equation}
which precisely vanishes when $k=k_{\rm instability}$. Therefore, we see that there exists a mode with zero frequency (static) that solves the equations of motion. Now, let us decrease the spatial momentum in the $y$-direction. The mode that solves the equations of motion now will have a non-trivial time evolution. Gregory and Laflamme \cite{Gregory:1993vy} studied the modes that are regulated on the horizon and are given by $h^{\mu\nu}=e^{\Omega t}e^{iky}H^{\mu\nu}(r)$ and showed that for $k<k_\text{GL}$, $\Omega$ has to be real and positive, which lead to time-divergent solutions. However, $k_\text{GL}=k_\text{instability}$! This agreement was explained by Real in \cite{Reall:2001ag}. To see why, note that the solutions are null functions of the operator $\partial_y^2+\Delta_L$. Decreasing $k$ from $k_\text{instability}$ will increase the eigenvalue which must be countered by a real positive value $\Omega$ which arises from the second-order time derivative appearing in $\Delta_L^{(d)}$. Therefore, it is no coincidence that Gregory and Laflamme found a perturbative instability for any $k<k_\text{instability}$. Thus, we conclude that an instability of the Euclidean solution is equivalent to a GL instability in the Lorentzian signature for any radius satisfying $R^{-1}<k_\text{instability}$. This is a crucial observation that we will use to motivate the existence of higher-dimensional string stars later in section~\ref{sec:GL instability for HP}.

\section{GL instability for HP solutions: hints towards saddles in $D>7$}\label{sec:3}

In the last section, we discussed how the GL instabilities of the lower-dimensional black holes signal the existence of higher-dimensional black holes that are localized in the extra dimensions. This is the temperature at which the lower-dimensional black hole becomes perturbatively unstable and decays into a higher-dimensional black hole. In this section we discuss the evidence for a GL-type instability for HP solutions which was first pointed out in \cite{Urbach:2022xzw} for string star solutions in AdS spaces. Moreover, the transitions between a uniform HP-brane and a \textit{non-uniform} HP-brane (which is distinct from higher-dimensional string stars) resulted from the GL-instability has also been recently discussed in \cite{Chu:2024ggi,Emparan:2024mbp}. 

In spacetime dimensions $3<d<7$, there are HP solutions which mimic some of the properties of black holes and can be thought of as string-sized extensions of black holes. Suppose we have a $D$-dimensional spacetime with $d$ non-compact dimensions such that $3<d<D<7$. If the compact dimensions are large in string units, there are multiple classes of HP solutions based on the number of compact dimensions in which the solution is localized. For example, two classes of solutions are 1) the higher-dimensional solutions that are localized in all of the compact dimensions, and 2) the lower-dimensional solutions that are uniform along all of the compact directions. For a more complete study of other solutions see \cite{Chu:2024ggi}. These two saddles respectively mimic the higher- and lower-dimensional black holes. Therefore, it is natural to ask if there is a GL instability that leads the lower-dimensional HP solution to decay into the higher-dimensional one. 

In this section, we show that such an instability indeed exists and can be observed by studying the lower-dimensional solutions alone. In other words, similar to the GL instability for black holes, the lower-dimensional solution becomes unstable at high temperatures which signals the existence of a more stable saddle that is localized in the extra dimensions. If the number of total dimensions is less than $7$, the higher-dimensional saddles are known. However, if $D\geq 7$, the instability is predicting the existence of an unknown saddle which would be the higher-dimensional counterpart of HP saddle\footnote{To remain consistent with existing literature, we will refer to these solutions in higher dimensions as string stars which are broadly defined as localized profiles of the winding mode that are not only restricted to $3<d<7$ Euclidean space, e.g., string stars in AdS$_3$ \cite{Agia:2023skp}.}. Moreover, as we explore in section~\ref{sec:expected properties of counter HP}, this argument allows us to predict some of the thermodynamic properties of these new saddles which we independently verify in later sections.

Since the HP saddles are Euclidean solutions, it is more natural to discuss the Euclidean instabilities. As explained in section~\ref{sec:GL instabilities for black holes}, the Euclidean instabilities of a solution are closely related to the GL instabilities of the Lorentzian solution assuming that there are extra compact dimensions. We use this connection to find the GL instabilities of the wrapped ``HP-brane'' solutions and show that when $D<7$ the instability occurs close to the temperature at which the higher-dimensional HP solution is more thermodynamically favored. 

\subsection{Euclidean instabilities of HP solutions}

In \cite{Chen:2021dsw}, the authors argued that the HP solution has a Euclidean instability since a rescaling of the solution decreases the Euclidean action in the first subleading term. Here, we use that argument to find an analytic lower bound on the eigenvalue associated with the instability. 

After integrating $\varphi$ out via the EoM \eqref{eq:radion EoM}, we can have a non-local action for $\chi$. We will study the instabilities of this action. First, consider a general perturbation $\delta\chi$ of the solution. Suppose the perturbation of $\chi$ changes the action as 
\begin{align}
    \delta S=-\frac{1}{2\kappa}\int \dd^{d-1} x\sqrt{g}\delta\chi^*\mathcal{D} \delta \chi\,,
\end{align}
where $\mathcal{D}$ is a linear non-local second-order differential operator associated with the EoM for the perturbations. The eigenfunctions of $\mathcal{D}$ with real eigenvalue $\lambda$ will satisfy 
\begin{equation*}
    \cD\delta \chi=\lambda\delta\chi\,,\qquad \rightarrow \qquad \frac{\delta S}{\frac{1}{2\kappa}\int \dd^{d-1}x\sqrt{g}|\delta\chi|^2}=-\lambda\,,
\end{equation*}
where $\lambda$ has units of $[\mathrm{mass}]^2$.
Note, in the above expansions, we explicitly only keep terms up to quadratic order in the perturbation $\cO(|\delta\chi|^2)$. Now, suppose $k_{\text{instability}}^2$ is the supremum of all the eigenvalue of $\mathcal{D}$. Then, for every perturbation $\delta\chi$ we have
\begin{align}
\label{eq:smallest k instability}
        \frac{\delta S}{||\delta \chi||^2}\geq -k_{\text{instability}}^2\,,
\end{align}
where we defined the normalized $L^2$-norm of $\delta \chi$ as
\begin{equation*}
    ||\delta \chi||^2:=\frac{1}{2\kappa}\int\dd^{d-1}x \sqrt{g}|\delta\chi|^2\,.
\end{equation*}
Now we consider a specific perturbation of the solution which is an infinitesimal rescaling
\begin{align}
    \delta\chi= \epsilon \chi\,.
\end{align}
This perturbation can be regarded as part of the rescaling scheme \eqref{eq:rescaling parameters} where we identify $\iota=1+\epsilon$. From equations \eqref{RSE} and \eqref{eq:scaling conclusion for HP}, we find that the change in the Euclidean action is 
\begin{align}
\label{eq:variation of action}
    \delta S= \frac{-8\epsilon^2}{d-3}S_{\text{kinetic}}\,,
\end{align}
where $S_{\text{kinetic}}$ is the kinetic contribution of $\chi$ to the total action. From equation~\eqref{eq:scaling conclusion for HP} we also know that the contribution of $\chi$'s mass term to the action is proportional to $S_{\text{kinetic}}$,
\begin{equation}\label{PROP}
    S_{\rm mass}=\frac{7-d}{d-3}S_{\text{kinetic}}\,,
\end{equation}
which leads to
\begin{align}
    ||\delta\chi||^2\equiv \frac{1}{2\kappa}\int d^{d-1} x \sqrt{g}|\delta\chi|^2=\frac{\epsilon^2}{m_\infty^2}S_{\rm mass}=\frac{(7-d)\epsilon^2}{(d-3)m_\infty^2}S_{\rm kinetic}\,.
\end{align}
 
Now, using the above relation between the leading-order perturbation of the action due to the perturbed winding mode, namely \eqref{eq:variation of action}, we can deduce the following relation
\begin{align}
    \frac{\delta S}{||\delta \chi||^2}=-\frac{8}{7-d}m_\infty^2\,.
\end{align}
Recall our inequality \eqref{eq:smallest k instability}, then from the above relation, we can precisely bound the smallest eigenvalue of $\cD$ as
\begin{align}\label{UBI}
    k_{\text{instability}}^2\geq \frac{8}{7-d}m_\infty^2\,.
\end{align}
Therefore, from this lower bound on $k_{\rm instability}^2$, we can see that for $d<7$, $-k_{\rm instability}^2$ must be a negative value. Hence, indeed perturbations of the winding mode leads to a decrease in the Euclidean action which indicates that the HP solutions have a Euclidean instability.

\subsection{GL instability for HP solutions and evidence for higher-dimensional solutions}
\label{sec:GL instability for HP}

As pointed out in \cite{Urbach:2022xzw}, similar to the black hole case, a Euclidean instability $\delta\chi$ can be used to construct a solution to the Lorentzian equations of motion for the Horowitz--Polchinski solution that extends in a parallel direction $y$. The solution is simply 
\begin{equation*}
    \delta\chi_{\rm combined}=\delta\chi\cdot \exp(i k_{\text{instability}}y)\,.
\end{equation*}
Again, similar to the black hole case, if we look at solutions with lower wavenumber along the parallel direction, they must have imaginary frequency so that they can keep satisfying the equation of motion 
\begin{equation*}
    \cD\delta\chi_{\rm combined}=(\Box+\dots) \delta\chi_{\rm combined}=0\,,
\end{equation*}
where $\dots$ involves the relevant partial derivatives that correspond to the action \eqref{eq:HP initial action} and the perturbation $\delta\chi$.
Therefore, if the $y$-direction is compact, any KK momentum smaller than $ k_{\text{instability}}$ will lead to an instability, i.e.,
\begin{align*}
    R_\text{KK}^{-1}<k_{\text{instability}}\,,\qquad \rightarrow \qquad \text{GL instability}\,.
\end{align*}
Using the inequality \eqref{UBI}, we find that the lower-dimensional ($d<7$) HP solution is unstable for 
\begin{align}
\label{eq:unstable HP solution}
    R_{\rm KK}m_{\infty}>\sqrt\frac{7-d}{8}\,.
\end{align}
The above analysis can be extended to having an arbitrary number of compact dimensions. Therefore, with the above lower bound and spacetime topology as $\mathbb{R}^{d-1}\times T^n\times S^1_{\beta}$ where $d<7$ and $n>0$ 
$R_{\rm KK}$ which denotes the radius of $T^n$ can be tuned to any value. Specifically, in the decompactification limit of $T^n$, we will necessarily encounter the unstable HP solution in $\mathbb{R}^{d-1}\times S^1_{\beta}$ which satisfies \eqref{eq:unstable HP solution}.

It is important to note that the HP solutions are inherently Euclidean and in the Lorentzian signature they are not well-understood (see \cite{Chen:2021dsw,Kutasov:2005rr}). However, as also suggested in \cite{Urbach:2022xzw}, motivated by the connection between the Euclidean negative modes and the GL instabilities for black holes, we use the Euclidean negative modes to learn about the GL instabilities of the Lorentzian string stars. Note that this is in addition to the argument that when $D<7$, there are localized higher-dimensional solutions that are known to be more thermodynamically favorable and therefore the lower-dimensional HP solutions are unstable. To elaborate on this point, in the region where \eqref{eq:unstable HP solution} is satisfied and with $3<d<D<7$, we have
\begin{equation*}
    \frac{F^{(D)}}{F^{(d)}}\sim \left(R_{\rm KK}m_{\infty}\right)^{d-D}\ll 1\,.
\end{equation*}
Hence, from the perspective of free energy, we also see that the aforementioned inequality indeed leads to a more thermodynamically stable higher-dimensional saddle. 
Additionally, we can also show that the free energy of HP solutions in $3<d<D<7$ are consistent with the our intuition of GL transitions for black holes. Recall the size of the HP solutions in these dimensions are proportional to $l\sim m_{\infty}^{-1}$. Then, the maximum free energy one can obtain from fitting higher-dimensional HP solutions in the extra dimensions is achieved by considering $(R_{\rm KK}m_{\infty})^{D-d}$ copies where the radius of the $(D-d)$-dimensional torus is denoted as $R_{\rm KK}$. To this end, the total free energy of this configuration is 
\begin{equation*}
    F\propto (R_{\rm KK}m_{\infty})^{D-d}M_{\pl,D}^{D-2}M_s^{-4}m_{\infty}^{7-D}\propto M_{\rm pl,d}^{d-2}M_s^{-4}m_{\infty}^{7-d}\propto F^{(d)}\,.
\end{equation*}
Thus, the free energy equation for higher- and lower-dimensional HP solutions are non-trivially consistent with each other to ensure the following identity.
\begin{align}
    F^{(d)}\sim \left(\frac{l_d}{l_D}\right)^nF^{(D)}\,,
\end{align}
where $l_d$ and $l_D$ are respectively the sizes of the lower and higher-dimensional solutions. This identity also holds for higher and lower-dimensional black holes which gives us more confidence in the existence of a GL instability for GL solutions. In particular, if $D\geq 7$, all of this is strongly suggestive of the existence of higher-dimensional string star solutions.

\subsection{Expected properties of the higher-dimensional counterparts to HP}
\label{sec:expected properties of counter HP}

The previous subsection hints that in spacetime where the number of non-compact dimensions is less than 7 but the total number of spacetime dimensions is greater than 7, there exists higher-dimensional saddles that are more stable than lower-dimensional HP solutions. In this section we assume such saddles exist and study their expected properties. As we see, the higher-dimensional counterparts of HP solutions differ from the lower-dimensional solutions in some key aspects.

Let us start with the free energy. The lower-dimensional solutions have vanishing free energy at the Hagedorn temperature. In other words, close to the Hagedorn temperature, the HP solutions have parametrically smaller free energies than the ones near the phase transition temperature to black holes $\Lambda_{\rm BH}$. Near the transition temperature, the free energy of the HP solutions is of the same order as the free energy of a black hole with temperature $\sim M_s$ which is
\begin{equation*}
    F_{\rm HP}(\Lambda_{\rm BH})^{(d)}\sim M_s^{d-3}M_{\pl,d}^{d-2}\,.
\end{equation*}
Now let us compare this free energy to the free energy of the HP solution at temperature $T$ such that $T-T_{\rm H}\ll M_s$. 
\begin{equation*}
    \frac{F_{\rm HP}^{(d)}(T)}{F_{\rm HP}^{(d)}(\Lambda_{\rm BH})}\sim \left(\frac{m_{\infty}}{M_s}\right)^{7-d}\ll 1\,,
\end{equation*}
where we recall our regime of validity for HP solutions \eqref{eq:HP regime of good control}. Again, this parametric difference arises in $d<7$. This parametric separation arises due to the temperature dependence of the HP free energy which vanishes in the limit $T\rightarrow T_{\rm H}$
\begin{align}
    F_{\rm HP}^{(d)}\propto (T-T_{\rm H})^{7-d}\,,
\end{align}
where we have substituted the definition of $m_{\infty}$ into the free energy.
In $d=7$, the free energy is independent of temperature as the exponent vanishes. The authors in \cite{Balthazar:2022hno} were able to find perturbative saddles in $7+\epsilon$ dimensions by utilizing the higher-order terms in the action which were previously ignored. Now, assuming the properties of black hole transitions proposed in \cite{Bedroya:2024uva} where we expect, in the weakly-coupled string limit of quantum gravity, black holes to transition into HP solutions, then one can see that the power-law behavior must break down for $d>7$ since the free energy cannot be divergent at the Hagedorn temperature. Otherwise, the saddle will have higher free energy than black holes and will no longer be more stable than the black hole saddle. The likely scenario is that the leading power-law behavior remains constant for $d>7$, so that the free energy does not diverge at the Hagedorn temperature and there is a continuous transition between $d>7$ and $d<7$. In particular, this means that for saddles in $d>7$ which the $d$-dimensional black holes transition into at high temperatures must exhibit the following properties:
\begin{itemize}
    \item they have non-vanishing free-energy at the Hagedorn temperature;
    \item the free energy at Hagedorn temperature is of the same order as the free energy of the smallest $d$-dimensional black holes.
\end{itemize}
\begin{figure}[H]
    \centering
\tikzset{every picture/.style={line width=0.75pt}} 

\begin{tikzpicture}[x=0.75pt,y=0.75pt,yscale=-.75,xscale=.8]

\draw    (38,418) -- (551,421.98) ;
\draw [shift={(553,422)}, rotate = 180.45] [color={rgb, 255:red, 0; green, 0; blue, 0 }  ][line width=0.75]    (10.93,-3.29) .. controls (6.95,-1.4) and (3.31,-0.3) .. (0,0) .. controls (3.31,0.3) and (6.95,1.4) .. (10.93,3.29)   ;
\draw    (68,448) -- (69,41) ;
\draw [shift={(69,39)}, rotate = 90.14] [color={rgb, 255:red, 0; green, 0; blue, 0 }  ][line width=0.75]    (10.93,-3.29) .. controls (6.95,-1.4) and (3.31,-0.3) .. (0,0) .. controls (3.31,0.3) and (6.95,1.4) .. (10.93,3.29)   ;
\draw [color={rgb, 255:red, 65; green, 117; blue, 5 }  ,draw opacity=1 ][line width=2.25]  [dash pattern={on 2.53pt off 3.02pt}]  (281,38) -- (282,419) ;
\draw [color={rgb, 255:red, 74; green, 144; blue, 226 }  ,draw opacity=1 ][line width=2.25]  [dash pattern={on 2.53pt off 3.02pt}]  (421,38) -- (421,419) ;
\draw [color={rgb, 255:red, 208; green, 2; blue, 27 }  ,draw opacity=1 ][line width=2.25]  [dash pattern={on 2.53pt off 3.02pt}]  (485,40) -- (487,419) ;
\draw [line width=2.25]    (99,72) .. controls (125,131) and (179.5,211) .. (280.5,270) ;
\draw  [draw opacity=0][fill={rgb, 255:red, 155; green, 155; blue, 155 }  ,fill opacity=0.77 ] (280.5,270) -- (420.5,270) -- (420.5,352) -- (280.5,352) -- cycle ;

\draw (57,11.4) node [anchor=north west][inner sep=0.75pt]    {$F$};
\draw (569,412.4) node [anchor=north west][inner sep=0.75pt]    {$T$};
\draw (265,426.4) node [anchor=north west][inner sep=0.75pt]  [color={rgb, 255:red, 65; green, 117; blue, 5 }  ,opacity=1 ]  {$\Lambda _{\rm BH}$};
\draw (411,426.4) node [anchor=north west][inner sep=0.75pt]  [color={rgb, 255:red, 74; green, 144; blue, 226 }  ,opacity=1 ]  {$T_{\rm H}$};
\draw (478,426.4) node [anchor=north west][inner sep=0.75pt]  [color={rgb, 255:red, 208; green, 2; blue, 27 }  ,opacity=1 ]  {$M_{\pl,d}$};
\draw (166,144.4) node [anchor=north west][inner sep=0.75pt]    {$F_{\text{BH}}$};
\draw (285,300) node [anchor=north west][inner sep=0.75pt]   [align=left] {$\displaystyle d >7$ solutions};

\end{tikzpicture}
    \caption{The free energy of higher-dimensional counterparts of HP solutions is expected to be in the grey region where the order of the free energy does not change. The black holes undergo a phase transition to these saddles at temperature $\Lambda_{BH}$.}
    \label{HDFE}
\end{figure}
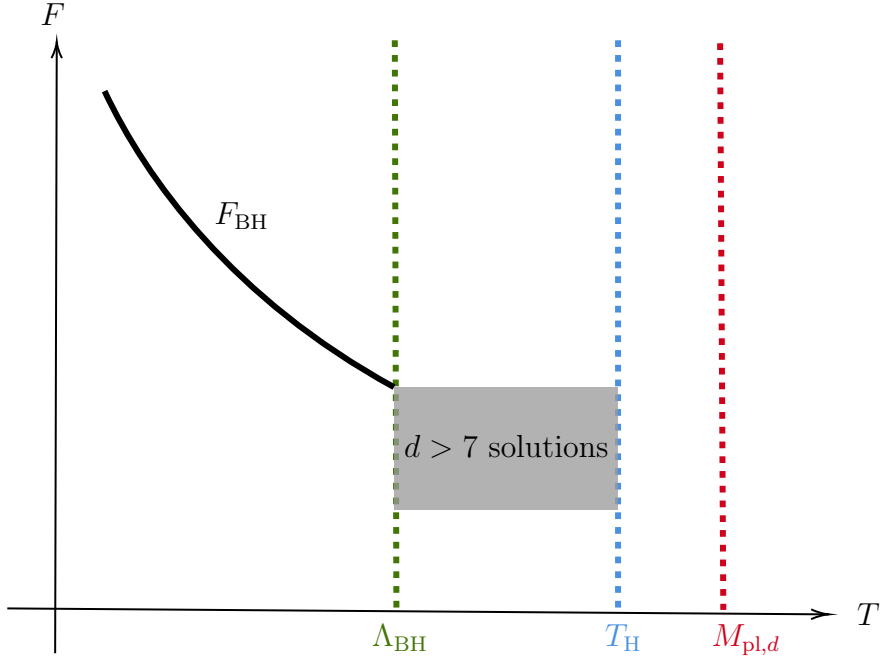

Interestingly, the saddles found in \cite{Balthazar:2022hno} for dimensions $d=7+\epsilon$ satisfy both of these properties. Now with the above listed properties of the saddle of interest, we can estimate the size of the solution. The HP solutions can have very large sizes as we get closer to the Hagedorn temperature. This is due to the fact that the solutions are getting diluted since the solution in the limit $T\rightarrow T_{\rm H}$ is identically vanishing. As the solution gets more diluted, it is less gravitationally bound. Therefore, it increases in size. In this case however, there is no parametric separation between the solution at the Hagedorn temperature and the solution at the point of transition with black holes. Therefore, we expect the sizes of these solution to also remain of the same order, which means they are all string-sized. In other words, the higher-dimensional saddles, are all truly stringy saddles. This sounds like bad news for a perturbative approach to finding these solutions since the $\alpha'$-corrections will become important for stringy solutions. We will address these concerns in section~\ref{sec:worldsheet description}. 
Now, with this, we can estimate the free energy of these solutions to be of the same order as those for a higher-dimensional black hole at string temperature
\begin{align}
    F\sim \frac{M_s}{g_s^2}\,,
\end{align}
where $g_s$ is the string coupling.

Finally, we can discuss the point of transition between these saddles and the lower-dimensional HP saddles. Suppose the transition between the $d$-dimensional HP solution and the $D$-dimensional saddle happens when the lower-dimensional HP solution has size $l_*$. (This is the point where the free energies match.) Then, we have the following relation
\begin{align*}
    M_sg_s^{-2}\sim l_*^{d-7}M_s^{D-6}R_{\rm KK}^{D-d}g_s^{-2}\,,
\end{align*}
where $R_{\rm KK}$ is the size of the compact dimensions. Therefore, we find
\begin{align}
    \frac{l_*}{l_s}\sim \left(\frac{R_{\rm KK}}{l_s}\right)^\frac{D-d}{7-d}\,.
\end{align}
Now let us compare this length scale to the length scale $R_{\rm GL}\sim R_{\rm KK}$ where the lower-dimensional HP solution becomes perturbatively unstable due to the GL instability. For this transition to be authentic, we need the lower-dimensional solution to be perturbatively stable. Therefore, we need $l_*\gtrsim R_{\rm KK}$. This is precisely achieved by $D-d\geq 7-d$ which occurs for $D\geq 7$.

Let us revisit the argument in section~\ref{sec:HP review} that conventional HP solutions cannot exist in dimensions $d\geq 7$ and see how the new solutions can avoid it. The argument showed that if one keeps the leading terms in the action, there is a proportionality relation between different terms which violates the positivity of the kinetic term for $d>7$. However, if the higher-order terms are considered and are relevant, the scaling argument no longer applies. This is how we can avoid the no-go arguments for these higher-dimensional saddles. To be more concrete, let us consider the following action (only included the relevant terms) near Hagedorn temperature, i.e., $m_{\infty}\ll M_s$,
\begin{equation}
\label{eq:action with general potential}
    I=\pi M_{\pl}^{d-2}R\int \dd^{d-1}x\sqrt{g}\left[\left(\nabla\varphi\right)^2+(\nabla\chi)^2+V_0(\varphi,\chi)+V_1(\varphi,\chi)\right]\,.
\end{equation}
Here, we restrict our attention to general higher-order non-derivative interactions between $\chi,\varphi$ which takes on the following form
\begin{equation*}
    V_0(\varphi,\chi,\chi^*)=\sum_{n,m\in\mathbb{Z_{+}}}a_{n,m}\varphi^n|\chi|^{2m}\,.
\end{equation*}
and $V_1$ is the leading contribution to the potential in $m_{\infty}^2$ which encodes all the temperature dependence. We now explain the relation between \( V_0 \) and \( V_1 \). The mass of the winding state is determined by the asymptotic value of the radius of the Euclidean time circle. We define the field \( \varphi \) such that its asymptotic value vanishes. However, one could alternatively redefine the variables so that the radius of the circle is given by \( R_H^2 \exp(2\varphi) \), in which case the asymptotic value of \( \varphi \) becomes
\[
\varphi_\infty = \ln\left(\frac{R}{R_H}\right) = \frac{1}{2} \ln\left(1 + \frac{m_\infty^2 \alpha'^2}{R_H^2}\right).
\]
This leads to the following identity, which simply follows from a change of variables:
\begin{align}
    \text{Bosonic/Type II}&:~~\mathcal{L}(\varphi,\chi,\chi^*;m_\infty) = \mathcal{L}\left(\varphi + \frac{1}{2} \ln\left(1 + \frac{m_\infty^2 \alpha'^2}{R_H^2}\right), \chi, \chi^*; 0 \right)\nonumber\\
    \text{Heterotic}&:~~\mathcal{L}(\varphi,\chi,\chi^*;m_\infty) = \mathcal{L}\left(\varphi + \frac{1}{2} \ln\left(\frac{\eta+\sqrt{\eta^2-\alpha'^2}}{2R_H^2}\right), \chi, \chi^*; 0 \right)\nonumber\\
    &~~~~~\eta=m_\infty^2\alpha'^2+R_H^2\alpha'+\frac{\alpha'^2}{4R_H^2}\,.
\end{align}
This implies that the mass dependence of the effective action at temperatures away from the Hagedorn temperature can be derived from its \( \varphi \)-dependence at the Hagedorn point. A leading-order expansion in \( m_\infty^2 \) then yields the identity 
\begin{equation}
\label{eq:general potential correction}
    V_1(\varphi,\chi,\chi^*;m_{\infty})=\left(\frac{m_{\infty}^2\alpha'}{\kappa}\right)\cdot\frac{\partial V_0(\varphi,\chi)}{\partial \varphi}\,.
\end{equation}

We defer a more detailed analysis of derivative interactions to future studies. Now, suppose we have an on-shell solution $\chi_*,\varphi_*$, then let us apply the scaling argument. Consider the rescaling from before, which we will rewrite here
\begin{equation*}
    \chi=\iota\chi_*(x/\gamma)\,,\qquad \varphi=\rho\varphi_*(x/\gamma)\,.
\end{equation*}
Applying this rescaling to the action \eqref{eq:action with general potential}, we have
\begin{equation}
    I=\rho^2\gamma^{d-3}I_{1,*}+\iota^2\gamma^{d-3}I_{2,*}+\gamma^{d-1}\int\dd^{d-1}x\sum_{n,m\in\mathbb{Z_+}}\left(\iota^{2m}\rho^na_{n,m}\varphi_*^n|\chi_*|^{2m}+\frac{m_{\infty}^2\alpha'{}^2}{2R_{\rm H}^2}\iota^{2m}\rho^{n-1}na_{n,m}\varphi_*^{n-1}|\chi_*|^{2m}\right)\,.
\end{equation}
where $I_1=\int\dd^{d-1}(\nabla\varphi)^2$ and $I_2=\int\dd^{d-1}x|\nabla\chi|^2$. As $\chi_*,\varphi_*$ are on-shell, we have the following three constraints
\begin{equation}\label{Var}
    \left.\left.\left.\frac{\partial I}{\partial \rho}\right\vert_{\iota=\rho=\gamma=1}=\frac{\partial I}{\partial \iota}\right\vert_{\iota=\rho=\gamma=1}=\frac{\partial I}{\partial \gamma}\right\vert_{\iota=\rho=\gamma=1}=0\,.
\end{equation}
Since the coefficients in the effective action are real and \( \varphi \) is real, any solution to the equations of motion will have \( \chi \) with a uniform phase. Consequently, \( \chi \) can be written as a constant phase multiplied by a real function. The situation would be different if we considered solutions with a source for the \( U(1) \) gauge symmetry under which \( \chi \) is charged. The gauge potential associated with this symmetry arises from the dimensional reduction of the NS--NS two-form gauge field on the Euclidean time circle.

However, since we are only interested in smooth solutions, we focus on neutral configurations. These can be made real via a field redefinition that absorbs the constant phase. Therefore, from this point onward, we assume that \( \chi \) is real. As a result, the real derivative \( \partial_\chi \) will replace \(\partial_{\chi^*}+ \partial_\chi\). For real solutions, the equations \eqref{Var} take the following form.
\begin{align*}
    2I_{1,*}+\int\dd^{d-1}x\varphi\frac{\partial V_0}{\partial\varphi}+\varphi\frac{\partial V_1}{\partial \varphi}&=0\,,\\
    2I_{2,*}+\int\dd^{d-1}x\chi\frac{\partial V_0}{\partial\chi}+\chi\frac{\partial V_1}{\partial \chi}&=0\,,\\
    (d-3)(I_{1,*}+I_{2,*})+(d-1)\int\dd^{d-1}x\left(V_0+V_1\right)&=0\,.
\end{align*}
These jointly give us the following constraint for the general polynomial potential
\begin{equation*}
    (d-3)\int\dd^{d-1}\left[\chi\frac{\partial V_0}{\partial \chi}+\varphi\frac{\partial V_0}{\partial \varphi}+\chi\frac{\partial V_1}{\partial \chi}+\varphi\frac{\partial V_1}{\partial \varphi}\right]-2(d-1)\int\dd^{d-1}x(V_0+V_1)=0\,.
\end{equation*}
Plugging in \eqref{eq:general potential correction}, we arrive at
\begin{equation}\label{SGI}
    m_{\infty}^2=\frac{\kappa}{\alpha'}\cdot \frac{\int\dd^{d-1}x(3-d)\left(\chi\partial_{\chi}V_0+\varphi\partial_{\varphi}V_0\right)+2(d-1)V_0}{\int\dd^{d-1}x(d-3)\left(\chi\partial_{\chi}\partial_{\varphi}V_0+\varphi\partial_{\varphi}^2V_0\right)-2(d-1)\partial_{\varphi} V_0}\geq0\,.
\end{equation}
Note that we assumed that $\mathcal{O}(m_\infty^4)$ terms are negligible, which is a valid assumption sufficiently close to the Hagedorn temperature.  Moreover, at the Hagedorn temperature, we have the condition that the general potential must satisfy
\begin{equation}
\label{eq:scaling solution}
    \int\dd^{d-1}x(3-d)\left(\chi\partial_{\chi}V_0+\varphi\partial_{\varphi}V_0\right)+2(d-1)V_0=0\,.
\end{equation}
The positivity bound \eqref{SGI} is powerful. For instance, if one considers only the cubic interaction \( V_0 = (\kappa/\alpha') \varphi \chi^2 \), the denominator on the right-hand side of equation \eqref{SGI} becomes negative, while the numerator is given by \( (7 - d) \int \dd^{d-1}x\, \varphi \chi^2 \). Since \( \varphi \) is always negative, this implies that the positivity bound \eqref{SGI} cannot be satisfied for \( d \geq 7 \), in agreement with the results of~\cite{Chen:2021dsw}. However, the positivity bound can indeed be satisfied for more general potentials that include higher-order interactions beyond the cubic term.

\section{In search of a worldsheet description in $d\geq 7$}
\label{sec:worldsheet description}

In the previous section we gave an argument that hints towards the existence of counterparts to HP solutions in spacetime dimensions greater than $6$, or higher-dimensional string stars. Moreover, as we argued, such solutions are expected to have finite non-zero free energy at Hagedorn temperature $T_{\rm H}$. This prompts us to study the existence of a solution at the Hagedorn temperature where the winding state becomes massless and the equations simplify. For that, we would first review an observation mentioned in \cite{Balthazar:2022szl} for the existence of an enhanced symmetry in the equations of motion at the Hagedorn temperature.

\subsection{$SU(2)$ symmetry at Hagedorn temperature}

Let us first review what typically happens when a massive vector field which is a string excitation becomes massless. In that case, the gauge symmetry is typically enhanced and the gauge symmetry is realized as a global symmetry on the worldsheet. To see why, consider the string worldsheet in the planar configuration as a defect in spacetime. A gauge transformation in the spacetime will have a non-zero action on the string worldsheet given by the anomaly inflow. However, since the gauge field only lives in the spacetime, the symmetry action on the worldsheet will be a global symmetry (see section II.7.3 of \cite{Agmon:2022thq} for a lightning review). Moreover, the generators of the global symmetry in the worldsheet CFT generate a current algebra and the generators are the vertex operators corresponding to the gauge bosons. For example, in Bosonic string theory at radius $R=l_s$, states with KK and winding numbers $\pm1$ with the appropriate string excitation, form vector bosons in the bulk. Given the non-zero winding and momentum, these states have non-zero left and right moving momenta in the compact direction
\begin{align}
\begin{split}\label{WSM}
    P_L&=\frac{n}{R}+\frac{wR}{\alpha'}\,,\\
    P_R&=\frac{n}{R}-\frac{wR}{\alpha'}\,.
\end{split}
\end{align}
Therefore, the corresponding vertex operators are 
\begin{align}
    \exp(iP_L\cdot X_L-iP_R\cdot X_R)=:\exp(\pm 2iX_L/l_s): \text{ and } :\exp(\pm 2iX_R/l_s):\,.
\end{align}
The $::$ represents the conventional normal ordering. In addition to the four vectors that become massless at the critical radius, there are two gauge bosons that are massless at every radius. These correspond to the dimensional reduction of the metric in the two T-dual frames. The vertex operators for these gauge bosons takes the form 
\begin{align}
    \frac{i}{l_s}\partial X_L\,,\qquad \frac{i}{l_s}\bar\partial X_R\,.
\end{align}
The operators we list below form a level one $SU(2)_L\times SU(2)_R$ current algebra on the worldsheet (see section~15.6.1 of \cite{DiFrancesco:1997nk}) 
\begin{alignat}{2}
\label{eq:generators of su2}
    &J^\pm=\exp(\pm 2iX_L/l_s)\,,\qquad  &&J^3=\frac{i}{l_s}\partial X_L\,,\nonumber\\
    &\bar J^\pm=\exp(\pm 2iX_R/l_s)\,,\qquad &&\bar J^3=\frac{i}{l_s}\bar \partial X_R\,.
\end{alignat}

This worldsheet symmetry is the manifestation of the enhanced gauge symmetry $SU(2)_L\times SU(2)_R$ in the bulk at the special radius $R=l_s$. The above current algebra exists for any radius and does not depend on the radius. However, at radius $R=l_s$ these operators are in the spectrum and are consistent with the quantization of the momentum lattice. At other radii, the operators are not mutually local with respect to the rest of the spectrum which means that the OPE's have branch cuts. 

Now let us go back to the thermal setup where a winding state becomes massless at a different radius $R=2l_s$. Note that this state is a scalar field and not a vector field. Therefore, there is no enhanced gauge symmetry in the spacetime and no current algebra in the spectrum of the CFT. For example, the operator $J^+$ corresponds to a state with $(P_L,P_R)=(2/l_s,0)$. If we set $R=2l_s$ in equation~\eqref{WSM}, we can see that the state corresponding to this vertex operator has even KK momentum and half winding. Since the winding number is not integer, this state does not belong to the spectrum. Alternatively, one can say that this state is not mutually local with states with odd KK momentum. The mutual locality is the condition that requires the integrality of the winding lattice. However, if we combine an even number of generators \eqref{eq:generators of su2}, they will belong to the spectrum. In fact, the vertex operator of the massless winding states ($n=0$, $w=\pm1$) are $J^+{\bar J} ^-$, $J^-{\bar J} ^+$.

The equations of motion for the fields that correspond to operators that are mutually local with respect to the generators \eqref{eq:generators of su2} will satisfy the $SU(2)_L\times SU(2)_R$ symmetry. These are states with even KK number. Therefore, if we turn off the odd KK backgrounds, the equations will have the $SU(2)_L\times SU(2)_R$ symmetry. However, this does not mean the solutions to those equations are also symmetric. The key observation made in \cite{Balthazar:2022szl} was that even though a background of radions and winding states will break the symmetry $SU(2)_L\times SU(2)_R$, the diagonal combination can still survive. The worldsheet Lagrangian in a background of $\chi$ and $\varphi$ is modified as 
\begin{align}\label{DAB}
  \delta \mathcal{L}\propto -2\sqrt{2}\varphi J^3\bar J^3+\chi J^+{\bar J}^-+\chi^* J^-{\bar J}^+\,.
\end{align}

Consider a background that satisfies $\chi =-\sqrt{2}\varphi$. For this background, the above deformation of the action will be invariant under the diagonal $SU(2)$  subgroup of $SU(2)_L\times SU(2)_R$. Therefore, there is an unbroken $SU(2)$ current algebra on the worldsheet in the background satisfying $\chi =-\sqrt{2}\varphi$. This $SU(2)$ symmetry is expected to remain unbroken under $\alpha'$ corrections. The spacetime equations of motion are generated by the vanishing of the $\beta$-function of a theory with $SU(2)$ symmetry. Therefore, the corrections must satisfy the $SU(2)$ symmetry as well. This does not guarantee the existence of an $SU(2)$ symmetric solution since such a symmetry can be spontaneously broken. However, we will narrow our search to such solutions which is to look for 2d CFTs at the Hagedorn temperature that preserve the $SU(2)$ symmetry. This assumption greatly simplifies the equations given that it relates the equations of motion of $\chi$ and $\varphi$ to each other by setting 
\begin{equation}
\label{eq:SU2 symmetry exchange}
    \chi=-\sqrt{2}\varphi\,.
\end{equation}

Similar to Bosonic string worldsheet, the type II worldsheet theory also has an unbroken $SU(2)$ symmetry given equation~\eqref{eq:SU2 symmetry exchange} is satisfied. This might sound surprising given that the Hagedorn radius for type II string is different from the Bosonic string. However, what makes the radius $\sqrt{2}l_s$ special is that a compact scalar at that radius is equivalent to two real fermions because of bosonization (section 12.6.1. of \cite{DiFrancesco:1997nk}).
\begin{align}
    &\psi_1+i\psi_2=e^{i\sqrt{2}X_L}\nonumber\\
    &\bar\psi_1+i\bar\psi_2=e^{i\sqrt{2}X_R}\,.
\end{align}
Suppose $\psi_3,\bar\psi_3$ are the worldsheet fermionic partners of $X_L, X_R$ in the RNS formalism. Then, the generators of a level two $SU(2)_L\times SU(2)_R$ Affine algebra are given by $J^a=\epsilon^{abc}\psi_b\psi_c$ and $\bar J^a=\epsilon^{abc}\bar\psi_b\bar\psi_c$ \cite{Balthazar:2022szl}. The deformed worldsheet action in the background of $\varphi$ and $\chi$ is the same as \eqref{DAB} with the new definitions for the current operators. Therefore, the condition \eqref{eq:SU2 symmetry exchange} again simplifies the equations and allows us to search for solutions with unbroken $SU(2)$ symmetries. 

Note that in the bosonic case, the ladder operators for the $SU(2)$ symmetry are given by 
\begin{align}
    J^{\pm} \propto \exp\left(\pm \frac{2iX_L}{l_s}\right),
\end{align}
which corresponds to $P_L = \pm 2$. In the type II case, the ladder operators take the form 
\begin{align}
    J^{\pm} \propto \psi_3 \exp\left(\pm i\frac{\sqrt{2}X_L}{l_s}\right),
\end{align}
which corresponds to $P_L = \pm \sqrt{2}$. Therefore, we conclude that if the left-(right-)moving part of the massless tachyon vertex operator at the Hagedorn temperature has $P_L(R) = \pm 2$, we obtain a level-1 $SU(2)$ symmetry, and if $P_L(R) = \pm \sqrt{2}$, we obtain a level-2 $SU(2)$ symmetry. 

Now, let us consider the case of the heterotic string. The Hagedorn radius in the heterotic case differs from both bosonic string theory and type II string theory. This might suggest that the $SU(2)$ symmetry does not exist at the Hagedorn temperature. However, the winding state is known to carry a half-integer momentum to satisfy level matching. Combining these two facts, we find a level-1 $SU(2)$ symmetry on the left-moving sector and a level-2 $SU(2)$ symmetry on the right-moving sector, which aligns with the heterotic string theory being a combination of a bosonic theory on the left-moving side and a type II theory on the right-moving side. 

The mass formula for a state with zero charge in the gauge charge lattice and zero excitation reads
\begin{align}
    \frac{m^2}{2} = \frac{1}{2} P_L^2 - 2 = \frac{1}{2} P_R^2 - 1.
\end{align}
We see that for the winding state to be massless, we require $P_L = 2$ and $P_R = \sqrt{2}$, which occurs when we substitute $w = 1$, $n = \frac{1}{2}$, and $R = (1 + 1/\sqrt{2}) l_s$ into \eqref{WSM}. Therefore, we conclude that there is a left-moving level-1 $SU(2)$ symmetry, similar to that of the bosonic string, and a right-moving level-2 $SU(2)$ symmetry, similar to that of the type II string. 

The symmetry generators are not mutually local with states whose KK number differs by an odd integer relative to the ground state, and therefore do not correspond to physical operators. Nonetheless, the diagonal $SU(2)\in SU(2)_L\times SU(2)_R$ can still be used to derive an $SU(2)$ symmetry at the level of the equations of motion for the winding state. When the diagonal $SU(2)$ symmetry acts on the Lagrangian \eqref{DAB}, it generates new interaction vertices. This implies that, in general, the $SU(2)$ symmetry is not preserved unless the equations of motion are extended to include the condensation of additional states. However, when the condition $\chi = -\sqrt{2}\varphi$ is imposed, the action of the $SU(2)$ symmetry maps the Lagrangian \eqref{DAB} to itself. Therefore, the $SU(2)$ symmetry is preserved for field configurations satisfying $\chi = -\sqrt{2}\varphi$.

In \cite{Balthazar:2022szl}, the authors showed that $SU(2)$ symmetric backgrounds with finite free energies exist if one infinitesimally changes the dimension to non integer values $7+\epsilon$. In the following, we show that $SU(2)$ symmetric backgrounds exist for every $d>6$ and moreover, they can be tuned to have finite free energy thanks to higher order corrections. 

\subsection{New perturbative worldsheet theories with infinite free energy}

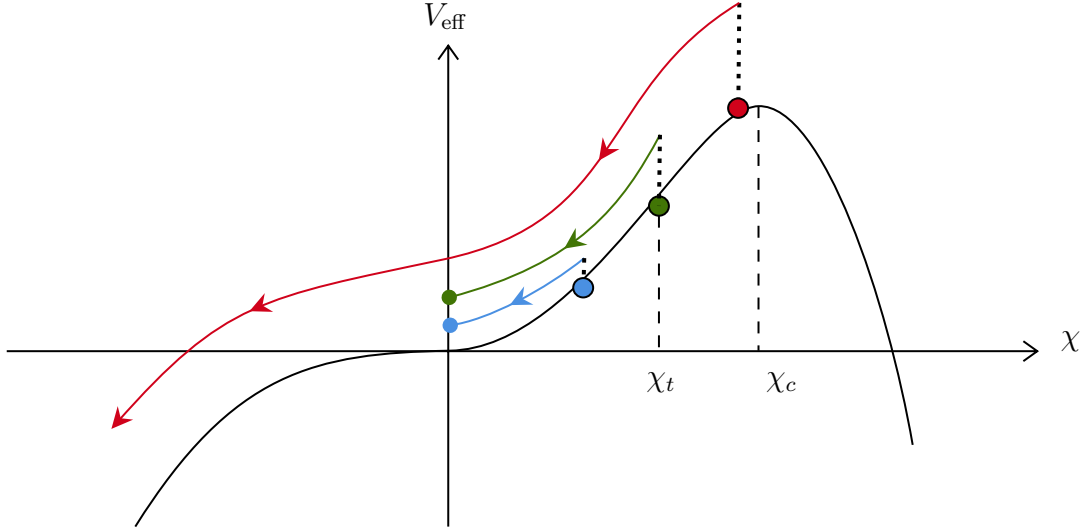
\begin{figure}
    \centering

\tikzset{every picture/.style={line width=0.75pt}} 

\begin{tikzpicture}[x=0.75pt,y=0.75pt,yscale=-1,xscale=1]

\draw [fill={rgb, 255:red, 208; green, 2; blue, 27 }  ,fill opacity=1 ][line width=1.5]  [dash pattern={on 1.69pt off 2.76pt}]  (441.52,16.28) -- (441,71) ;
\draw  (74,191) -- (591.02,191)(295.5,37.72) -- (295.5,279) (584.02,186) -- (591.02,191) -- (584.02,196) (290.5,44.72) -- (295.5,37.72) -- (300.5,44.72)  ;
\draw    (138.5,279) .. controls (187.5,201) and (228.5,192) .. (296.29,190.75) .. controls (364.09,189.51) and (418.74,69) .. (451.15,68.08) .. controls (483.56,67.17) and (516.01,166.34) .. (528.5,238) ;
\draw  [dash pattern={on 4.5pt off 4.5pt}]  (451.15,68.08) -- (451.15,190.75) ;
\draw [color={rgb, 255:red, 74; green, 144; blue, 226 }  ,draw opacity=1 ][line width=0.75]    (296.5,178) .. controls (315.23,175.57) and (344.56,159.89) .. (363.72,144.32) ;
\draw [shift={(326.33,167.89)}, rotate = 333.23] [fill={rgb, 255:red, 74; green, 144; blue, 226 }  ,fill opacity=1 ][line width=0.08]  [draw opacity=0] (10.72,-5.15) -- (0,0) -- (10.72,5.15) -- (7.12,0) -- cycle    ;
\draw [shift={(296.5,178)}, rotate = 352.61] [color={rgb, 255:red, 74; green, 144; blue, 226 }  ,draw opacity=1 ][fill={rgb, 255:red, 74; green, 144; blue, 226 }  ,fill opacity=1 ][line width=0.75]      (0, 0) circle [x radius= 3.35, y radius= 3.35]   ;
\draw [color={rgb, 255:red, 65; green, 117; blue, 5 }  ,draw opacity=1 ][line width=0.75]    (296,164) .. controls (361,147) and (382.99,117.38) .. (401.72,82.47) ;
\draw [shift={(353.65,139.56)}, rotate = 324.81] [fill={rgb, 255:red, 65; green, 117; blue, 5 }  ,fill opacity=1 ][line width=0.08]  [draw opacity=0] (10.72,-5.15) -- (0,0) -- (10.72,5.15) -- (7.12,0) -- cycle    ;
\draw [shift={(296,164)}, rotate = 345.34] [color={rgb, 255:red, 65; green, 117; blue, 5 }  ,draw opacity=1 ][fill={rgb, 255:red, 65; green, 117; blue, 5 }  ,fill opacity=1 ][line width=0.75]      (0, 0) circle [x radius= 3.35, y radius= 3.35]   ;
\draw [color={rgb, 255:red, 208; green, 2; blue, 27 }  ,draw opacity=1 ][line width=0.75]    (128.99,227.16) .. controls (183.06,165.95) and (201.97,164.14) .. (293,145) .. controls (385.41,125.57) and (369.98,59.66) .. (441.52,16.28) ;
\draw [shift={(195.77,170.94)}, rotate = 335.94] [fill={rgb, 255:red, 208; green, 2; blue, 27 }  ,fill opacity=1 ][line width=0.08]  [draw opacity=0] (10.72,-5.15) -- (0,0) -- (10.72,5.15) -- (7.12,0) -- cycle    ;
\draw [shift={(371.03,95.4)}, rotate = 305.16] [fill={rgb, 255:red, 208; green, 2; blue, 27 }  ,fill opacity=1 ][line width=0.08]  [draw opacity=0] (10.72,-5.15) -- (0,0) -- (10.72,5.15) -- (7.12,0) -- cycle    ;
\draw [shift={(126.5,230)}, rotate = 311.19] [fill={rgb, 255:red, 208; green, 2; blue, 27 }  ,fill opacity=1 ][line width=0.08]  [draw opacity=0] (10.72,-5.15) -- (0,0) -- (10.72,5.15) -- (7.12,0) -- cycle    ;
\draw  [dash pattern={on 4.5pt off 4.5pt}]  (401.2,123.04) -- (401.2,192.57) ;
\draw  [fill={rgb, 255:red, 208; green, 2; blue, 27 }  ,fill opacity=1 ] (435.91,69.08) .. controls (435.91,66.4) and (438.15,64.23) .. (440.91,64.23) .. controls (443.67,64.23) and (445.9,66.4) .. (445.9,69.08) .. controls (445.9,71.77) and (443.67,73.94) .. (440.91,73.94) .. controls (438.15,73.94) and (435.91,71.77) .. (435.91,69.08) -- cycle ;
\draw  [fill={rgb, 255:red, 65; green, 117; blue, 5 }  ,fill opacity=1 ] (396.2,118.18) .. controls (396.2,115.5) and (398.44,113.33) .. (401.2,113.33) .. controls (403.96,113.33) and (406.19,115.5) .. (406.19,118.18) .. controls (406.19,120.87) and (403.96,123.04) .. (401.2,123.04) .. controls (398.44,123.04) and (396.2,120.87) .. (396.2,118.18) -- cycle ;
\draw [fill={rgb, 255:red, 208; green, 2; blue, 27 }  ,fill opacity=1 ][line width=1.5]  [dash pattern={on 1.69pt off 2.76pt}]  (401.72,82.47) -- (401.2,118.18) ;
\draw [fill={rgb, 255:red, 208; green, 2; blue, 27 }  ,fill opacity=1 ][line width=1.5]  [dash pattern={on 1.69pt off 2.76pt}]  (363.72,144.32) -- (363.2,164.04) ;
\draw  [fill={rgb, 255:red, 74; green, 144; blue, 226 }  ,fill opacity=1 ] (358.2,159.18) .. controls (358.2,156.5) and (360.44,154.33) .. (363.2,154.33) .. controls (365.96,154.33) and (368.19,156.5) .. (368.19,159.18) .. controls (368.19,161.87) and (365.96,164.04) .. (363.2,164.04) .. controls (360.44,164.04) and (358.2,161.87) .. (358.2,159.18) -- cycle ;

\draw (601.26,179.51) node [anchor=north west][inner sep=0.75pt]    {$\chi $};
\draw (281.54,14.76) node [anchor=north west][inner sep=0.75pt]    {$V_{\text{eff}}$};
\draw (453.52,200) node [anchor=north west][inner sep=0.75pt]    {$\chi _{c}$};
\draw (393.45,200) node [anchor=north west][inner sep=0.75pt]    {$\chi _{t}$};

\end{tikzpicture}
    \caption{The equation of motion for $\chi(r)$ mimic the motion of a particle in the potential $V_{\rm eff}$ with a time dependent friction term. \textbf{Red: } If the particle starts with too much energy the friction will not be enough to stop the motion at $\chi=0$. \textbf{Blue: } Particles with small initial energy on the other hand, manage to avoid rolling off. These are bounded solutions. \textbf{Green: } At the point of transition between the two behaviors, we find a solution that has the highest energy and the fastest descent among the bounded solutions.}
    \label{fig:potential}
\end{figure}

In this section, we show how the higher order corrections to the equations of motion lead to new string star solutions in $d>7$ spacetime dimensions. From the arguments outlined in sec.~\ref{sec:expected properties of counter HP}, we expect to find a non-zero solution at the Hagedorn temperature. Therefore, we use the $SU(2)$ symmetry enhancement reviewed in the previous section and narrow our search to such $SU(2)$ preserving saddles which satisfy $\chi=-\sqrt{2}\varphi$. We expect the string star to be a solution to the equations of motion, and also have finite free energy. In this section, we present a new one-parameter family of perturbative backgrounds that exist in any spacetime dimension $d>7$. However, these backgrounds do not generically have finite free energy. In the next subsection, we provide  evidence that a particular solution in this one-parameter family indeed has finite free energy and therefore corresponds to a higher-dimensional string star.

Let us focus on the interactions up to quartic order in $\chi,\chi^*,\varphi$. The cubic interaction was previously computed in \cite{Horowitz:1997jc} while the quartic interactions have been computed in \cite{Schulgin:2011zb} for the heterotic string theories and in \cite{Brustein:2021ifl} for the bosonic and type II string theories. They, respectively, take on the following explicit form
\begin{align}
\begin{split}
    I_{\rm het}&=\pi M_{\pl}^{d-2}R\int \dd^{d-1}x\sqrt{g}e^{-2\phi}\left[-\cR-4(\nabla\phi)^2+(\nabla\varphi)^2+|\nabla\chi|^2\right.\\
    &\qquad\qquad\qquad\qquad\qquad \left.+\left(m_{\infty}^2+\frac{\kappa}{\alpha'}\varphi+\frac{3\sqrt{2}\kappa}{4\alpha'}\varphi^2\right)|\chi|^2+\frac{3\sqrt{2}\kappa}{16\alpha'}|\chi|^4\right]\,,
\end{split}\\
\begin{split}
    I_{\text{bos, type II}}&=\pi M_{\pl}^{d-2}R\int \dd^{d-1}x\sqrt{g}e^{-2\phi}\left[-\cR-4(\nabla\phi)^2+(\nabla\varphi)^2+|\nabla\chi|^2\right.\\
    &\qquad\qquad\qquad\qquad\qquad \left.+\left(m_{\infty}^2+\frac{\kappa}{\alpha'}\varphi+\frac{\kappa}{\alpha'}\varphi^2\right)|\chi|^2+\frac{\kappa}{4\alpha'}|\chi|^4\right]\,.
\end{split}
\end{align}
Hence, as expected from our previous subsection, after imposing the condition $\chi=-\sqrt{2}\varphi$, the equation of motion for $\varphi$ and $\chi$ become proportional for all string theories. Therefore, we only need to find a bounded solution for $\varphi$ or $\chi$. We further impose spherical symmetry for simplicity. Then, the equation of motion for $\chi$ takes the form
\begin{equation}\label{eom1}
    \nabla^2\chi=\frac{\kappa}{\alpha'}\left(-\frac{1}{\sqrt{2}}\chi^2+\tilde{\kappa}\chi^3\right)\,,
\end{equation}
where we have introduced the following constant
\[\tilde{\kappa}=
\begin{cases}
1\,,&\text{Bosonic and Type II}\,,\\
\frac{3\sqrt{2}}{4}\,,&\text{Heterotic}\,.
\end{cases}
\]

The above equation~\eqref{eom1} can be written as
\begin{align}\label{eq:EoM}
    \partial_r^2\chi+\frac{d-2}{r}\partial_r\chi=-V'_{\rm eff}(\chi)\,,
\end{align}
where,
\begin{equation}\label{QTP}
    V_{\rm eff}(\chi)=\frac{\kappa}{\alpha'}\left(\frac{1}{3\sqrt{2}}\chi^3-\frac{\tilde{\kappa}}{4}\chi^4\right)+\mathcal{O}(\chi^5)\,,
\end{equation}
We are interested in differentiable ($\partial_r \chi(0)=0$) and bounded solutions. The only remaining condition needed to solve for $\chi(r)$ everywhere is $\chi(r=0)$. The ``potential'' $V_{\rm eff}$ is illustrated in fig.~\ref{fig:potential} where the analogy between the behavior of the (spherically symmetric) winding mode as a function of the radius $r$ and the position of a particle moving in the potential $V_{\rm eff}$ as a function of time can be made. Hence, the second term on the left in equation~\eqref{eq:EoM}, arising from the Laplacian defined on $S_r^{d-1}$, can be viewed as a friction term. Without the friction term, any solution with a small initial condition for $\chi(0)$ would eventually roll off to negative $\chi$ values leading and will not necessarily be bounded from below by zero. We do not have a global picture of the potential, i.e., the knowledge of arbitrarily high $\alpha$'-corrections to the action, to understand what happens to these solutions at large radii. However, in our present considerations which involves interactions up to quartic order between $\chi,\varphi$, real solutions that cross the $\chi=0$ threshold runaway to $\chi=-\infty$. Hence, here on after, we will refer to the solutions that become negative somewhere in space as unbounded solutions. In contrast to these solutions, if the friction term is sufficiently strong, we can hope that the solution will asymptote to $\chi=0$ as $r\rightarrow \infty$. This is exactly what happens when $d>7$. As one can see in the equation~\eqref{eq:EoM}, $d$ controls the strength of the friction term. To understand this ODE better, it is helpful to redefine our variables. As was noted in \cite{Mckane:1978me}, the equation~\eqref{eq:EoM} is nothing other than the (generalized) Emden--Fowler equations and can be rewritten as

\begin{equation}
\label{eq:EoM change of variables}
    \partial_x^2\mathcal{X}+(d-7)\partial_x\mathcal{X}=2(d-5)\mathcal{X}-\frac{1}{\sqrt{2}}\mathcal{X}^2+\tilde{\kappa}e^{-2x}\mathcal{X}^3+\mathcal{O}(\mathcal{X}^4)\,.
\end{equation}
where $\mathcal{X}=e^{2x}\chi$ and $r\sqrt{\kappa/\alpha'}=e^x$. Let us first only consider the first two terms on the right. This corresponds to only keeping the cubic interactions which is a good approximation for solutions $\chi\ll1$. For such solutions, the above equation can be written as 
\begin{equation}\label{EOM2}
    \partial_x^2\mathcal{X}+(d-7)\partial_x\mathcal{X}=-\mathcal{V}_{\rm eff}'(\mathcal{X})\,,
\end{equation}
where
\begin{align}
\mathcal{V}_{\rm eff}(\mathcal{X})=-(d-5)\mathcal{X}^2+\frac{1}{3\sqrt{2}}\mathcal{X}^3\,.
\end{align}

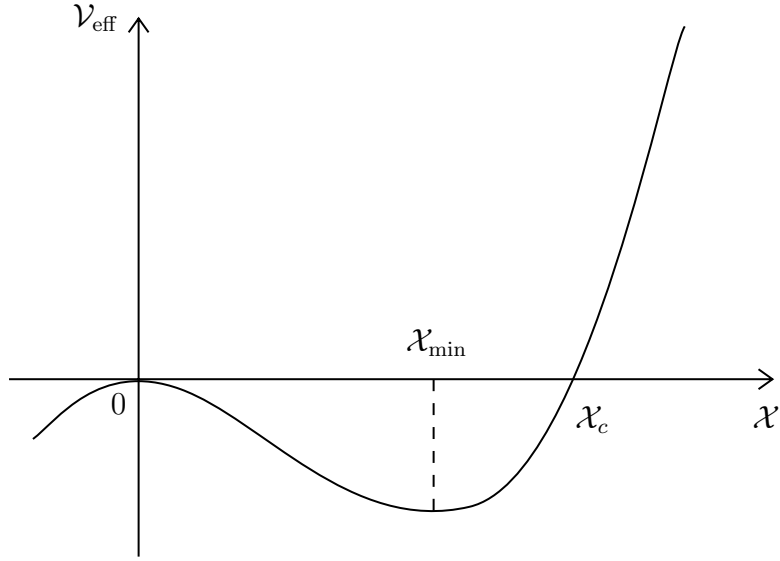
\begin{figure}[H]
\label{VEFF}
\centering
\tikzset{every picture/.style={line width=0.75pt}} 

\begin{tikzpicture}[x=0.75pt,y=0.75pt,yscale=-1,xscale=1]

\draw [line width=0.75]  (137.5,198) -- (520.5,198)(202.5,17) -- (202.5,287) (513.5,193) -- (520.5,198) -- (513.5,203) (197.5,24) -- (202.5,17) -- (207.5,24)  ;
\draw [line width=0.75]    (149.5,228) .. controls (161.26,219.18) and (182.5,188) .. (223.5,203) .. controls (264.5,218) and (308.5,276) .. (368.5,262) .. controls (428.5,248) and (467.5,35) .. (476.5,21) ;
\draw  [dash pattern={on 4.5pt off 4.5pt}]  (350.5,198) -- (350.5,266) ;

\draw (168,8.4) node [anchor=north west][inner sep=0.75pt]    {$\mathcal{V} _{\text{eff}}$};
\draw (509,210) node [anchor=north west][inner sep=0.75pt]    {$\mathcal{X} $};
\draw (187,203.4) node [anchor=north west][inner sep=0.75pt]    {$0$};
\draw (420,210) node [anchor=north west][inner sep=0.75pt]    {$\mathcal{X} _{c}$};
\draw (335,171.4) node [anchor=north west][inner sep=0.75pt]    {$\mathcal{X} _{\min}$};

\end{tikzpicture}
\caption{The effective potential determining the ODE for the redefined variable $\mathcal{X}=\chi r^2\kappa/\alpha'$.}
\end{figure}

All of the solutions of interest have the boundary condition $\lim_{x\rightarrow-\infty}\mathcal{X}=0$. They all roll away from the local maximum of $\mathcal{V}$ at $\mathcal{X}=0$ toward positive $\mathcal{X}$. 
The solutions start from the asymptotic boundary condition $\mathcal{X}=0$ at $x\rightarrow -\infty$ and roll away toward positive $\mathcal{X}$. In spacetime dimensions $5<d<7$, there is an anti friction term in the equation which will pump energy into the solution causing it to overshoot the bump at $\mathcal{X}=0$ at $x\rightarrow \infty$. Therefore, there is no bounded solution at Hagedorn temperature in spacetime dimensions $d<7$.

For spacetime dimensions $d>7$, the friction term takes away energy from the solution causing it to localize in the valley at $\mathcal{X}=\mathcal{X}_{\min}$. Recalling the relation between $\mathcal{X}$ and $\chi(r)$, we have
\begin{align}\label{CIO}
    \lim_{x\rightarrow\infty }\mathcal{X}=\mathcal{X}_{\min}\qquad \Rightarrow \qquad \text{at large $r$:  }\chi(r)\sim \frac{1}{r^2}\,.
\end{align}
Therefore, for any sufficiently small value of $\chi(0)$, there exist a bounded solution that decays as $1/r^2$ in the large radius limit.

A natural question is what happens if we keep increasing the initial condition $\chi(0)$? For large values of $\chi(0)$, the higher-order $\alpha'$ corrections will become important. In Appendix \ref{sec:unbounded}, we show that if the effective potential $V_{\rm eff}$ defined in \eqref{eq:EoM} has a local maximum or the higher-order correction lead to any polynomial behavior $\sim \chi^n$ with $n\neq 3$, solutions with sufficiently large $\chi(0)$ will be unbounded. For such potentials, there is always an upper bound $\chi_t$ for the initial condition $\chi(0)$ that extends to a bounded and decaying  solution at $r\rightarrow\infty$. We will come back to this upper bound later in the next subsection. We conclude that in $d>7$ there are bounded solutions which are parametrized by the boundary condition $\chi(0)\in [0,\chi_t)$. These solutions are always under perturbative control at large radius, however, for $\chi(0)$ that are $\mathcal{O}(1)$, the higher-order $\alpha'$ corrections will become important at small radii. For sufficiently large $\chi(0)$ ($\chi(0)>\chi_t$), the solution will be unbounded. We do not have a global picture of the potential, i.e., the knowledge of arbitrarily high $\alpha$'-corrections to the action, to understand what happens to these solutions at large radii. However, in our present considerations which involves interactions up to quartic order between $\chi,\varphi$, real solutions that cross the $\chi=0$ threshold runaway to $\chi=-\infty$. Hence, here on after, we will refer to the solutions that become negative somewhere in space as unbounded solutions. We will return to the special solution $\chi(0)=\chi_t$ that corresponds to the transition between the two behaviors in the next subsection.

In spacetime dimension $d=7$, there is no friction term and energy is conserved. Therefore, the solution will asymptote to $\mathcal{X}=0^+$ as $x\rightarrow\infty$. In fact, the exact solution for $\chi$ is known \cite{Balthazar:2022szl,Mckane:1978me} and is given by 
\begin{equation}
\label{eq:7d exact cubic solution}
    \chi(r)=\frac{\chi(0)}{\left(1+\frac{\kappa}{24\sqrt{2}\alpha'}\chi(0)r^2\right)^2}\,.
\end{equation}

To conclude, we found a class of solutions for $d>7$ that are parametrized by $\chi(0)$ and are bounded for sufficiently small $\chi(0)$. Note that these solutions are qualitatively different from the solution \eqref{eq:7d exact cubic solution} found in \cite{Balthazar:2022szl}. In particular, the large-radius behavior of the solutions we find are typically given by $\sim 1/r^2$. Let us have a closer look at all the possible large-radius behaviors for bounded solutions to the equations of motion. Suppose that the behavior at long distance for $\chi(r)$ takes on the form of $a\cdot r^{b}$. Then, plugging this ans\"atz into the equations of motion \eqref{eq:EoM}, we obtain the following solutions
\begin{equation}
\label{eq:solutions to EoM}
   r\gg 1:~~ a\cdot b\left(b+d-3\right)r^{b-2}\simeq-\frac{\kappa}{\alpha'}\frac{a^2}{\sqrt{2}}r^{2b}\qquad \rightarrow\qquad \chi(r)=\begin{cases}
   \frac{2\sqrt{2}(d-5)\alpha'}{\kappa}r^{-2}\,, & b=-2\,\\
   \text{or}\,\\
    a\cdot r^{3-d}\,,& b=3-d\,.
    \end{cases}
\end{equation}
In the top case, the two sides are of the same order and cancel each other out ($b-2=2b$). In the bottom case, however, $r^{b-2}$ is parametrically larger than $r^{2b}$ and, therefore, its coefficient is set to zero. The second option ($b=3-d$) can only arise in spacetime dimensions $d>5$ where $b-2>2b$. Both asymptotic behaviors satisfy $b<1$ and, thus, $\chi(r)\ll 1$ in the large-$r$ regime. Note that the above options are the possible bounded asymptotic behavior, however, they are not guaranteed to be realized for a solution that is differentiable at $r=0$. However, if such solutions can be extended to a differentiable solution at $r=0$, they would be under perturbative control at large radius given that $\lim_{r\rightarrow\infty}\chi=\lim_{r\rightarrow\infty}\phi=0$.

The new bounded solutions that we found in the beginning of this subsection for $d>7$ are compatible with the classification of \eqref{eq:solutions to EoM} since they decay as $r^{-2}$.

\begin{figure}
    \centering
    \begin{subfigure}{.49\textwidth}
        \centering
        \includegraphics[width=\linewidth]{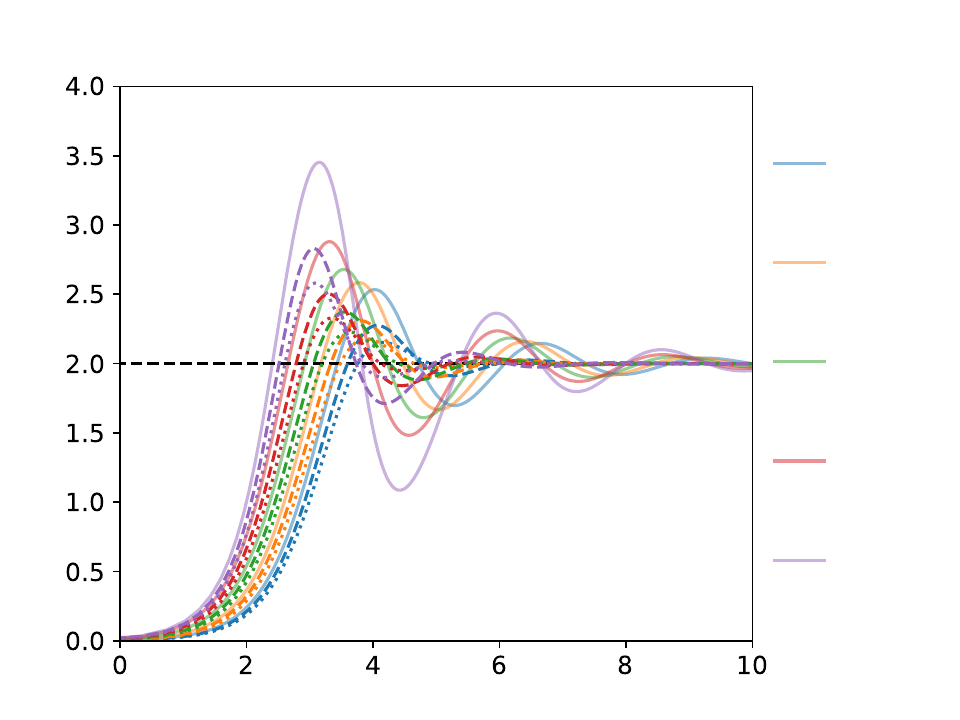}
            \begin{picture}(0,0)\vspace*{-1.2cm}
            \put(75,165){\footnotesize $\ln\chi(0)$}
            \put(90,150){\footnotesize $-3$}
            \put(90,125){\footnotesize $-2.5$}
            \put(90,101){\footnotesize $-2$}
            \put(90,76){\footnotesize $-1.5$}
            \put(90,52){\footnotesize $-1$}
            \put(-120,190){\footnotesize $-\frac{\dd \ln(\chi(r))}{\dd \ln(r)}$}
            \put(-20,10){\footnotesize $\ln(r)$}
            \end{picture}\vspace*{-0.5cm}
            \caption{Bosonic/Type II}
    \end{subfigure}
    \hfill
    \begin{subfigure}{.49\textwidth}
        \centering
        \includegraphics[width=\linewidth]{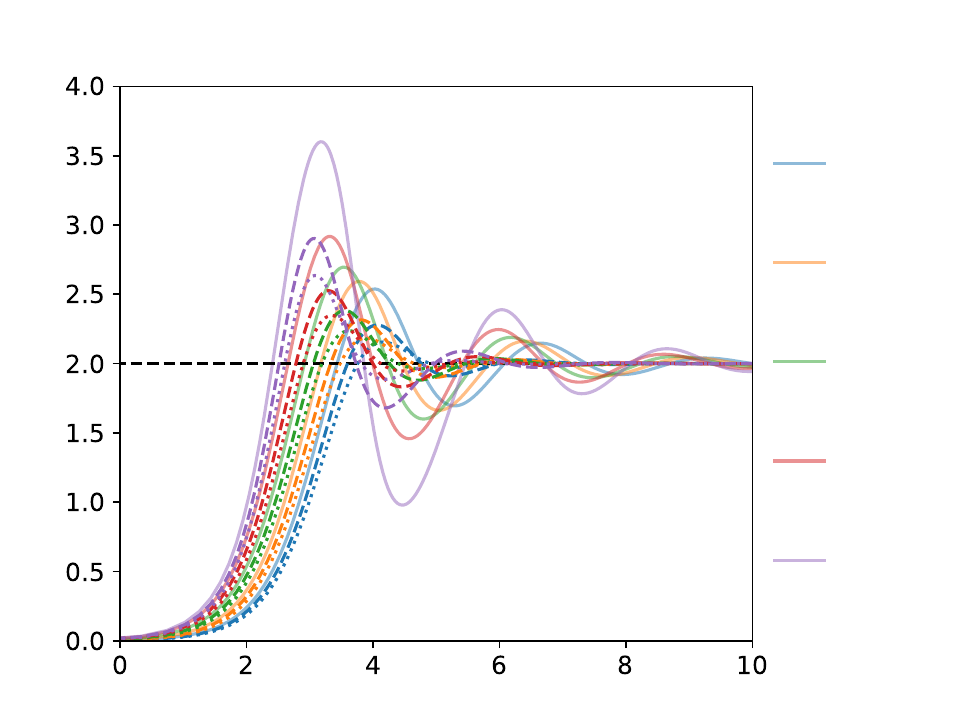}
            \begin{picture}(0,0)\vspace*{-1.2cm}
            \put(75,165){\footnotesize $\ln\chi(0)$}
            \put(90,150){\footnotesize $-3$}
            \put(90,125){\footnotesize $-2.5$}
            \put(90,101){\footnotesize $-2$}
            \put(90,76){\footnotesize $-1.5$}
            \put(90,52){\footnotesize $-1$}
            \put(-120,190){\footnotesize $-\frac{\dd \ln(\chi(r))}{\dd \ln(r)}$}
            \put(-20,10){\footnotesize $\ln(r)$}
            \end{picture}\vspace*{-0.5cm}
            \caption{Heterotic}
    \end{subfigure}
    \caption{The large-distance behavior of perturbative solutions to \eqref{eq:EoM} in various dimensions. Here, the solid, dashed, dotted line respectively represent the behavior in $d=8,9,10$ with the indicated boundary condition. Furthermore, $b$ denotes the negative exponent of $\chi(r)$.}
    \label{fig:large-distance behavior of pert sol}
\end{figure}
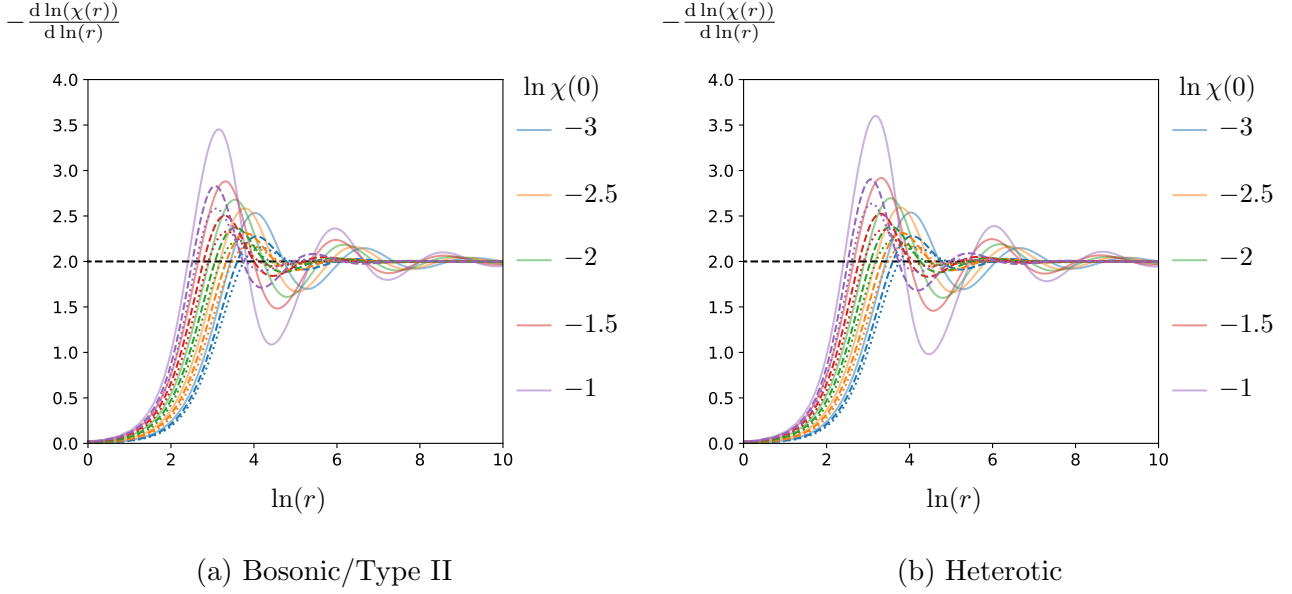

Finding an analytic solution to the equations of motion globally is a difficult task. But nevertheless, it is illuminating to check the behavior of the new solution by numerically integrating the above second-order ODE as a function of the Dirichlet boundary condition at the origin of space $\chi(r=0)$, with the Neumann boundary condition $\partial_r\chi\vert_{r=0}=0$ fixed. This can be implemented straightforwardly with, e.g., solving with the Runge--Kutta method via \verb+scipy.integrate+ in \verb+python+. The results, for small boundary conditions at the origin, i.e., $\chi(r=0)\in[e^{-3},e^{-1}]$, across dimensions are shown in fig.~\ref{fig:large-distance behavior of pert sol}. 

There are two note-worthy features of the numerical results. First, for these small boundary conditions imposed at the origin for $\chi(r)$, we can see that at large distance, all of these solutions display a $r^{-2}$ behavior for sufficiently large $r$ which is the behavior as predicted by the argument in the beginning of this subsection. Note that $-\log(\chi)/\log(r)$ oscillates before converging to $2$, and as we increase the boundary condition $\chi(r=0)$, the amplitude of this oscillation grows. This observation will become relevant in the next subsection. 

It is natural to ask what is the physics behind these solutions, i.e., what are some thermodynamic properties of these bounded profiles of the winding mode and the radion. To this end, we can compute the free energy of these solutions as
\begin{equation*}
    F=\frac{I_{\rm EFT}}{\beta}\,.
\end{equation*}

The action simplifies for the new solutions at the Hagedorn temperature. Firstly, there is no mass term since the winding state becomes massless at the Hagedorn temperature. Secondly, the radion and the winding state are proportional ($\chi=-\sqrt{2}\varphi$) which allows us to write the action in terms of $\chi$ alone. Thirdly, the solutions are spherically symmetric. These properties allow us to express the free energy as follows.
\begin{equation}
\label{eq:free energy at Hagedorn}
    F\vert_{T=T_{\rm H}}=
    \frac{\omega_{d-2}M_{\rm pl}^{d-2}}{2}\int \dd r\, r^{d-2}\left[\frac32\left(\partial_r\chi(r)\right)^2+\frac{\kappa}{\alpha'}\left(-\frac{1}{\sqrt{2}}\chi(r)^3+\frac{3}{4}\tilde{\kappa}\chi(r)^4+\mathcal{O}(\chi^5)\right)\right]\,.
\end{equation}
We would like to know if the free energy is finite. Since our solution behaves as $r^{-2}$ at large radii, the convergence of the solution is determined by the kinetic term and the cubic term which become leading at large radii. Now, suppose that we are in the large-$r$ regime of space. Then, the dominating terms in this regime in the above integral scale as
\begin{equation*}
    \propto \left[6-2(d-5)\right]\int_{\delta r}^{\infty} \dd r r^{d-2}r^{-6}\,,
\end{equation*}
for some sufficiently large value of $\delta r$. The above integral is always divergent when $d\geq 7$. However, the prefactor vanishes for $d=8$. Therefore, in the case of $d=8$, we have to look at the subleading contribution. We can assume that the analytic expression for $\chi(r)$ for large $r$ can be approximated with the following perturbative correction
\begin{equation}
\label{eq:approx chi large r}
    \chi(r)=\frac{2\sqrt{2}(d-5)\alpha'}{\kappa}r^{-2}+\delta\chi(r)\,,
\end{equation}
where we assume $\delta\chi(r)\ll 1$. Plugging this ans\"atz back into the equation of motion, we obtain
\begin{equation}\label{eq:perteom}
    \partial_r^2\delta\chi(r)+\frac{d-2}{r}\partial_r\delta \chi(r)=-\frac{\kappa}{\alpha'}\frac{2a}{\sqrt{2}}\delta\chi(r)\,\qquad\rightarrow\qquad  c^2+c(d-3)+4(d-5)=0\,.
\end{equation}
Thus, we have the subleading correction to $\chi(r)$ is
\begin{equation}
\label{eq:subleading}
    \delta\chi(r)\simeq a r^{(3 - d)/2-\sqrt{89-22d+d^2}}+a^* r^{(3 - d)/2+\sqrt{89-22d+d^2}}\,,
\end{equation}
Here, \( a\) is a complex parameter, and the quantity \( 89 - 22d + d^2 \) is negative for \( 5 < d < 17 \), which includes the dimensions of interest in this work—namely, \( d = 8, 9, 10 \). Therefore, the corresponding term is oscillatory. Note that the equation \eqref{eq:perteom} assumes the quartic term in the action remains negligible even at subleading order. This assumption can be checked \emph{a posteriori}. The quartic interaction contributes to the equation of motion as \( \chi(r)^3 \sim r^{-6} \). For this term to be negligible relative to the oscillatory term, we require \( 6 > (d - 3)/2 \), which holds for all \( d < 15 \). Thus, for any spacetime dimension \( d < 15 \)—which includes all unitary superstring theories—the subleading behavior is well approximated by the oscillatory term \eqref{eq:subleading}.

If we substitute this subleading correction back into equation~\eqref{eq:free energy at Hagedorn} to compute the action, we find that the leading-order contribution to the integrand scales as \( 1/\sqrt{r} \). Consequently, the action still diverges in \( d = 8 \).

We conclude that the bounded solutions that we found in dimensions $d>7$ always have infinite free energy due to their $r^{-2}$ at large radii (in combination with the subleading corrections when relevant). These solutions correspond to boundary condition $\chi(0)<\chi_t$, where $\chi_t$ separates bounded solutions from unbounded solutions. The solution with $\chi(0)=\chi_t$ can have a different asymptotic behavior. Let us see if that solution can possibly have finite free energy. As we showed in equation~\eqref{eq:solutions to EoM}, the only large-$r$ behaviors permitted by the equations of motion are $\chi(r) \sim r^{-2}, r^{3-d}$. In fact, the $r^{3-d}$ solution would have finite free energy given that both the kinetic term and the mass term in the Euclidean action will converge for dimensions of interest ($d>7$), i.e.,
\begin{align}
&\int d^{d-1} \chi^2 \sim \int_{\delta r}^\infty dr r^{d-2} r^{6-d}<\infty\,, \nonumber\\
&\int d^{d-1} (\partial\chi)^2 \sim \int_{\delta r}^\infty dr r^{d-2} r^{4-d}<\infty\,.
\end{align}
The mass term is absent at the Hagedorn temperature but is present at temperatures close to the Hagedorn temperature. Therefore, the remaining question is, could increasing the initial condition $\chi(0)$ to $\chi_t$ change the decay power from $-2$ to $3-d$? In the next subsection, we argue that the answer is yes!

\subsection{Evidence for a worldsheet theory with finite free energy}
\label{sec:evidence}

Now, let us examine what can happen when we increase the Dirichlet boundary condition at $r=0$, namely $\chi(r=0)$. In other words, we will now proceed to continue beyond the perturbative regime in which $\chi(r=0)\ll 1$. For larger values of $\chi(0)$, we have to consider higher-order corrections to the equations of motion. For now, we focus on the potential that captures all dynamics up to quartic interactions between the winding mode and the radion. We only do this to demonstrate a point about the nature of the solutions, but as we explain, we expect our findings to be generic and hold for the fully corrected potential. We will elaborate on this important point at the end of this subsection.

\begin{figure}[htp!]
    \centering
    \hspace*{-.5in}
    \begin{subfigure}{.5\textwidth}
        \centering
        \includegraphics[width=\linewidth]{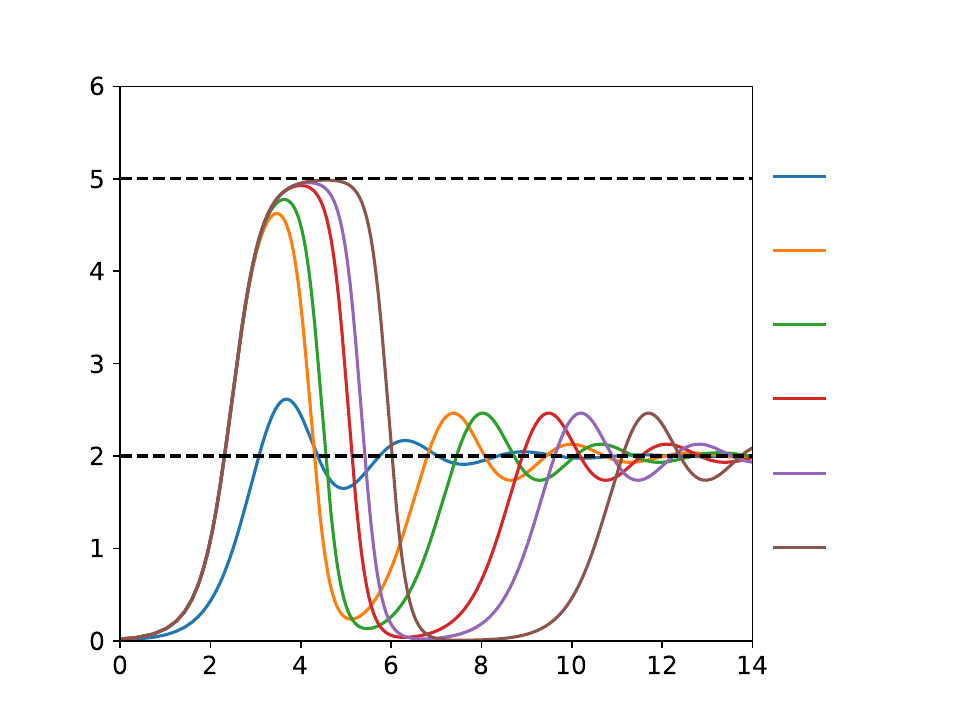}
        \begin{picture}(0,0)\vspace*{-1.2cm}
        \put(83,165){\footnotesize $\chi(0)$}
        \put(90,148){\footnotesize $0.1$}
        \put(90,130){\footnotesize $0.46$}
        \put(90,112){\footnotesize $0.462$}
        \put(90,93){\footnotesize $0.4627$}
        \put(90,74){\footnotesize $0.46273$}
        \put(90,55){\footnotesize $0.462739$}
        \put(-143,100){\footnotesize $-\frac{\dd\ln(\chi(r))}{\dd\ln(r)}$}
        \put(-20,10){\footnotesize $\ln(r)$}
        \end{picture}\vspace*{-0.6cm}
        \caption{$d=8$}
    \end{subfigure}
    \hfill
    \begin{subfigure}{.5\textwidth}
        \centering
        \includegraphics[width=\linewidth]{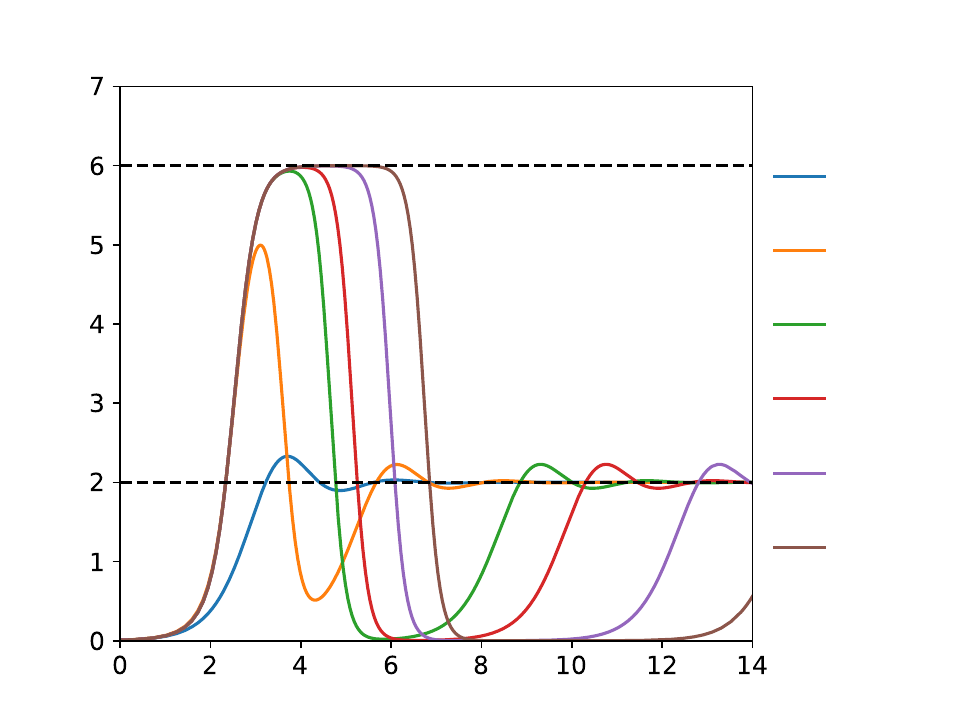}
        \begin{picture}(0,0)\vspace*{-1.2cm}
        \put(83,165){\footnotesize $\chi(0)$}
        \put(90,148){\footnotesize $0.1$}
        \put(90,130){\footnotesize $0.6$}
        \put(90,112){\footnotesize $0.609$}
        \put(90,93){\footnotesize $0.609614$}
        \put(90,74){\footnotesize $0.60961481$}
        \put(90,55){\footnotesize $0.6096148154$}
        \put(-140,100){\footnotesize $-\frac{\dd\ln(\chi(r))}{\dd\ln(r)}$}
        \put(-20,10){\footnotesize $\ln(r)$}
        \end{picture}\vspace*{-0.6cm}
        \caption{$d=9$}
    \end{subfigure}
    \hfill
    \begin{subfigure}{.5\textwidth}
        \centering
        \includegraphics[width=\linewidth]{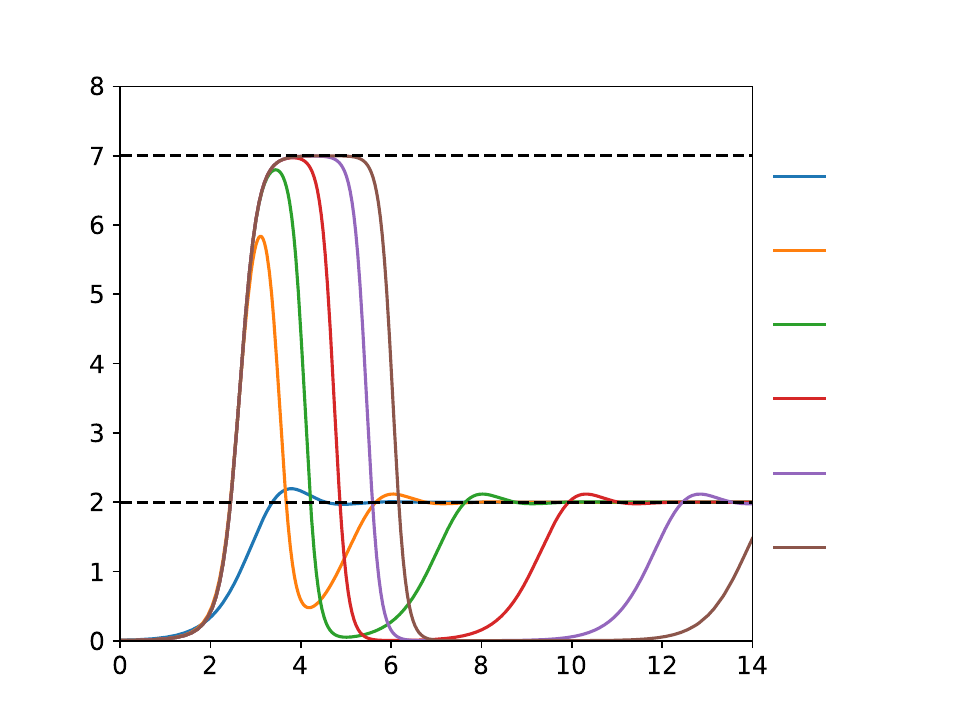}
        \begin{picture}(0,0)\vspace*{-1.2cm}
        \put(83,165){\footnotesize $\chi(0)$}
        \put(90,148){\footnotesize $0.1$}
        \put(90,130){\footnotesize $0.66$}
        \put(90,112){\footnotesize $0.6647$}
        \put(90,93){\footnotesize $0.664784$}
        \put(90,74){\footnotesize $0.66478486$}
        \put(90,55){\footnotesize $0.6647848655$}
        \put(-145,100){\footnotesize $-\frac{\dd\ln(\chi(r))}{\dd\ln(r)}$}
        \put(-20,10){\footnotesize $\ln(r)$}
        \end{picture}\vspace*{-0.6cm}
        \caption{$d=10$}
    \end{subfigure}\vspace{-0.2cm}
    \caption{The large-distance behavior of bounded, non-normalizable solutions in various dimensions for Bosonic/Type II string theories.}
    \label{fig:large-distance behavior}
\end{figure}
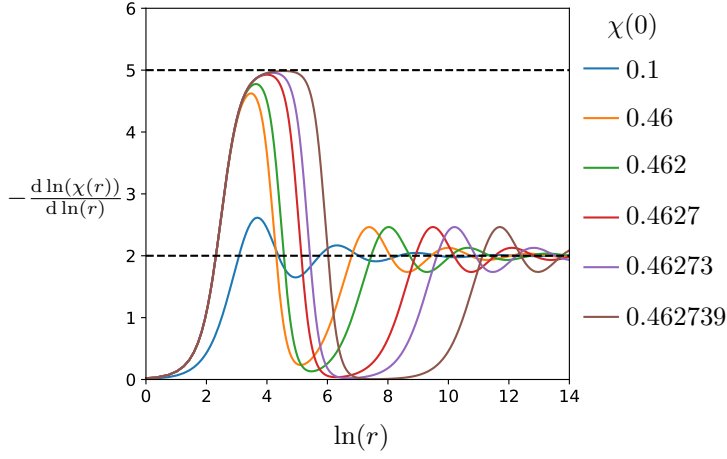
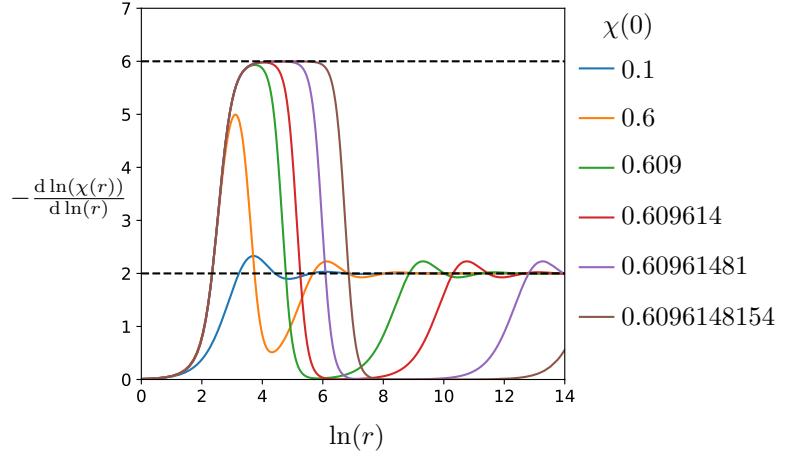
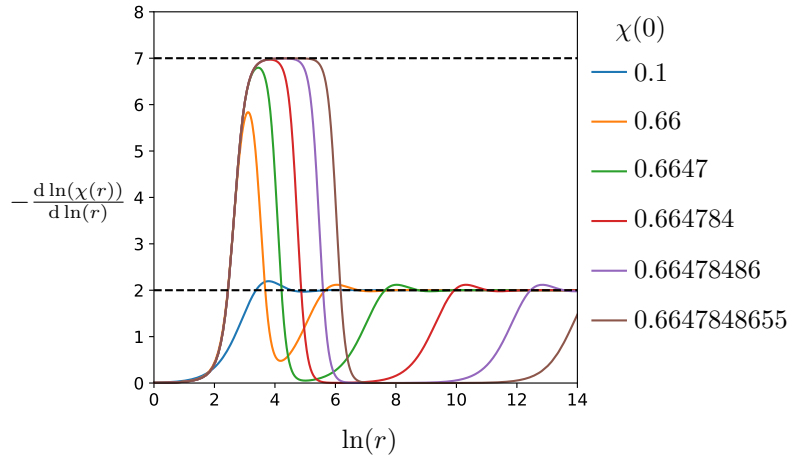

The numerical results for larger $\chi(r=0)$ in different dimensions are shown in fig.~\ref{fig:large-distance behavior} for Bosonic and Type II string theories. The case of heterotic string theory is qualitatively similar to that of Bosonic and Type II. Similar to the previous small-$\chi(r=0)$ regime, we notice that the $\chi(r)\sim r^{-2}$ behavior persists even as we increase the value of $\chi(0)$. However, an interesting feature that arises across dimensions is that as we increase $\chi(0)$, solutions behave as $r^{3-d}$ for some region in space. As we explained in the previous subsection, if the asymptotic behavior of these solution remained as $\sim r^{3-d}$, they would have a finite free energy. However, all of these solutions eventually revert back to the $\sim r^{-2}$ behavior. Nonetheless, this is suggestive that if we chose the largest boundary value that would give us a bounded solution, $\chi(0)=\chi_t$ (fig.~\ref{fig:potential}), the region of space where $\chi\sim r^{3-d}$ might extend to infinity and the solution become not only bounded, but also normalizable. The numerical solutions suggest that this is exactly what happens.

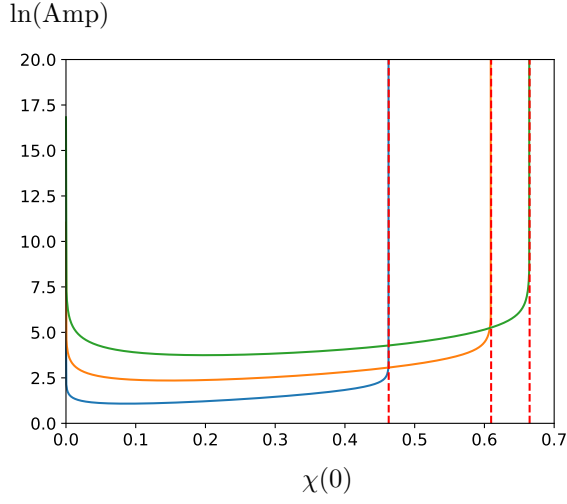
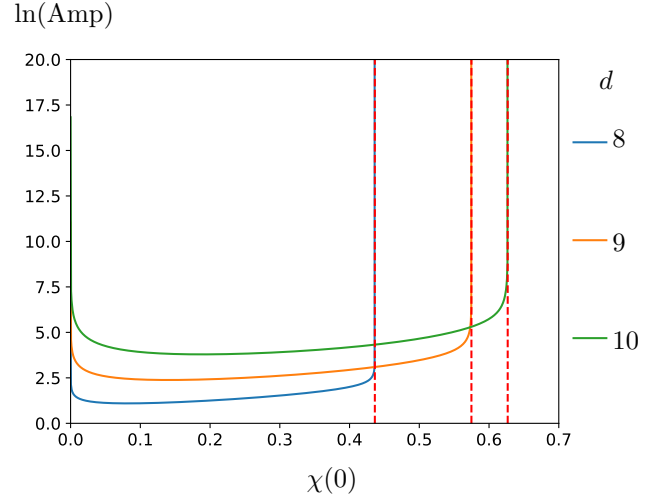
\begin{figure}
    \centering
    \begin{subfigure}{.49\textwidth}
    \centering
    \includegraphics[width=\linewidth]{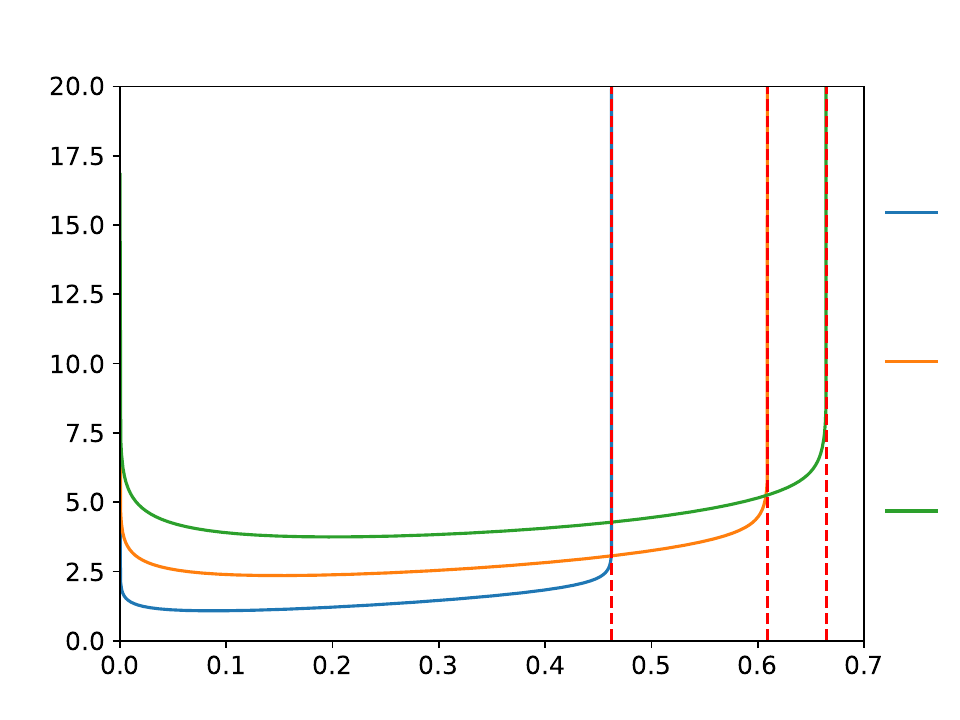}
        \begin{picture}(0,0)\vspace*{-1.2cm}
        \put(110,160){\footnotesize $d$}
        \put(115,138){\footnotesize $8$}
        \put(115,99){\footnotesize $9$}
        \put(115,62){\footnotesize $10$}
        \put(-110,185){\footnotesize $\ln(\mathrm{Amp})$}
        \put(0,10){\footnotesize $\chi(0)$}
        \end{picture}\vspace*{-0.5cm}
        \caption{Bosonic/Type II}
    \end{subfigure}
    \hfill
    \begin{subfigure}{.49\textwidth}
        \centering
        \includegraphics[width=\linewidth]{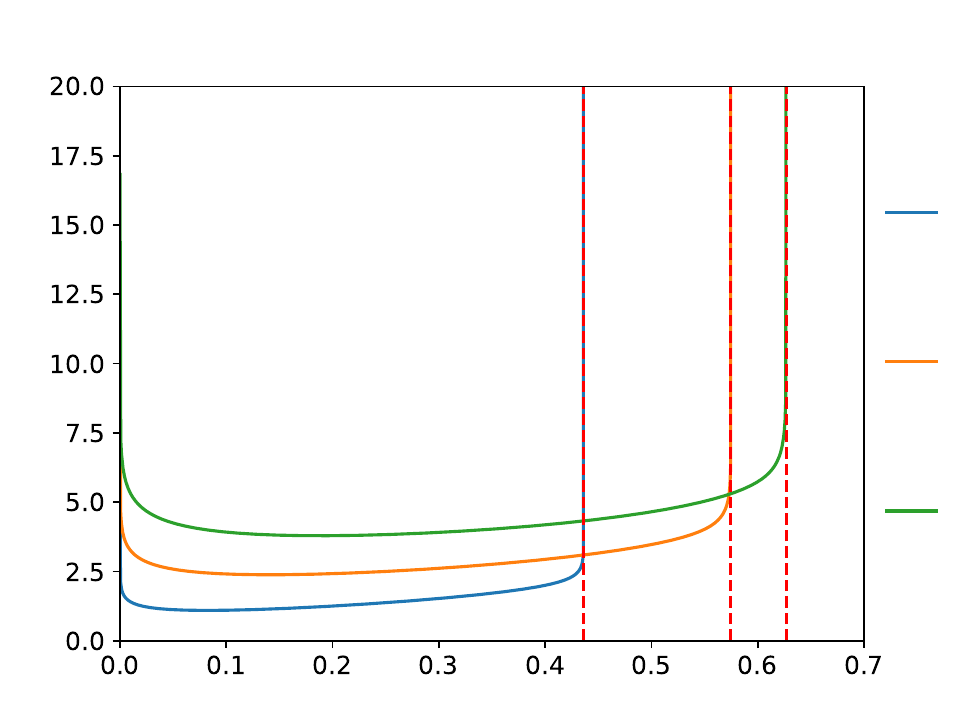}
        \begin{picture}(0,0)\vspace*{-1.2cm}
        \put(110,160){\footnotesize $d$}
        \put(115,138){\footnotesize $8$}
        \put(115,99){\footnotesize $9$}
        \put(115,62){\footnotesize $10$}
        \put(-110,185){\footnotesize $\ln(\mathrm{Amp})$}
        \put(0,10){\footnotesize $\chi(0)$}
        \end{picture}\vspace*{-0.5cm}
        \caption{Heterotic}
    \end{subfigure}
    \caption{The coefficient of the subleading large-radius oscillatory term $|a|$ for the bounded solutions in various dimensions. Here, the red dashed vertical line indicates the location of divergence with respect to each dimension which coincide with the transition values $\chi_t$.}
    \label{fig:asymptotic divergence}
\end{figure}

\begin{figure}[htp!]
    \centering
    \begin{subfigure}{.49\textwidth}
        \centering
        \includegraphics[width=\linewidth]{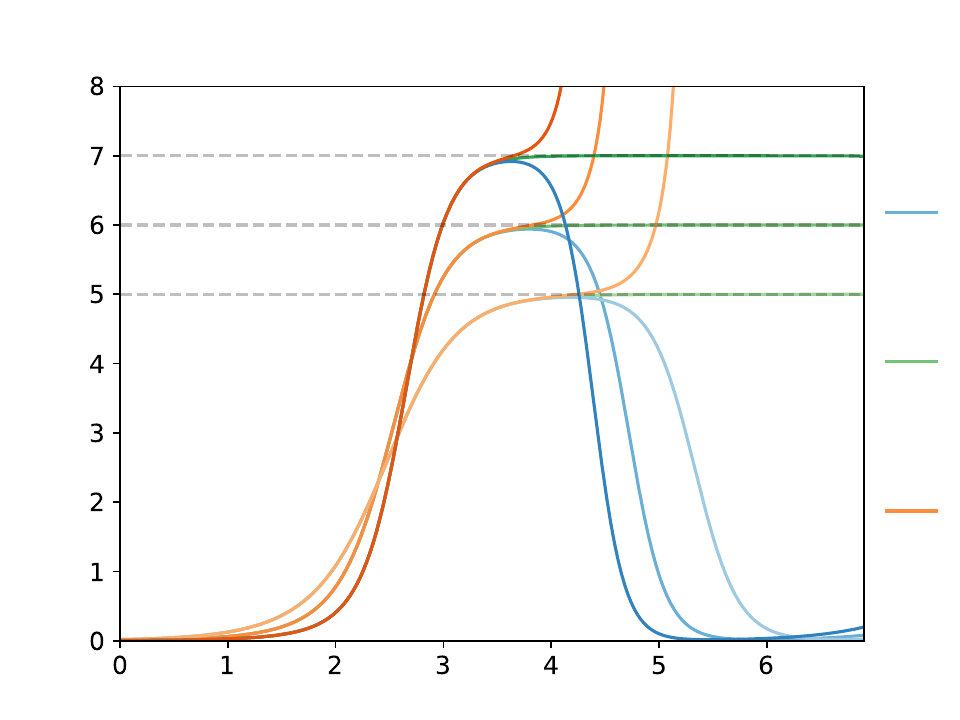}
            \caption{Bosonic/Type II}
    \end{subfigure}
    \hfill
    \begin{subfigure}{.49\textwidth}
        \centering
        \includegraphics[width=\linewidth]{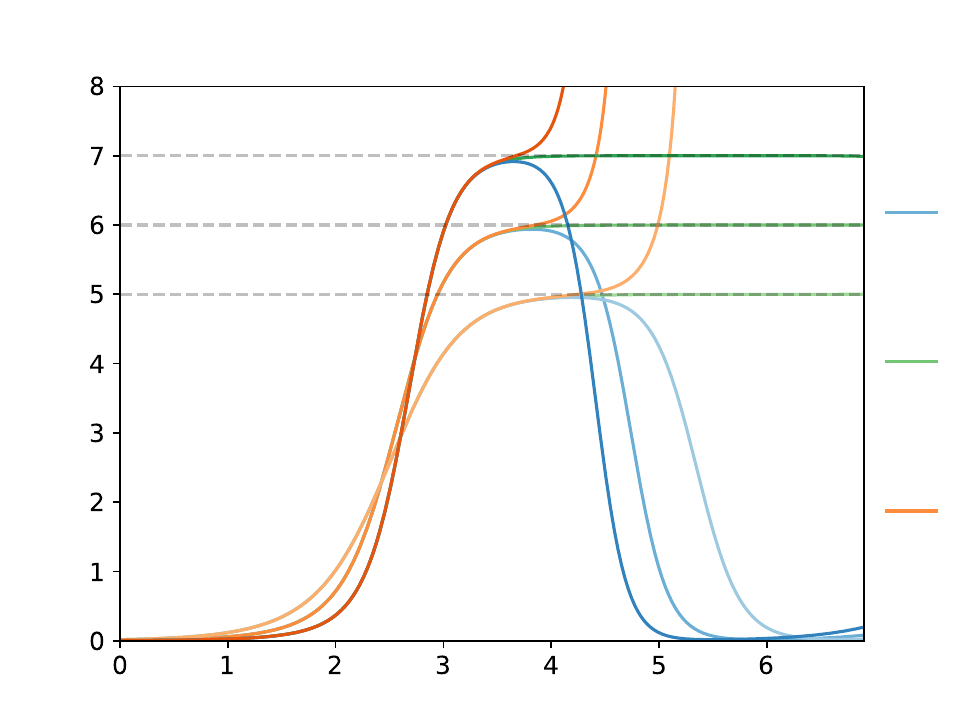}
            \begin{picture}(0,0)\vspace*{-1.2cm}
        \put(105,160){\footnotesize $\chi(0)$}
        \put(115,139){\footnotesize $\chi_t-\varepsilon$}
        \put(115,102){\footnotesize $\chi_t$}
        \put(115,65){\footnotesize $\chi_t+\varepsilon$}
        \put(-120,190){\footnotesize $-\frac{\dd\ln(\chi(r))}{\dd\ln(r)}$}
        \put(0,10){\footnotesize $\ln(r)$}

        \put(-143,160){\footnotesize $\chi(0)$}
        \put(-133,139){\footnotesize $\chi_t-\varepsilon$}
        \put(-133,102){\footnotesize $\chi_t$}
        \put(-133,65){\footnotesize $\chi_t+\varepsilon$}
        \put(-368,190){\footnotesize $-\frac{\dd\ln(\chi(r))}{\dd\ln(r)}$}
        \put(-248,10){\footnotesize $\ln(r)$}
        \end{picture}\vspace*{-0.5cm}
            \caption{Heterotic}
    \end{subfigure}
    \caption{The behavior of the exponent where $\chi\sim r^{-b}$ as a function of $r$ near the critical transitional value $\chi_t$ where $\varepsilon=10^{-5}$. From the lightest to the darkest shades, these each correspond to solutions to the EoM in $d=8,9,10$, respectively. The numerical value of $\chi_t$ differs across dimensions and their specific values are recorded in table~\ref{tab:all chi_t's}.
    The faded black dashed line indicates the exponent we expect from the fundamental solution to the $(d-1)$ dimensional Laplacian in each dimension.}
    \label{fig:chit}
\end{figure}

The numerical value of $-\frac{\dd \ln(\chi)}{\dd \ln(r)}$ has been plotted for various values of $\chi(0)$ in $d=8,9$, and $10$ in fig.~\ref{fig:chit}. We observe that there exist a critical boundary condition $\chi_t$ in each dimension such that for $\chi(0)$ values below $\chi_t$, $\lim_{r\to+\infty}\ln(\chi)/\ln(r)=-2$, the solution is bounded yet non-normalizable. For $\chi(0)$ values above $\chi_t$, the solution is unbounded. However, for the special solution $\chi(0)=\chi_t$, the solution is both bounded and normalizable as it decays as $\sim r^{3-d}$. The critical values of $\chi_t$ are shown in table~\ref{tab:all chi_t's}. Note that although these values are not parametrically small, they are all less than one, which allows possibility that higher-order corrections $\mathcal{O}(\chi^6)$ and other derivative corrections are subleading.

\begin{table}[htp!]
    \centering
    \begin{tabular}{c|c|c}
        $d$ &  bosonic/type II & heterotic\\
        \hline
        8 & 0.46273947 & 0.43627487 \\
        9 & 0.60961481 & 0.57475035 \\
        10 & 0.66478487 & 0.62676516
    \end{tabular}
    \caption{The value of $\chi_t$ in various dimensions for all string theories.}
    \label{tab:all chi_t's}
\end{table}

Let us understand the transition between the two asymptotic behaviors $\sim r^{-2}$ and $r^{3-d}$ better. As we analytically showed in the previous subsection, the two leading terms for the solutions with $\chi(0)<\chi_t$ are
\begin{align}
    \chi(r)\simeq \frac{2\sqrt{2}(d-5)\alpha'}{\kappa}r^{-2}+|a|\left[\frac{a}{|a|} r^{(3 - d)/2-\sqrt{89-22d+d^2}}+\frac{a^*}{|a|} r^{(3 - d)/2+\sqrt{89-22d+d^2}}\right]\,.
\end{align}
The coefficient of the $r^{-2}$ term is fixed due to the non-linearity of the equation of motion for $\chi(r)$. Therefore, changing $\chi(0)$ and choosing values closer to $\chi_t$ does not make the $r^{-2}$ contribution smaller. The other possibility to change the leading order behavior is that the coefficient of the second term, which is oscillatory, increases for smaller values of $\varepsilon=\chi_t-\chi(0)$ and diverges in the limit $\varepsilon\rightarrow 0$. For any fixed radius $r=R$, there is an $\varepsilon_R>0$ such that for $\chi(0)>\chi_t-\epsilon_R$, the $r^{-2}$ term is no longer dominant at radius $r=R$. The order of limits here is important. At any fixed $\varepsilon>0$, the large-$r$ limit is controlled by $r^{-2}$, however, if we first take the limit of $\varepsilon\rightarrow 0$ and then the limit $r\rightarrow\infty$, the leading term is controlled by an $\sim r^{3-d}$ term. If this is correct, we must see a divergence of the coefficient $|a(\varepsilon)|$  in the limit of $\chi(0)\rightarrow\chi_t$. We numerically verified that this is indeed what happens (see fig.~\ref{fig:asymptotic divergence}).

\begin{figure}
    \centering
    \begin{subfigure}{\linewidth}
    \centering
    \begin{subfigure}{.475\textwidth}
    \centering
    \includegraphics[width=\linewidth]{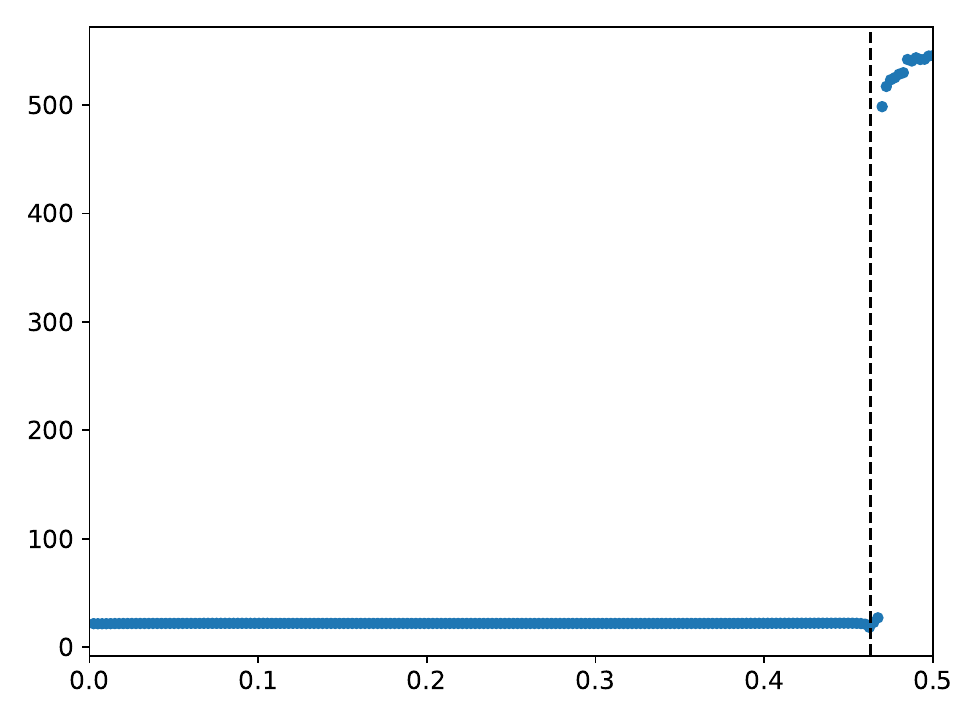}
    \begin{picture}(0,0)\vspace*{-1.2cm}
    \put(-10,10){\footnotesize $\chi(0)$}
    \put(90,10){\footnotesize $\chi_t$}
    \put(-140,100){\footnotesize $\ln(I_r)$}
    \put(-50,60){\includegraphics[width=0.5\linewidth]{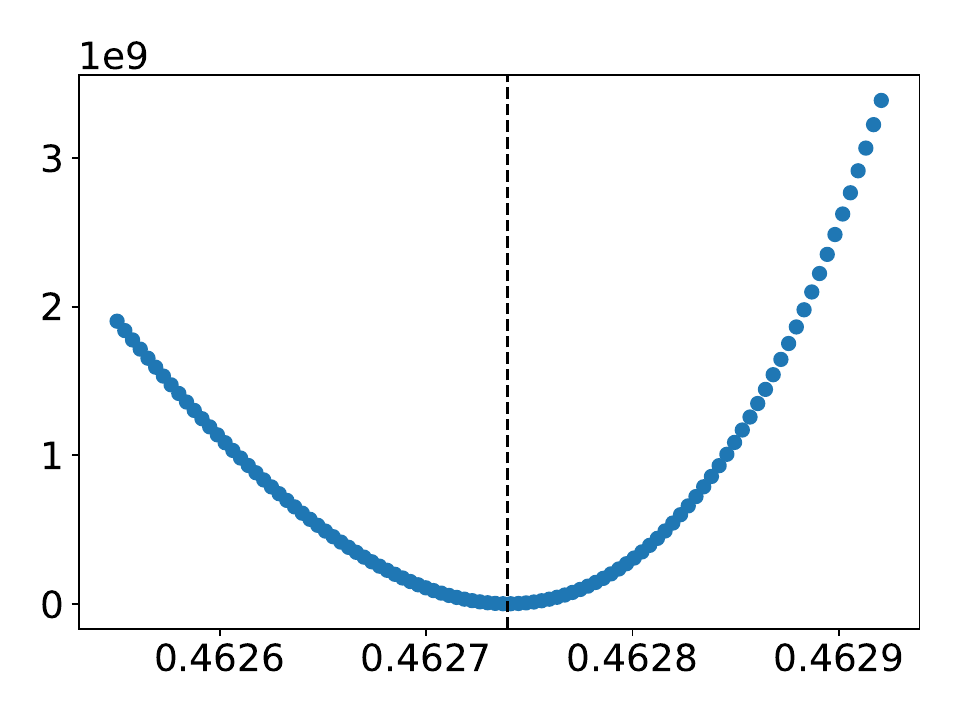}}
    \put(-60,103){\footnotesize $I_r$}
    \put(-35,63){\line(6,-1){125}}
    \put(60,135){\line(1,-3){30}}
    \end{picture}\vspace*{-0.75cm}
    \end{subfigure}
    \hfill
    \begin{subfigure}{.475\textwidth}
    \centering
    \includegraphics[width=\linewidth]{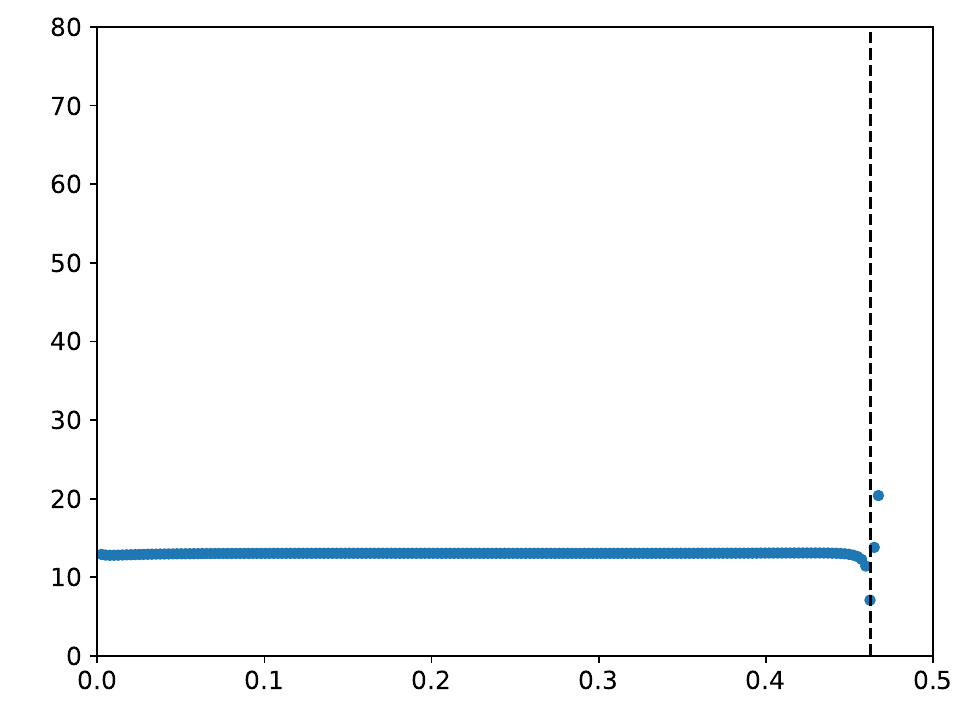}
    \begin{picture}(0,0)\vspace*{-1.2cm}
    \put(-10,10){\footnotesize $\chi(0)$}
    \put(90,10){\footnotesize $\chi_t$}
    \put(-133,100){\footnotesize $\ln(|I_s|)$}
    \put(-50,80){\includegraphics[width=0.5\linewidth]{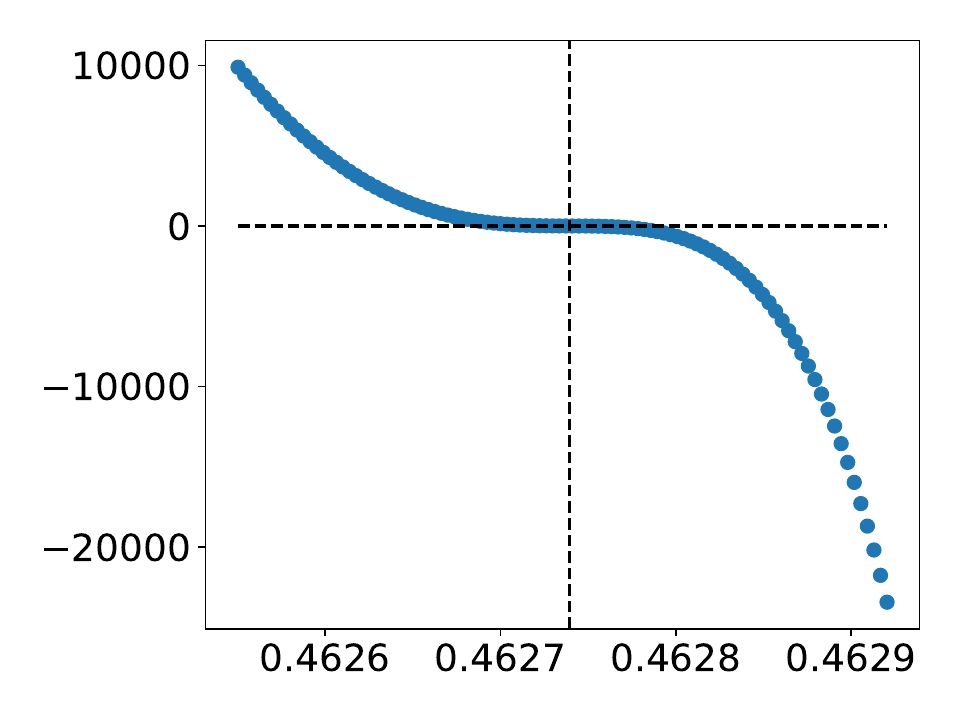}}
    \put(-60,120){\footnotesize $I_s$}
    \put(-35,80){\line(6,-1){125}}
    \put(63,150){\line(1,-3){30}}
    \end{picture}
    \vspace*{-0.75cm}
    \end{subfigure}
    \caption{$d=8$.}
    \label{fig:d8integrals}
    \end{subfigure}
    \hfill
    \begin{subfigure}{\linewidth}
    \centering
    \begin{subfigure}{.475\textwidth}
    \centering
    \includegraphics[width=\linewidth]{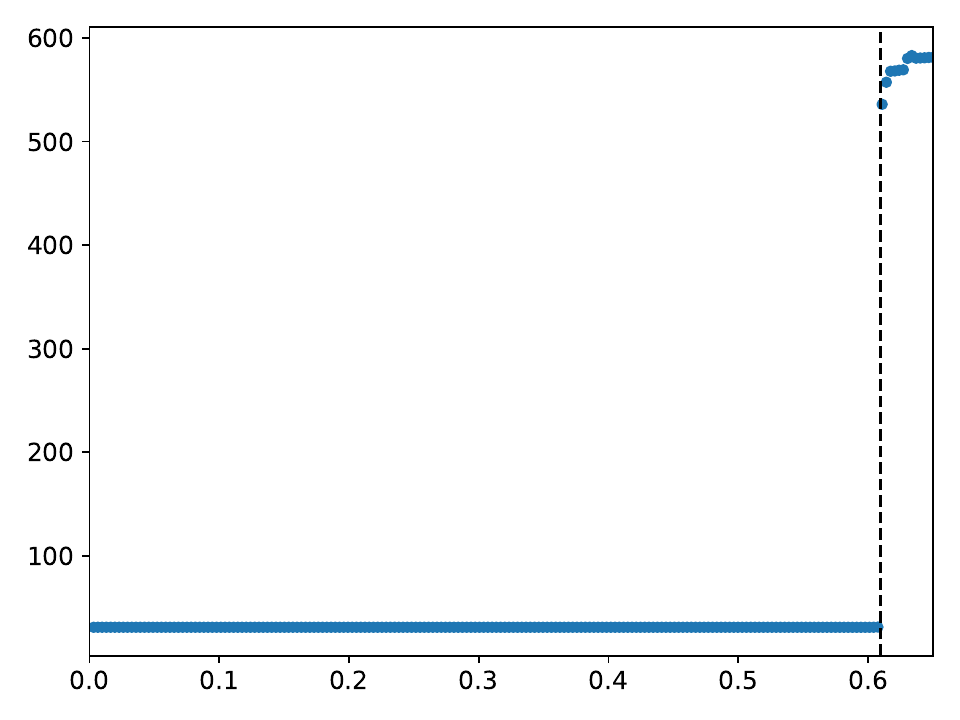}
    \begin{picture}(0,0)\vspace*{-1.2cm}
    \put(-10,10){\footnotesize $\chi(0)$}
    \put(90,10){\footnotesize $\chi_t$}
    \put(-140,100){\footnotesize $\ln(I_r)$}
    \put(-47,57){\includegraphics[width=0.5\linewidth]{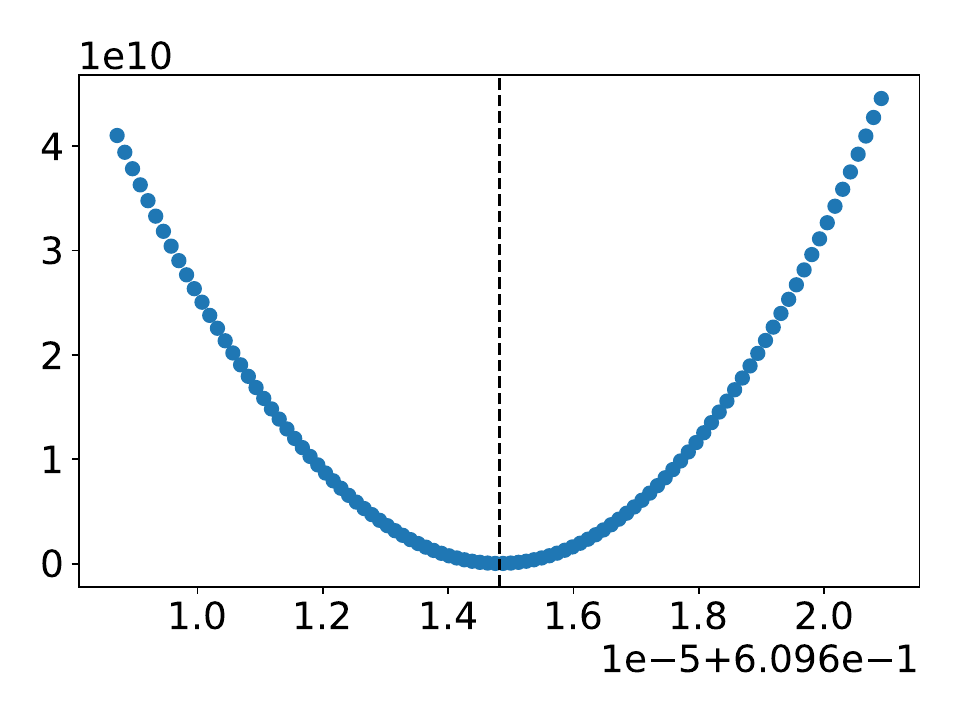}}
    \put(-60,103){\footnotesize $I_r$}
    \put(-33,63){\line(6,-1){125}}
    \put(63,135){\line(1,-3){30}}
    \end{picture}\vspace*{-0.75cm}
    \end{subfigure}
    \hfill
    \begin{subfigure}{.475\textwidth}
    \centering
    \includegraphics[width=\linewidth]{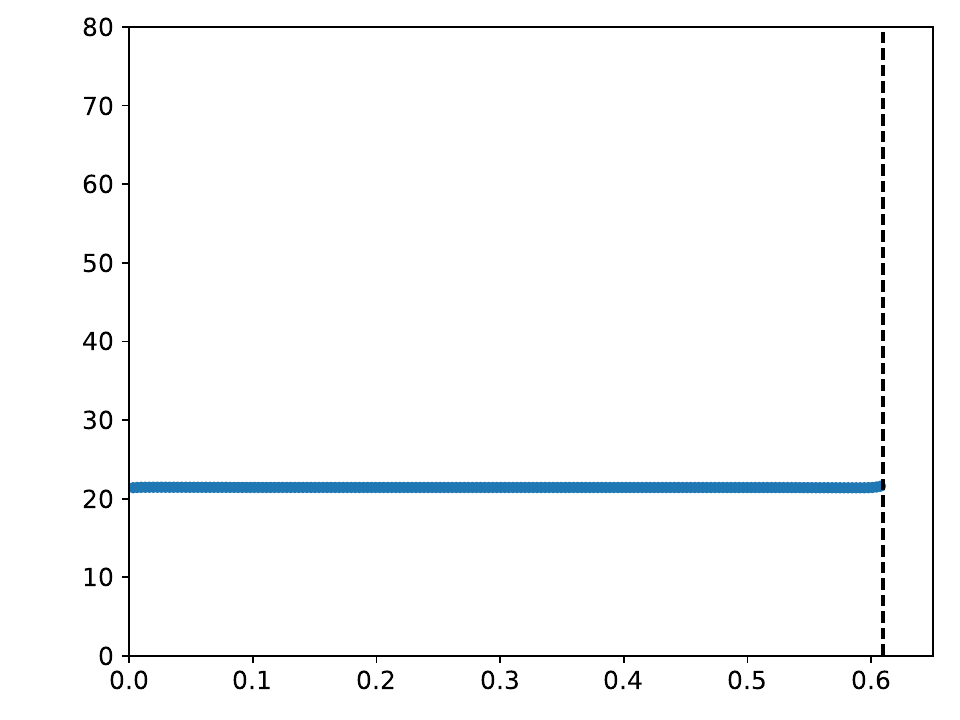}
    \begin{picture}(0,0)\vspace*{-1.2cm}
    \put(-10,10){\footnotesize $\chi(0)$}
    \put(90,10){\footnotesize $\chi_t$}
    \put(-130,100){\footnotesize $\ln(|I_s|)$}
    \put(-45,88){\includegraphics[width=0.5\linewidth]{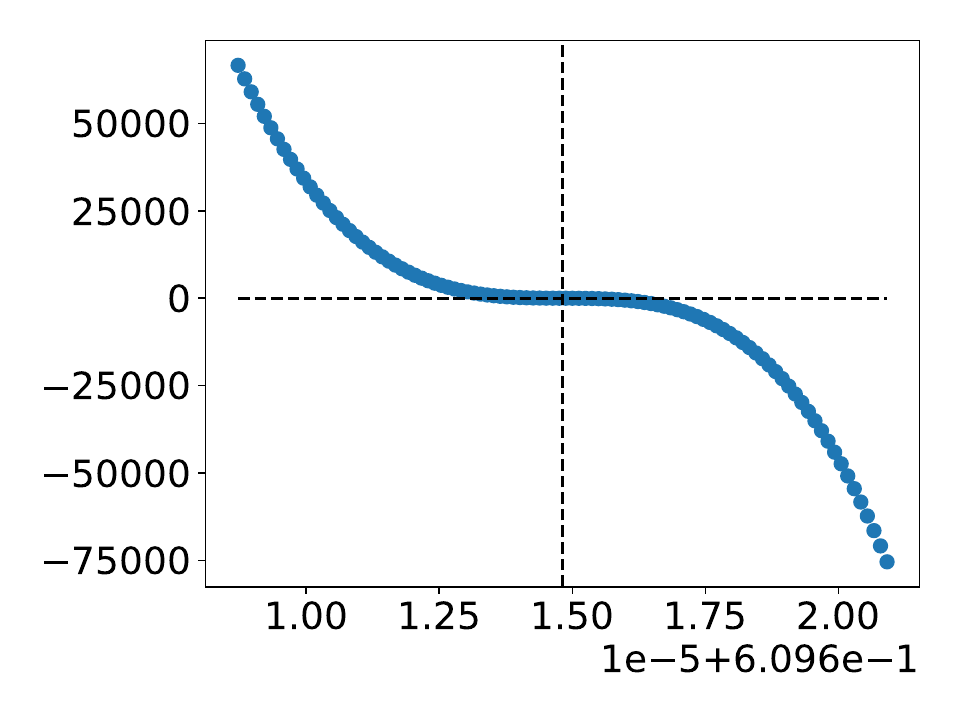}}
    \put(-55,140){\footnotesize $I_s$}
    \put(-33,98){\line(6,-1){125}}
    \put(65,170){\line(1,-3){30}}
    \end{picture}
    \vspace*{-0.75cm}
    \end{subfigure}
    \caption{$d=9$.}
    \label{fig:d9integrals}
    \end{subfigure}
    \hfill
    \begin{subfigure}{\linewidth}
    \centering
    \begin{subfigure}{.475\textwidth}
    \centering
    \includegraphics[width=\linewidth]{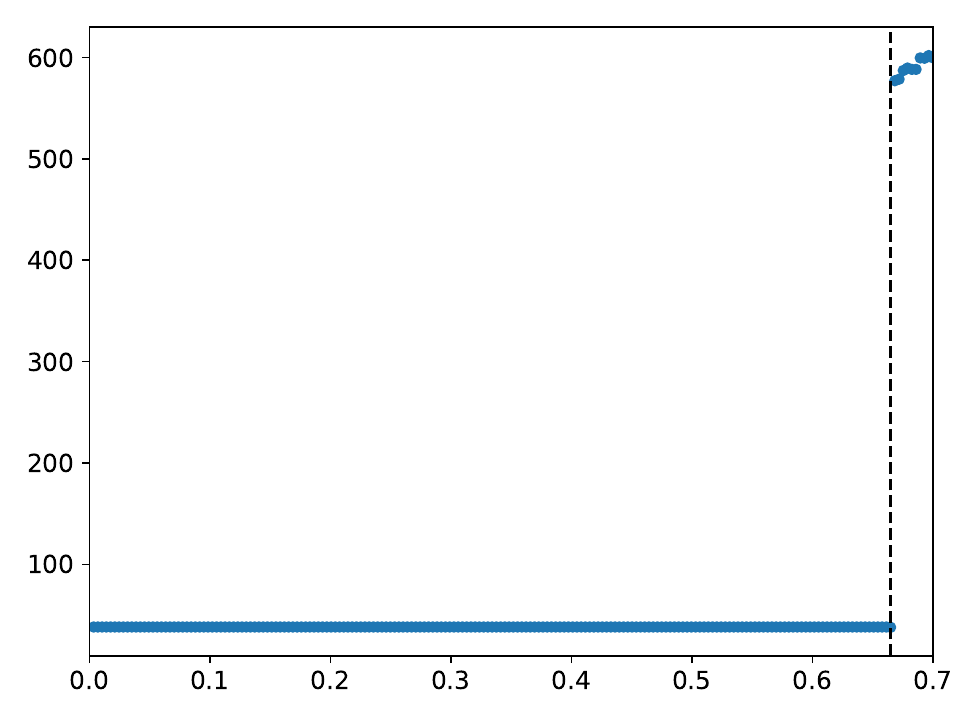}
    \begin{picture}(0,0)\vspace*{-1.2cm}
    \put(-10,10){\footnotesize $\chi(0)$}
    \put(95,10){\footnotesize $\chi_t$}
    \put(-140,100){\footnotesize $\ln(I_r)$}
    \put(-45,57){\includegraphics[width=0.5\linewidth]{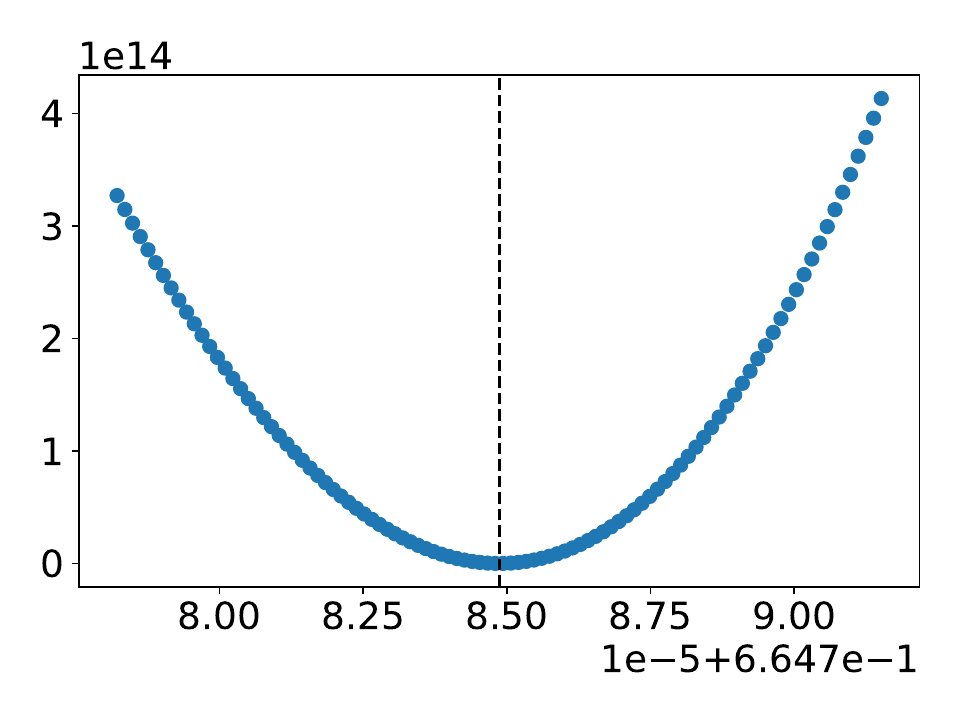}}
    \put(-60,103){\footnotesize $I_r$}
    \put(-33,63){\line(6,-1){125}}
    \put(65,135){\line(1,-3){30}}
    \end{picture}\vspace*{-0.75cm}
    \end{subfigure}
    \hfill
    \begin{subfigure}{.475\textwidth}
    \centering
    \includegraphics[width=\linewidth]{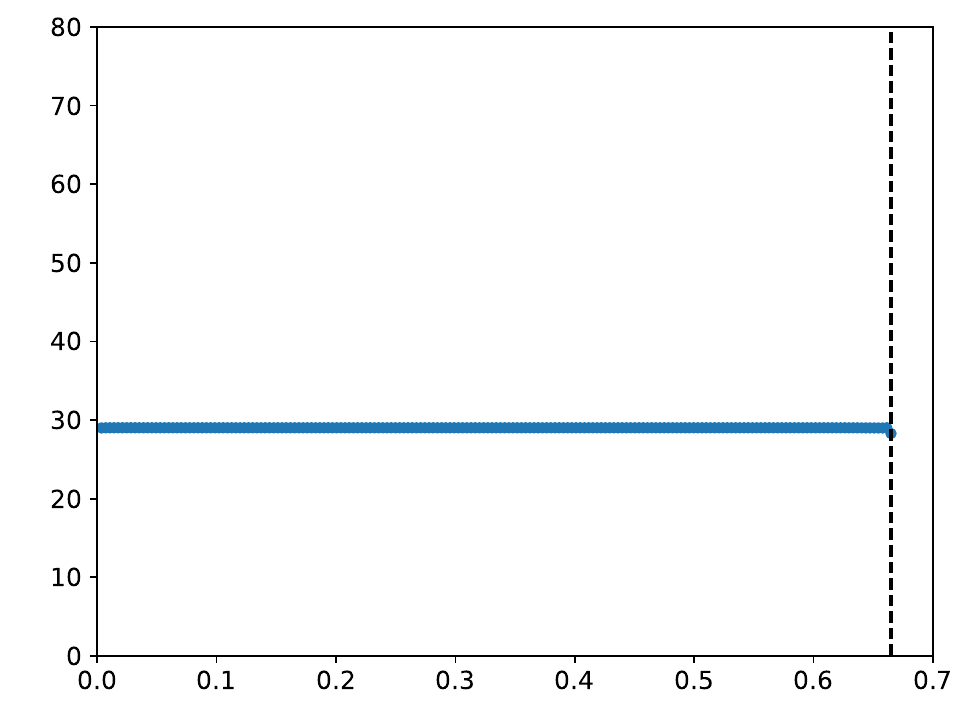}
    \begin{picture}(0,0)\vspace*{-1.2cm}
    \put(-10,10){\footnotesize $\chi(0)$}
    \put(95,10){\footnotesize $\chi_t$}
    \put(-133,100){\footnotesize $\ln(|I_s|)$}
    \put(-40,93){\includegraphics[width=0.5\linewidth]{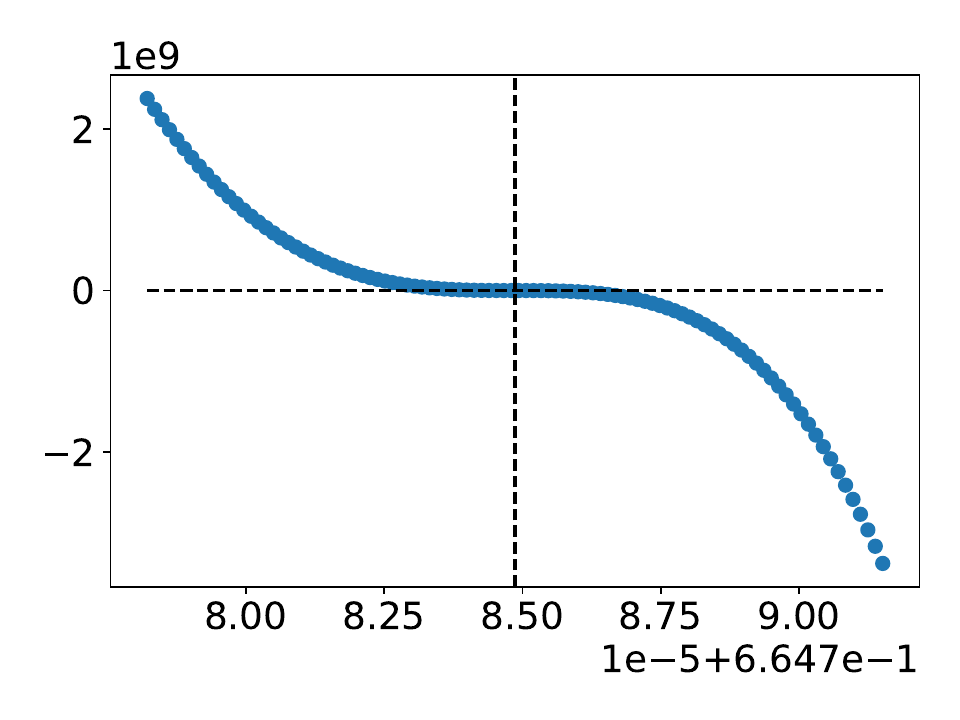}}
    \put(-50,140){\footnotesize $I_s$}
    \put(-33,107){\line(6,-1){125}}
    \put(67,180){\line(1,-3){30}}
    \end{picture}
    \vspace*{-0.75cm}
    \end{subfigure}
    \caption{$d=10$.}
    \label{fig:d10integrals}
    \end{subfigure}
    \caption{The integral constraints for $\chi(0)\in(0,0.7]$ in various spacetime dimensions with $r_{\rm max}=10^3\,l_s$ for Bosonic/Type II string theories. In particular, the dotted line indicates $\chi_t^{(d)}$. The interior figures indicate a zoomed in view near $\chi_t$. Additionally, as $I_s$ can be negative, to see how the magnitude changes as we vary $\chi_t$, the logarithms are taken with respect to the absolute value of $I_s$. However, this is not needed for $I_r$ as the $L^2$-norm is automatically positive. At last, for $\chi(0)>\chi_t$, the scaling integral immediately diverges, hence not shown in the above figures.}
    \label{fig:integral constraints}
\end{figure}

We have numerically motivated the existence of the normalizable solution by plotting the large $r$ behavior of $-\frac{\dd\ln(\chi) }{\dd\ln(r)}$ in figure \ref{fig:chit} and by verifying the expected behavior for the coefficient of the subleading term in \ref{fig:asymptotic divergence}. Below, we provide another non-trivial test for the existence of this solution. At the end of sec. \ref{sec:expected properties of counter HP}, we generalized the scaling arguments presented in \cite{Chen:2021dsw,Balthazar:2022hno} to potentials of arbitrary high orders, and concluded that any normalizable solution to the action must satisfy \eqref{eq:scaling solution}. Applying this condition to the action that includes the quartic interactions implies that the following integral must vanish.
\begin{align}\label{eq:vsq}
I_s:=\int_{0}^{\infty}\dd^{d-1}x\,\sqrt{2}(d-7)\chi^3+3(5-d)\tilde{\kappa}\chi^4=0\,.
\end{align}
Moreover, we expect the norm of $\chi$ to be finite for the normalizable solution. Note that both the norm of $\chi$ and the integral $I_s$ diverge for the non-normalizable solutions. To see how they become finite at $\chi(0)=\chi_t$ we would like to keep track of them as a function of $\chi(0)$. For that, we define two regulated parameters, the regularized norm of $\chi$ which we denote by $I_r^{\rm IR}$ and the regulated integral $I_s$ which we denote by $I_s^{\rm IR}$.
\begin{align}
\begin{split}
    I^{\rm IR}_r&=\int_0^{r_{\rm max}}\dd^{d-1}x\,|\chi|^2\,,\\
    I^{\rm IR}_s&=
    \int_{0}^{r_{\rm max}}\dd^{d-1}x\,\sqrt{2}(d-7)\chi^3+3(5-d)\tilde{\kappa}\chi^4\,.
\end{split}
\end{align}
We have imposed an IR cut-off of $r_{\max}=10^{3}\, \sqrt{\alpha'/\kappa}\gg l_s$. We have chosen $r_{\max}$ to be much greater than $l_s$ so that at $\chi(0)\simeq \chi_t$ we can estimate the free energy by looking at the profile of $\chi$ for $r<r_{\rm max}$.\footnote{In particular, in the appendix~\ref{sec:stability}, we show that the free energy converges quickly in space, e.g., its numerical value stabilizes for any $r_{\rm max}$ beyond $\sim \cO(10^2)$.} We expect to see $I^{\rm IR}_s$ to vanish at $\chi(0)\simeq \chi_t$ and the norm $I^{IR}_r$ to have a global minimum at $\chi(0)\simeq\chi_t$. This is exactly what we numerically observe. We have plotted $I^{\rm IR}_r$ and $I^{\rm IR}_s$ in spacetime dimensions $d=8$, $9$, and $10$, respectively in figure~\ref{fig:integral constraints} for the Bosonic and Type II string theories. Again, for these quantities, the heterotic string theory is qualitatively similar to the Bosonic/Type II string theories. Therefore, the equation \eqref{eq:scaling solution} which fails for non-normalizable solution, is satisfied for our normalizable solution. For $\chi(0)=\chi_t$ we can also integrate the Euclidean action to approximate the free energy. Their numerical values are shown in table~\ref{tab:higher dimension string stars free energies}. However, these numerical values are not precise since, as we explain later, the higher-order $\alpha'$ corrections are expected to correct the numerical values.
\begin{table}[htp!]
    \centering
    \begin{tabular}{c|c|c}
        Free energy & Bosonic/Type II & Heterotic \\
        \hline
        $F^{(8d)}$ & $11142.28\cdot \frac{\omega_6}{2}\left(\frac{\alpha'}{\kappa}\right)^{5/2}M_{{\mathrm pl}}^{6}$ & $11147.15\cdot \frac{\omega_6}{2}\left(\frac{\alpha'}{\kappa}\right)^{5/2}M_{{\mathrm pl}}^{6}$\\
        $F^{(9d)}$ & $419819.59\cdot \frac{\omega_7}{2}\left(\frac{\alpha'}{\kappa}\right)^{3}M_{{\mathrm pl}}^{7}$ & $419705.48\cdot \frac{\omega_7}{2}\left(\frac{\alpha'}{\kappa}\right)^{3}M_{{\mathrm pl}}^{7}$ \\
        $F^{(10d)}$ & $16972516.01\cdot \frac{\omega_8}{2}\left(\frac{\alpha'}{\kappa}\right)^{7/2}M_{{\mathrm pl}}^{8}$ & $16938303.47\cdot \frac{\omega_8}{2}\left(\frac{\alpha'}{\kappa}\right)^{7/2}M_{{\mathrm pl}}^{8}$
    \end{tabular}
    \caption{Approximate numerical value of the free energy for string stars in $d=8,9,10$ for the relevant string theories.}
    \label{tab:higher dimension string stars free energies}
\end{table}

In the previous section we motivated the existence of string stars in dimensions $d\geq 7$. Moreover, we argued that if they exist, in dimensions $d>7$ they are expected to not vanish at Hagedorn temperature. In this section we studied bounded solutions in dimensions $d>7$ at Hagedorn temperature. First, we used the emergence of an $SU(2)$ symmetry at $T=T_{\rm H}$ to narrow our search to symmetry preserving solutions which simplifies the equations. Then, we found a one-parameter family of bounded solutions parametrized by $\chi(0)$. All of these solutions decay as $r^{-2}$ at infinity which makes them bounded, but non-normalizable. Finally, in this subsection, with the help of numerical integration, we argued that the solution separating bounded from unbounded solution decays as $r^{3-d}$ which is the only permitted asymptotic behavior for bounded solutions. The fact that the profile of $\chi(r)$ for the transition solution decays as $r^{3-d}$ does not seem to depend on the details of $V_{\rm eff}$. One can intuitively understand this by thinking of the solutions of the equation of motion in terms of the motion of a particle friction in the potential $V_{\rm eff}$. The unbounded solutions reach $\chi=0$ at finite $r$ due to having too much initial energy at $r=0$. On the other hand, the bounded solutions reach $\chi=0$ at $r\rightarrow \infty$ thanks to the friction term. The solution at the transition point has higher energy compared to all the other bounded solutions and is therefore expected to roll to $\chi=0$ faster. But at the same time, it is a limit of bounded solutions and is itself bounded. Therefore, it reaches $\chi=0$ faster than all of the other bounded solutions via a faster decay rate. However, the only other permitted decay rate is $r^{3-d}$ which is why the solution behaves as $\sim r^{3-d}$ at large radii. A rigorous mathematical proof of this statement would be very interesting.

One may have objections regarding our analysis depending on the quartic potential which may potentially breakdown when the magnitude of the winding mode is of order-one. In this case, there are two possible sources of breakdown: 1) the perturbation in $\chi$ breaks down due to all interactions being of the same order, and 2) interactions with heavier states in the theory, e.g., string excitations, may become relevant as we reach string scale. We addressed the first reason in the heuristic argument above. Although the higher-order corrections will certainly change $V_{\rm eff}$, the normalizability of the solution that separates bounded from unbounded solutions does not seem to depend on the details of the potential. As for the concern regarding massive string excitations, let us point out an important feature of the bounded solutions: all of them are perturbative at large radii. Instead of parameterizing the solutions using $\chi(0)$, we could parametrize them using $\chi(R)$ where $R\gg l_s$. For sufficiently small $\chi(R)$, we know that there is a unique bounded solution with that value at $r=R$ that is differentiable at $r=0$, and for sufficiently large $\chi(R)$, we expect the friction term to not be strong enough to stop $\chi$ from rolling down $V_{\rm eff}$ leading to unbounded solutions. Therefore, the transition value $\chi(R)\ll1$, which is perturbatively reliable, would correspond to $\chi_t\sim \mathcal{O}(1)$ that separates the bounded from unbounded solutions. Even if we cannot trust the equations for small $r$, the existence of this background is trustable. Given the generality of the $r^{3-d}$ behavior for the transition solution combined with the heuristic argument we presented above, we think the massive string excitations will not change the nature of this solution. However, we have not proven the normalizability of this background.

\subsection{Revisiting $d=7$}
\label{sec:7d}

In the above analyses, we left out the analysis regarding normalizable solutions in 7d and their associated thermodynamics properties. In 7d, unlike higher dimensions, the action which includes the cubic interaction has normalizable solutions that have an asymptotic behavior of $\sim r^{3-d}\sim r^{-4}$ for any boundary condition for $\chi(0)$ (see \eqref{eq:7d exact cubic solution}). This is in contrast to $d>7$ where solutions with arbitrary but small values of $\chi(0)$ decay like $r^{-2}$ at infinity and only a particular $\mathcal{O}(1)$ value of $\chi(0)$ has $\sim r^{3-d}$ behavior at large $r$. As was studied in \cite{Balthazar:2022hno}, the addition of the quartic term makes it impossible to have non-zero normalizable solutions at $d=7$ at the Hagedorn temperature because the equation \eqref{eq:vsq} cannot be satisfied for any non-zero solution. To circumvent this problem, the authors of \cite{Balthazar:2022hno} considered moving away from $d=7$ and $m_\infty=0$ by formally changing the spacetime dimension ($d=7+\epsilon$) combined with a small change in $m_\infty$ which picks out a special solution among \eqref{eq:7d exact cubic solution} depending on the ratio $m_\infty/\epsilon$. For $m_\infty/\epsilon\gg1$, $\chi(0)\propto m_\infty^2$ while for $m_\infty/\epsilon\ll 1$, $\chi(0)\propto m_\infty$. Notably, the free energy of the solution \eqref{eq:7d exact cubic solution} is independent from $\chi(0)$ \cite{Balthazar:2022szl}
\begin{equation}
\label{eq:7d free energy}
F=\frac{\pi M_{\pl}^5R}{2\pi R}\cdot \frac{2\pi^3}{\Gamma[3]}\int_0^{\infty} \dd r\, r^5\left[\frac32(\nabla \chi)^2-\frac{\kappa}{\alpha'}\frac{1}{\sqrt{2}}\chi^3\right]=\frac{\pi^3}{2}\cdot \frac{1152}{5}\cdot\left(\frac{\alpha'}{\kappa}\right)^{2}M_{\pl}^5\,.
\end{equation}
Our methods also allow us to find normalizable solutions in dimensions $d=7+\epsilon$ at Hagedorn temperature in an independent way. We can see what happens to our normalizable solutions if we approach $d=7, m_\infty=0$ as we take $\epsilon$ to zero. We plot the dependence of the free energy on $\epsilon$ in fig.~\ref{fig:free energy in 7d} and we see that in fact the free energy of the solutions that we find at the Hagedorn temperature converge to the same value \eqref{eq:7d free energy}. 

\begin{figure}[H]
    \centering

    \includegraphics[width=.65\linewidth]{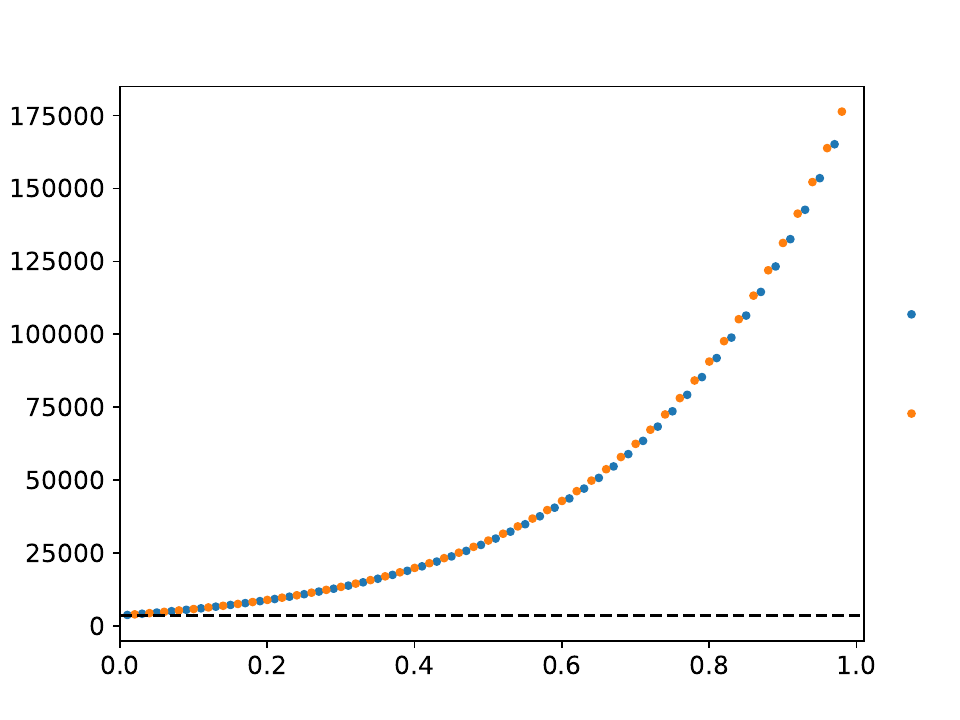}
    \begin{picture}(0,0)\vspace*{-1.2cm}
    \put(-375,120){\footnotesize $F\cdot\frac{\kappa^2}{\alpha'^2 M_{\rm pl}^5}$}
    \put(-150,-5){\footnotesize $\epsilon$}
    \put(-250,100){\includegraphics[width=0.25\linewidth]{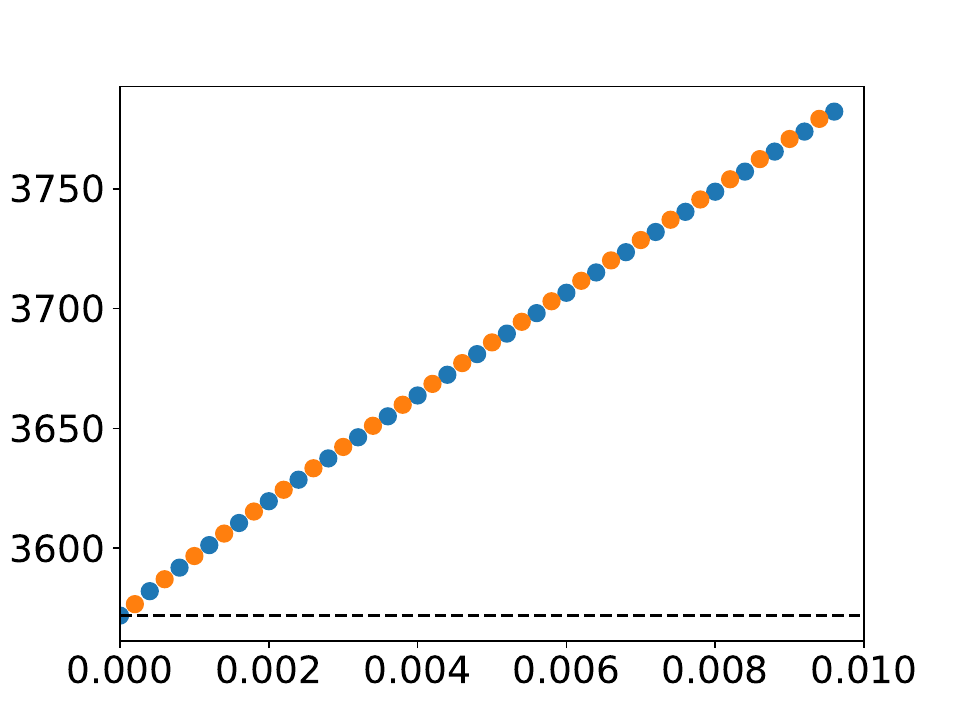}}
    \put(-30,33){\footnotesize $\frac{\pi^3}{2}\cdot \frac{1152}{5}$}
    \put(-10,97){\footnotesize Heterotic}
    \put(-10,130){\footnotesize Bosonic/Type II}
    \put(-275,40){\line(1,6){25}}
    \put(-260,40){\line(2,1){150}}
    \put(-265,35){\oval(27,10)}
    \end{picture}\vspace*{0.3cm}
    
    \caption{Free energy of string stars in $(7+\epsilon)$d at Hagedorn temperature. The dashed horizontal line indicates the free energy of the 7d HP solution \eqref{eq:7d free energy} in units of $(\alpha'/\kappa)^2M_{\pl,7}^5$. The behavior of the free energy at Hagedorn temperature for small $\epsilon$'s is shown in the zoomed in plot. }
    \label{fig:free energy in 7d}
\end{figure}

\begin{figure}
\centering
\hspace*{-.25in}
\begin{subfigure}{.48\textwidth}
        \centering
        \includegraphics[width=\linewidth]{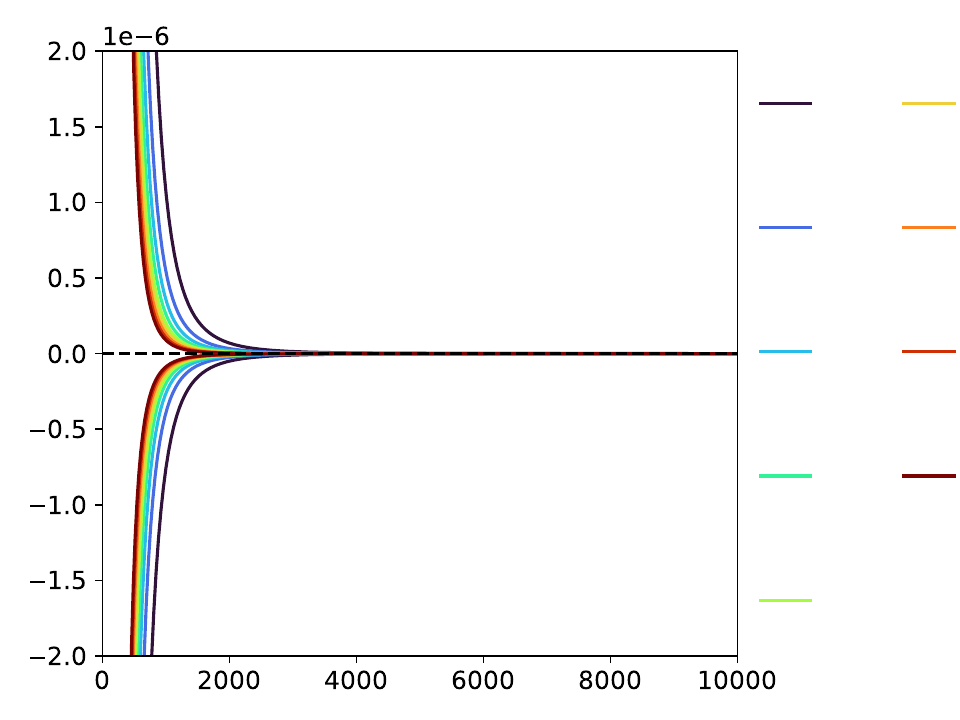}
        \begin{picture}(0,0)\vspace*{-1.2cm}
        \put(80,185){\footnotesize $\chi(0)$ ($\times 10^{-3}$)}
        \put(85,161){\footnotesize $1$}
        \put(85,131){\footnotesize $2$}
        \put(85,101){\footnotesize $3$}
        \put(85,71){\footnotesize $4$}
        \put(85,41){\footnotesize $5$}
        \put(125,161){\footnotesize $6$}
        \put(125,131){\footnotesize $7$}
        \put(125,101){\footnotesize $8$}
        \put(125,71){\footnotesize $9$}
        \put(-120,175){\footnotesize $\chi$}
        \put(-120,30){\footnotesize $\varphi$}
        \put(-30,5){\footnotesize $r$ ($\sqrt{\alpha'/\kappa}$)}
        \end{picture}\vspace*{-0.6cm}
        \caption{Bosonic/Type II: Spatial profile of $\chi$ and $\varphi$.}
        \label{fig:chi and phi away from hagedorn in 7d}
    \end{subfigure}
    \hfill
    \begin{subfigure}{.48\textwidth}
        \centering
        \includegraphics[width=\linewidth]{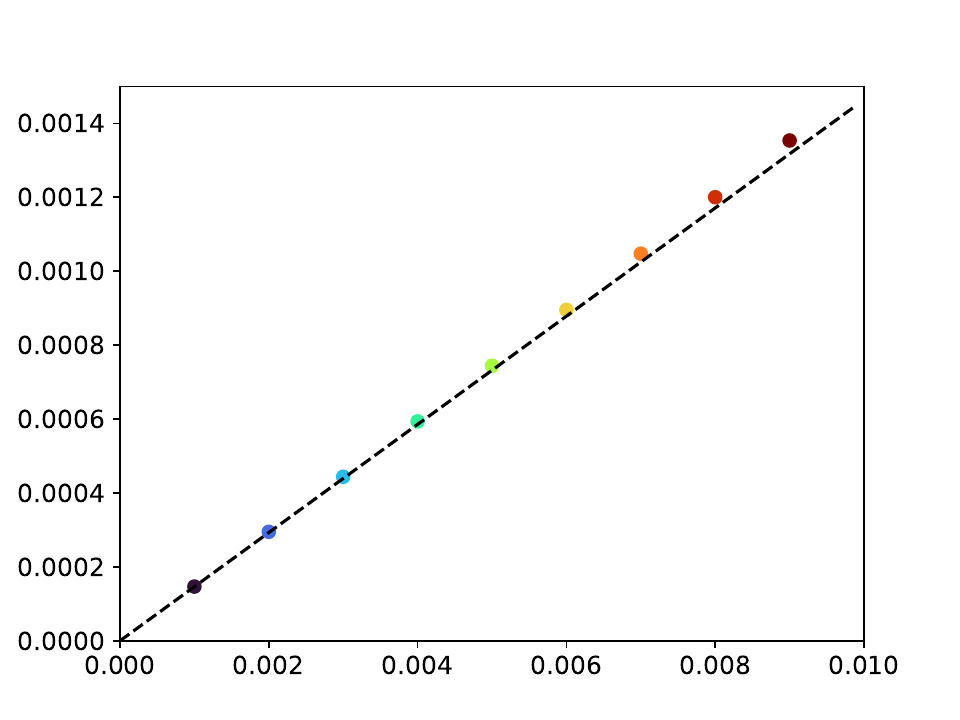}
        \begin{picture}(0,0)\vspace*{-1.2cm}
        \put(-100,180){\footnotesize $m_{\infty}$}
        \put(0,5){\footnotesize $\chi(0)$}
        \end{picture}\vspace*{-0.6cm}
        \caption{Bosonic/Type II: $m_{\infty}$ vs $\chi(0)$.}
        \label{fig:7d chi vs minf}
    \end{subfigure}
    \hspace*{-.25in}
    \begin{subfigure}{.48\textwidth}
        \centering
        \vspace{0.5cm}
        \includegraphics[width=\linewidth]{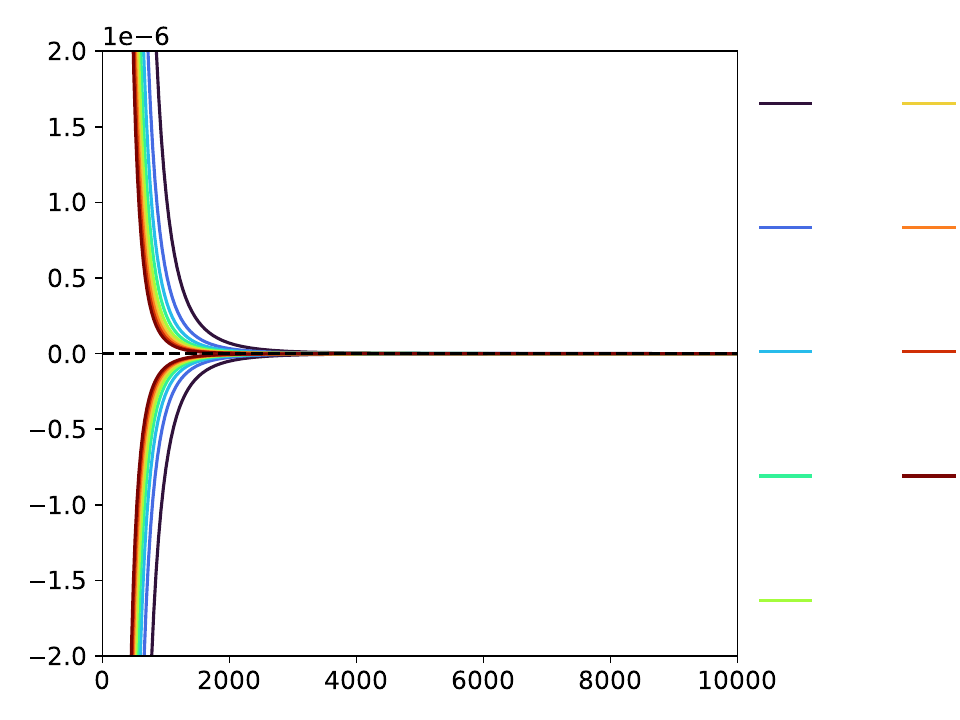}
        \begin{picture}(0,0)\vspace*{-1.2cm}
        \put(80,185){\footnotesize $\chi(0)$ ($\times 10^{-3}$)}
        \put(85,161){\footnotesize $1$}
        \put(85,131){\footnotesize $2$}
        \put(85,101){\footnotesize $3$}
        \put(85,71){\footnotesize $4$}
        \put(85,41){\footnotesize $5$}
        \put(125,161){\footnotesize $6$}
        \put(125,131){\footnotesize $7$}
        \put(125,101){\footnotesize $8$}
        \put(125,71){\footnotesize $9$}
        \put(-120,175){\footnotesize $\chi$}
        \put(-120,30){\footnotesize $\varphi$}
        \put(-30,5){\footnotesize $r$ ($\sqrt{\alpha'/\kappa}$)}
        \end{picture}\vspace*{-0.6cm}
        \caption{Heterotic: Spatial profile of $\chi$ and $\varphi$.}
        \label{fig:chi and phi away from hagedorn in 7d het}
    \end{subfigure}
    \hfill
    \begin{subfigure}{.48\textwidth}
        \centering
        \includegraphics[width=\linewidth]{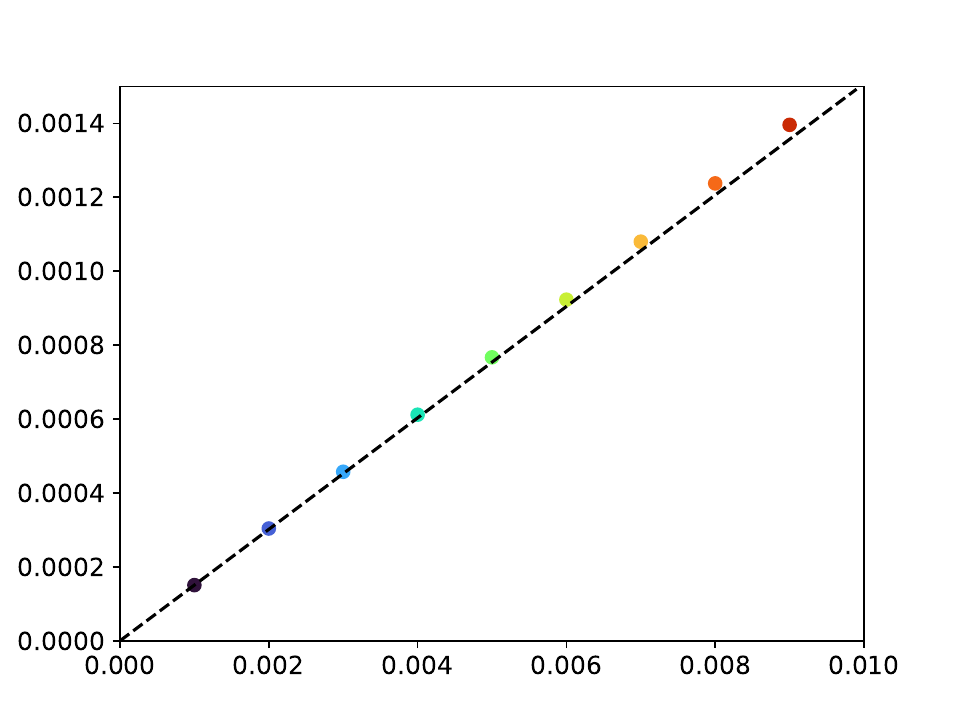}
        \begin{picture}(0,0)\vspace*{-1.2cm}
        \put(-100,180){\footnotesize $m_{\infty}$}
        \put(0,5){\footnotesize $\chi(0)$}
        \end{picture}\vspace*{-0.6cm}
        \caption{Heterotic: $m_{\infty}$ vs $\chi(0)$.}
        \label{fig:7d chi vs minf het}
    \end{subfigure}
    \caption{The thermodynamic properties of HP solutions away from Hagedorn in 7d for the various string theories. In~\ref{fig:chi and phi away from hagedorn in 7d} and~\ref{fig:chi and phi away from hagedorn in 7d het}, we vary $\chi(0)$ and determine $\varphi(0)$ such that $\chi(r)$ is bounded and normalizable. 
    Here, the pair $(\chi,\varphi)$ corresponding to a given $\chi(0)$ is distinguished by their colors. 
    In~\ref{fig:7d chi vs minf} and~\ref{fig:7d chi vs minf het}, $m_{\infty}$ is computed for given $\chi(0)$'s where the values of $\chi(0)$ are indicated by the same color scheme used in~\ref{fig:chi and phi away from hagedorn in 7d} and~\ref{fig:chi and phi away from hagedorn in 7d het}, respectively. The dashed line indicates the linear relation $m_{\infty}=\sqrt{\frac{3\tilde{\kappa}\kappa}{140\alpha'}}\chi(0)$ where $\tilde{\kappa}=1$.}
    \label{fig:7d away from hagedorn}
\end{figure}

This suggests that even though $\chi$ is converging to the zero solution at $d=7$, it does so in a way that the free energy converges to $F=\frac{1152\pi^3\alpha'^2}{10\kappa^2}M_{\pl}^5$. The fact that different ways of converging to $ d=7,m_\infty=0$ agree on the free energy, suggests that the free energy is a continuous function of $m_\infty$ and $d$ at $d=7, m_\infty=0$ as suggested in \cite{Balthazar:2022hno}. If that is the case, we must be able to approach this point by changing the temperature while keeping the dimension fixed at $d=7$. In other words, there must be HP solutions at $d=7$ when the temperature is strictly less than $T_{\rm H}$. The benefit of such solutions is that they will have a small profile for $\chi$ and will therefore be under perturbative control since we know $\chi$ will go to zero in the limit of $T\rightarrow T_{\rm H}$. Before presenting these solutions numerically, let us review the analysis of \cite{Balthazar:2022hno} which motivates their existence and understnad some aspects of them better analytically. Assuming these solutions almost have the $SU(2)$ symmetry at the Hagedorn temperature, we would imagine the solution to be close to one the solutions \eqref{eq:7d exact cubic solution} with a specific value of $\chi(0)$. As pointed out in \cite{Balthazar:2022hno}, the scaling equation \eqref{SGI}, leads to 
\begin{align}
m_\infty=\sqrt{\frac{3\tilde{\kappa}}{140}\cdot\frac{\kappa}{\alpha'}}\chi(0)\,,
\end{align}
for these solutions. We observe that the analytic estimate near the Hagedorn temperature shows excellent agreement with the numerical results, as illustrated in Figs.~\ref{fig:chi and phi away from hagedorn in 7d} and~\ref{fig:chi and phi away from hagedorn in 7d het}, corresponding to the Bosonic/Type~II and Heterotic theories, respectively.
The solution \eqref{eq:7d exact cubic solution} also has a characteristic length scale set by $L\propto \chi(0)^{-1/2}l_s$. Therefore, for these solutions, we find 
\begin{align}
L_{\rm 7d}\sim \sqrt{\frac{l_s}{m_{\infty}}}\,.
\end{align}
Note that this is different from the $d<7$ case where the size of the string star scales as $L_{d<7}\sim m_{\infty}^{-1}$. Prior to \cite{Balthazar:2022hno}, the quartic term was ignored in the study of HP solutions with the reasoning that for small values of $\chi$, they are subleading compared to the other terms.  However, this argument has a loophole which the 7d solutions use. At temperatures close to Hagedorn, the mass $m_\infty$ which multiplies the quadratic term is also small. Therefore, if $\chi(0)\sim m_\infty l_s$, the quartic term $(\kappa/\alpha')\chi^4$ can compete with the quadratic term $m_\infty^2\chi^2$ for small radii. This is exactly what happens in $d=7$. In contrast to $d<7$ where $\chi(0)\sim m_\infty^2\alpha'$, these solutions make the quartic relevant despite having a small $\chi\ll1$. Note that the even higher-order terms such as quintic interactions are always subleading compared to the quartic term, because the coefficient of the quartic term is of order one in string units and is not suppressed like the mass term. 

\begin{figure}
    \centering
    \includegraphics[width=.65\linewidth]{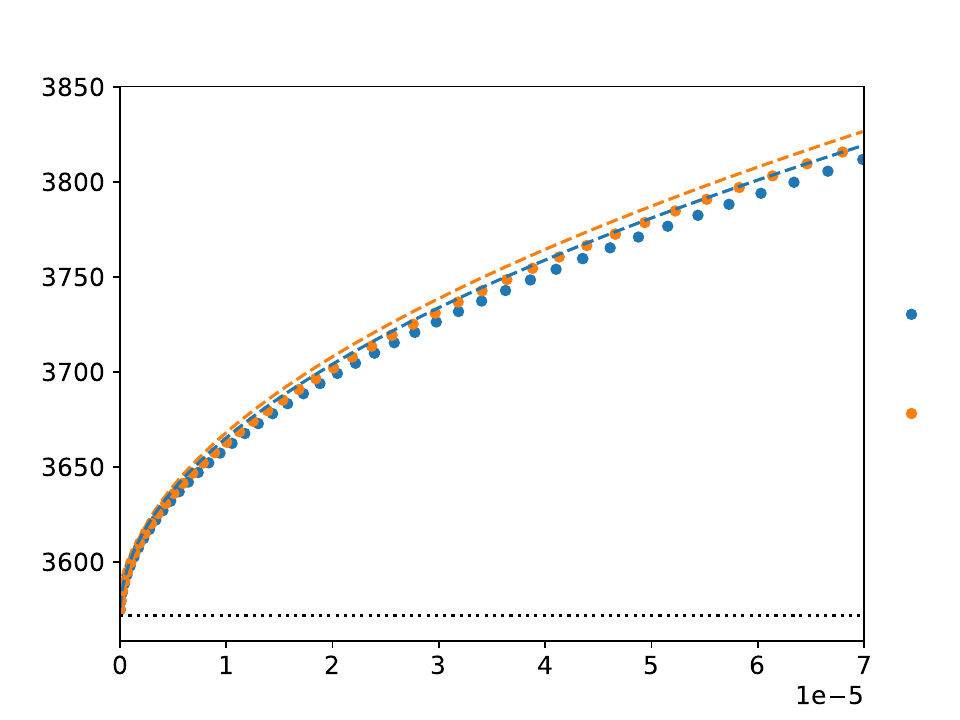}
    \begin{picture}(0,0)\vspace*{-1.2cm}
    \put(-375,120){\footnotesize $F\cdot\frac{\kappa^2}{\alpha'^2 M_{\rm pl}^5}$}
    \put(-170,-5){\footnotesize $\frac{T_{\rm H}-T}{T_{\rm H}}$}
    \put(-30,33){\footnotesize $\frac{\pi^3}{2}\cdot \frac{1152}{5}$}
    \put(-10,130){\footnotesize Bosonic/Type II}
    \put(-10,97){\footnotesize Heterotic}
    \end{picture}\vspace*{0.3cm}
    \caption{Free energy in $7d$ away from Hagedorn temperature. Here, the dashed line indicates the analytic estimate $C\cdot \sqrt{\frac{T_{\rm H}-T}{T_{\rm H}}}$ given in~\eqref{analyticest}.
    }
    \label{fig:free energy in 7d}
\end{figure}

Let us also discuss the validity of EFT for solutions close to the Hagedorn temperature. It was argued in \cite{Chen:2021dsw}, that for HP solutions in dimensions $d<7$, the change in the action for an $\mathcal{O}(1)$ rescaling of $\chi$ is given by 
\begin{align}
\label{CIA}
    \delta S \propto \beta M_{\rm pl}^{d-2} m_\infty^2\int d^{d-1}x~|\chi|^2\propto m_\infty^{7-d}/g_s^2\,,
\end{align}
which must be greater than one for stability against quantum fluctuations. Therefore, in $d<7$, at temperatures too close to Hagedorn such that $m_{\infty}^{7-d}\lesssim g_s^2$,  quantum fluctuations are too large and the EFT is not trusted. Note that in $d=7$, the RHS in \eqref{CIA} becomes temperature independent. Even though the above equation was derived for the HP solutions in $d<7$ spacetime dimensions, the conclusion remains correct. The key observation is that the action in $d=7$ does not converge to zero at Hagedorn temperature. Therefore, an order-one rescaling of the fields will lead to a change in the action which is of the same order as the action itself
\begin{align*}
    \delta I\sim I\sim \frac{1}{g_s^2}\,,
\end{align*}
which is small for small string couplings.

Now let us explicitly verify the above claims  via numerical analysis. In figure \ref{fig:chi and phi away from hagedorn in 7d} we can see the profile of normalizable solutions for $\chi$ and $\phi$ for a range of different initial conditions $\chi(0)$. As we see, the relation $m_\infty\simeq \sqrt{\frac{3\kappa}{140\alpha'}}\chi(0)$ which was derived analytically is satisfied by the numerical solutions. By finding these normalizable solutions for a range of temperatures, we arrive at figure~\ref{fig:free energy in 7d} for the dependence of the free energy on temperature. These results are also consistent with our expectation from section~\ref{sec:expected properties of counter HP} given that the free energy of these saddles is of the same order as the string sized black hole. Note that the change in the free energy away from the Hagedorn temperature scales as
\begin{align}
    \delta F\sim (T_{\rm H}-T)^{1/2}\sim m_\infty\,.
\end{align}
To understand this, remember that our solution is well approximated by the solution \eqref{eq:7d exact cubic solution} for a specific value of $\chi(0)\propto m_\infty$ plus a subleading correction. Both the correction and the mass term change the value of the free energy from its value at the Hagedorn temperature. Let us first look how the mass term changes the free energy
\begin{align}
\label{eq:7d free energy as func of temp}
    \delta F_m\simeq \frac{M_{\rm pl}^5}{2}\int \dd^6 x\,  m_\infty^2 |\chi^2|&\simeq \frac{M_{\rm pl}^5m_{\infty}^2}{2}\int \dd^6 x \left(\frac{\chi(0)}{\left(1+\frac{\kappa}{24\sqrt{2}\alpha'}\chi(0)r^2\right)^2}\right)^2\nonumber\\
    &=\frac{M_{\rm pl}^5m_\infty^2\cdot 2304\sqrt{2}\pi^3\alpha'^3}{\chi(0)\kappa^3}\nonumber\\
    &=\frac{2304\sqrt{3}\pi^3M_{\rm pl}^5 \alpha'^2}{\sqrt{70}\kappa^2}\sqrt\frac{T_{\rm H}-T}{T_{\rm H}}\,,
\end{align}
where in the last line we used $m_\infty\simeq \sqrt\frac{\kappa(T_{\rm H}-T)}{\alpha'T_{\rm H}}$. Note that this is not the entire change in the free energy, but rather provides us with the next-order corrections in the temperature $T$ to the free energy $F$ away from Hagedorn temperature $T_{\rm H}$. Namely, the equation \eqref{SGI} shows us that the rest of the free energy will also receive a correction that is proportional to the mass term. Therefore, the overall change in the free energy as we move away from the Hagedorn temperature is proportional to $(T_{\rm H}-T)^{1/2}$ for small temperature differences $T_{\rm H}-T\ll M_s$.

We can also compute the ADM mass associated to these solution. The explicit form of the metric along with the derivation of the ADM mass can be found in \cite{Chen:2021dsw}. In particular, the Einstein frame metric of these bounded normalizable solutions at large-$r$ region of space are
\[
\dd s^2\sim f\dd t^2+\frac{\dd r^2}{f}+r^2\dd\Omega_{d-2}^3\,,
\]
where we have $f\sim 1-\frac{\mu}{r^{d-3}}=1+2\frac{d-3}{d-2}\varphi$.
In $d=7$, at large radii and temperatures close to the Hagedorn, we still have an approximate $SU(2)$ symmetry as observed numerically in \cite{Balthazar:2022hno}.  Therefore, in the temperature window $0<T_{\rm H}-T\ll M_s$, we have the following approximate relations
\begin{align}
\varphi\simeq -\frac{\chi(r)}{\sqrt{2}}\simeq -\left(\frac{\alpha'}{\kappa}\right)^2\frac{576\sqrt{2}}{\chi(0)}r^{-4}\simeq -\left(\frac{\alpha'}{\kappa}\right)^\frac{3}{2}\frac{576\sqrt{3\tilde{\kappa}}}{\sqrt{70}m_\infty}r^{-4}
\end{align}
Plugging the bounded normalizable profile of $\varphi$ into the metric above, we can deduce the ADM mass as
\begin{align}
    \mu\simeq\frac{8}{5}\left(\frac{\alpha'}{\kappa}\right)^\frac{3}{2}\frac{576\sqrt{3\tilde{\kappa}}}{\sqrt{70}m_\infty}\qquad\rightarrow \qquad 
    M=\frac{(d-2)\omega_{d-2}M_{\rm pl}^{d-2}}{2}\mu=\frac{2304\omega_{5}\sqrt{3\tilde{\kappa}}}{\sqrt{70}m_\infty}\left(\frac{\alpha'}{\kappa}\right)^\frac{3}{2}M_{\pl}^5\,.
\end{align}
Suppose $F=F_0+C\cdot m_\infty+\mathcal{O}(m_\infty^2)$, the ADM mass becomes
\begin{align}\label{7dM}
    M\simeq \frac{C\kappa }{\alpha' 2m_\infty}\,,
\end{align}
where we have made use of the following derivation
\begin{align*}
    \beta\frac{dm_\infty}{d\beta}=-T\frac{dm_\infty}{dT}=\frac{T\sqrt{\kappa}}{2\sqrt{\alpha' T_H}\sqrt{T_H-T}}\simeq\frac{\kappa }{\alpha' 2m_\infty}\,.
\end{align*}
The equations of motion will force the mass computed using thermodynamics to agree with that of the ADM mass \cite{Chen:2021dsw}. Therefore, extending this expectation into the higher-dimensional string stars solutions, namely demanding the consistency between the two calculations of mass for string star solutions gives rise to a precise numerical value of the coefficient appearing in the next-order correction of the free energy 
\begin{align}\label{Analyticcoeff}
   C=\frac{3072\pi^3\sqrt{3\tilde{\kappa}}}{\sqrt{70}}\left(\frac{\alpha'}{\kappa}\right)^{5/2}M_{\pl}^5\,.
\end{align}
Therefore, near the Hageodorn temperature in $d=7$ we have the following leading order approximations
\begin{align}\label{analyticest}
    F&=M_{\pl}^5\left[\frac{1152\pi^3\alpha'^2}{10\kappa^2}+\left(\frac{3072\pi^3\sqrt{3\tilde{\kappa}}}{\sqrt{70}}\left(\frac{\alpha'}{\kappa}\right)^{5/2}\right)m_\infty\right]+\mathcal{O}(m_\infty^2)\nonumber\\
    M&=\frac{1536\pi^3\sqrt{3\tilde{\kappa}}M_{\pl}^5}{\sqrt{70}}\left(\frac{\alpha'}{\kappa}\right)^{3/2}m_\infty^{-1}+\mathcal{O}(m_\infty^0)\nonumber\\
    S&=\frac{3072\pi^4R_H\sqrt{3\tilde{\kappa}}M_{\pl}^5}{\sqrt{70}}\left(\frac{\alpha'}{\kappa}\right)^{3/2}m_\infty^{-1}+\mathcal{O}(m_\infty^0)\,,
\end{align}
where we used \eqref{7dM} for the mass and we used $S=(M-F)/T$ to find the entropy. The analytical approximation of the free energy near the Hagedorn temperature is plotted alongside the numerical evaluation of \( F \), and we observe remarkable agreement between the two, as shown in Fig.~\ref{fig:7d away from hagedorn}.

\subsection{Open string analoge}

The Horowitz--Polchinski solutions, originally constructed from closed strings wrapping the thermal circle, have an open string analogue, as discussed in~\cite{Chen:2021dsw,Balthazar:2022hno}. In this context, we consider a D-brane and an anti-D-brane separated by a distance \( L \) at tree level, while neglecting the one-loop contribution to the action that induces an attractive force between them. Since the brane/anti-brane system breaks supersymmetry, the mode \( \chi_{\rm op} \), corresponding to the open string stretched between the two branes, becomes tachyonic at a certain separation denoted by \( L_H \), in analogy to the closed string Hagedorn radius \( R_H \).

The equations of motion for the transverse mode \( \varphi_{\rm op} \) on the brane, which captures variations in the brane separation, are analogous to those for the radion field \( \varphi \) in the closed string case. In fact, the equations of motion for \( \varphi_{\rm op} \) and the open string tachyon \( \chi_{\rm op} \) are identical to those in the closed string setup, with the only difference being that the dimension-less coupling constants \( \kappa \) and \( \tilde{\kappa} \) must be replaced with different positive numbers. Moreover, it was shown in~\cite{Balthazar:2022hno} that an \( SU(2) \) symmetry persists in this open string system at the Hagedorn separation. 

For example, the effective action of the open string tachyon and the transverse mode has been studied in~\cite{Balthazar:2022hno} up to quartic order (for bosonic strings) and takes the form:
\begin{equation}
    S_{\rm op} = \frac{\tau_d L^2}{2} \int \dd^d x \left[ 
    (\nabla \varphi_{\rm op})^2 
    + |\nabla \chi_{\rm op}|^2 
    + \left( m_\infty^2 
    + \frac{\kappa}{2\alpha'} \varphi_{\rm op} 
    + \frac{\kappa}{2\alpha'} \varphi_{\rm op}^2 \right) |\chi_{\rm op}|^2 
    + \frac{\kappa}{8\alpha'} |\chi_{\rm op}|^4 
    \right],
\end{equation}
where \( \tau_d \) is the brane tension. At the special separation where \( m_\infty = 0 \), the system exhibits an enhanced \( SU(2) \) symmetry, under which \( \chi_{\rm op} = -\sqrt{2} \varphi_{\rm op} \).

Our new results for closed strings include the existence of non-normalizable but bounded \( SU(2) \)-preserving solutions, as well as the observation that the normalizable solution is the boundary representative of such non-normalizable configurations. Both of these conclusions rely solely on the presence of an \( SU(2) \) symmetry at \( m_\infty = 0 \) and a positive quartic coefficient for \( \chi \). 

Therefore, we conclude that these results extend naturally to the open string system, where an analogous condensation exists in spacetime dimensions \( d \geq 7 \), and has finite action. As in the case of string stars, the size of these solutions remains string-scale for all \( d \geq 8 \). However, in \( d = 7 \), the size diverges at \( m_\infty = 0 \), although the action remains finite.

\subsection{String stars vs black holes}

\begin{figure}[H]
    \centering

\tikzset{every picture/.style={line width=0.75pt}} 

\begin{tikzpicture}[x=0.75pt,y=0.75pt,yscale=-1,xscale=1]

\draw  (19,307) -- (299.5,307)(48.5,18) -- (48.5,374) (292.5,302) -- (299.5,307) -- (292.5,312) (43.5,25) -- (48.5,18) -- (53.5,25)  ;
\draw    (47.5,63) .. controls (94.5,78) and (193.5,67) .. (191.5,350) ;
\draw  [fill={rgb, 255:red, 0; green, 0; blue, 0 }  ,fill opacity=1 ] (179,215.25) .. controls (179,213.46) and (180.46,212) .. (182.25,212) .. controls (184.04,212) and (185.5,213.46) .. (185.5,215.25) .. controls (185.5,217.04) and (184.04,218.5) .. (182.25,218.5) .. controls (180.46,218.5) and (179,217.04) .. (179,215.25) -- cycle ;
\draw  [dash pattern={on 4.5pt off 4.5pt}]  (48.5,214) -- (182.25,215.25) ;
\draw  (355,309) -- (635.5,309)(384.5,20) -- (384.5,376) (628.5,304) -- (635.5,309) -- (628.5,314) (379.5,27) -- (384.5,20) -- (389.5,27)  ;
\draw    (511.5,37) .. controls (514.5,106) and (520.5,292) .. (519.5,354) ;
\draw  [fill={rgb, 255:red, 0; green, 0; blue, 0 }  ,fill opacity=1 ] (515,217.25) .. controls (515,215.46) and (516.46,214) .. (518.25,214) .. controls (520.04,214) and (521.5,215.46) .. (521.5,217.25) .. controls (521.5,219.04) and (520.04,220.5) .. (518.25,220.5) .. controls (516.46,220.5) and (515,219.04) .. (515,217.25) -- cycle ;

\draw (16,21.4) node [anchor=north west][inner sep=0.75pt]    {$d$};
\draw (284,331.4) node [anchor=north west][inner sep=0.75pt]    {$m_{\infty }$};
\draw (67.06,224.72) node [anchor=north west][inner sep=0.75pt]  [rotate=-0.41] [align=left] {String stars};
\draw (155,84) node [anchor=north west][inner sep=0.75pt]   [align=left] {Black holes};
\draw (185,189.4) node [anchor=north west][inner sep=0.75pt]    {$\sim M_{s}$};
\draw (2,204.4) node [anchor=north west][inner sep=0.75pt]    {$d=7$};
\draw (352,23.4) node [anchor=north west][inner sep=0.75pt]    {$d$};
\draw (614,326.4) node [anchor=north west][inner sep=0.75pt]    {$m_{\infty }$};
\draw (413.06,165.72) node [anchor=north west][inner sep=0.75pt]  [rotate=-0.41] [align=left] {String stars};
\draw (525,98) node [anchor=north west][inner sep=0.75pt]   [align=left] {Black holes};
\draw (523,191.4) node [anchor=north west][inner sep=0.75pt]    {$\sim M_{s}$};
\draw (153,378) node [anchor=north west][inner sep=0.75pt]   [align=left] {(a)};
\draw (512,379) node [anchor=north west][inner sep=0.75pt]   [align=left] {(b)};

\end{tikzpicture}
    \caption{Even if the black hole and string star are the same saddle, there are two possibilities for the curve separating the two phases.}
    \label{TP}
\end{figure}
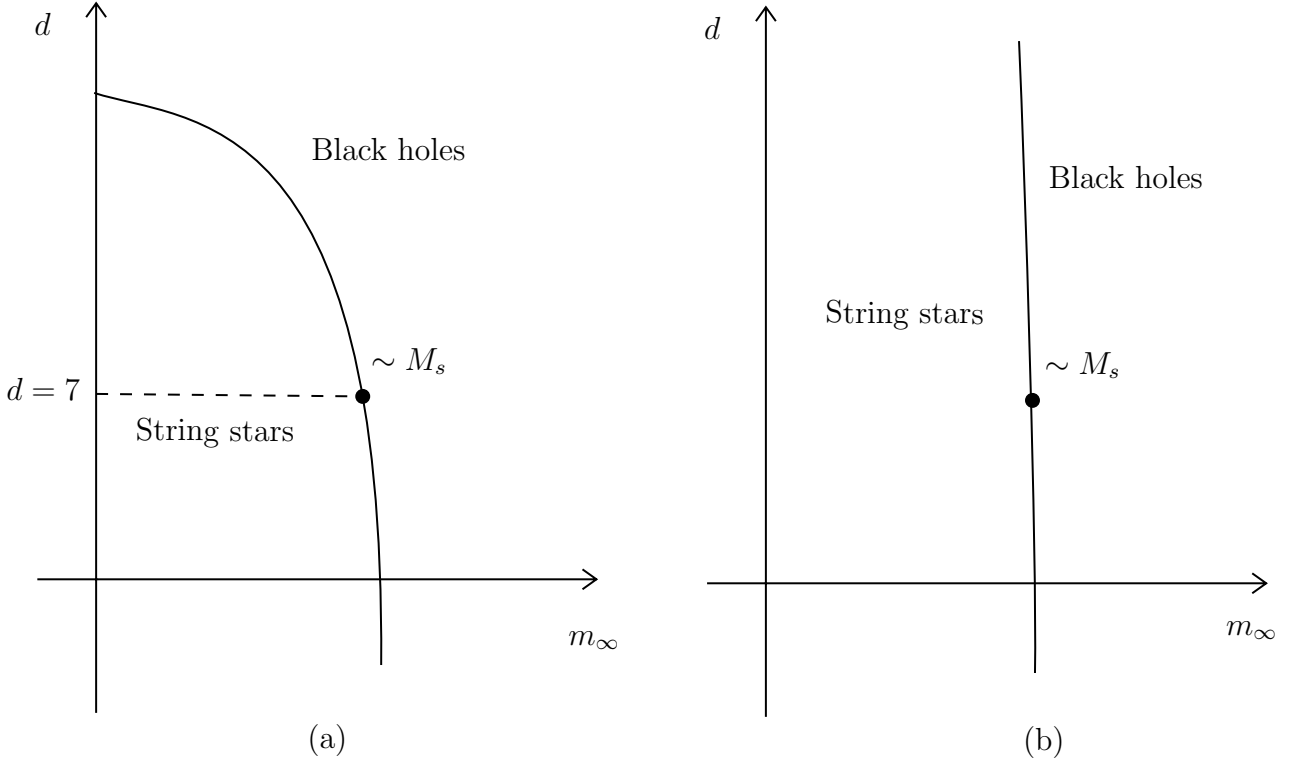

The string stars in $d\geq8$ have non-zero free energy at the Hagedorn temperature, are non-perturbative under $\alpha'$-expansion, and have sizes of the same order as black holes. Given these similarities between these string stars and black holes, one might speculate that these string stars are the Euclidean black holes\footnote{We thank Erez Urbach for bringing this discussion to our attention and we also thank Jinwei Chu and David Kutasov for very helpful comments.}. It is clear that neither of the descriptions can be described by a weakly coupled 2d CFT near the Hagedorn temperature, but the question remains if two different non-perturbative saddles coexist at the same temperature? It was pointed out in \cite{Chen:2021dsw} that in type II theories, there is a supersymmetric index that separates the two saddles and creates a sharp dividing line between the two saddles. If we formally extend the saddles to non-integer dimensions, the separations between the two saddles would become a curve in the plane of $(m_\infty,d)$. From the previous subsection, we know that the $d\geq 8$ string stars continuously connect to the $d=7$ string stars which are under perturbative control. This leaves us with two possibilities for the phase transition curve shown in figure \ref{TP} (a) there is not a phase transition above a certain spacetime dimension. If black hole saddles are better descriptions when the string stars become non-perturbative, this dimension would be somewhere between $7$ and $8$, (b) the phase transition extends to higher dimensions.

Under possibility (a), the string star is not only the same saddle as black holes, but also there is no phase transition in $d>8$. In this scenario there is a unique family of 2d CFTs that experiences a phase transition at $m_\infty=0$ along the dimension axis. However, unlike $d\rightarrow\infty$ \cite{Chen:2021emg}, it is unclear why the black hole description which is universal across dimensions would exhibit a critical behavior at a spacetime dimension between 7 and 8. Moreover, it was pointed out in \cite{Bedroya:2024uva}, that there is strong evidence that the lightest tower of states always accompany black holes phase transitions to saddles that requires those light states, such as how a KK tower and the GL transition coexist in the presence of an extra dimension. Similarly, since the tower of string excitations exist in weakly coupled string theories in $d>7$, we expect the string star phase transition to also exist.

\section{Conclusions}

In this work, we have presented evidence for the existence of string stars in spacetime dimensions $d\geq 7$. Moreover, we explicitly constructed such solutions at $d=7$ at different temperatures where they are under perturbative control. To that end, as was suggested in \cite{Urbach:2022xzw}, we showed that in the presence of large extra dimensions, the lower-dimensional string stars suffer from the GL instability just like the black holes. In $d>7$, we identified the corresponding string worldsheet CFT as a special strongly coupled member of a 1-parameter family of 2d CFTs which also has perturbative members. This family is parametrized by the boundary condition for the winding condensate. The worldsheet of the string star is strongly coupled because the $\alpha'$-corrections become important in a bounded region of space. However, the corresponding background is still under perturbative control at large radii. As a proof of concept, we numerically demonstrated certain features of these backgrounds after truncating the equations of motion to quartic order in the fields. However, higher-order $\alpha'$ corrections are relevant and are expected to alter the numerical values of the free energies. It would be very interesting to identify this CFT more precisely. A potential route for understanding this CFT is via string field theory. In particular, more recently, the HP solutions have been reconstructed in \cite{Mazel:2024alu} using Bosonic closed string field theory. The precise equations of motion of the HP solutions were recovered up to quartic interactions between fields. Hence, as the string field theory perspective contains analysis on the deformations of the 2d worldsheet CFT, this could offer additional insights into the peculiar behaviors of the 2d CFTs describing the bounded string stars with finite free energy we observed here in $d>7$.

The higher-dimensional string stars are different from the lower-dimensional HP solutions in a number of important ways. The higher-dimensional string stars in $d>7$ have non-vanishing free energy at Hagedorn temperature, are string-sized, and their free energy and mass are of the same order as those of string-sized black holes. The string star solutions remain connected to the thermal Minkowski saddle via a family of non-normalizable solutions.  It is challenging to determine whether these string stars are more stable than black holes because the $\alpha'$ corrections in the calculation of free energy are not under perturbative control in either case. However, it is reasonable to believe the string stars are different from black holes given that the thermal circle does not shrink to zero size in these backgrounds. This feature leads to a non-trivial supersymmetric index in type II theories that can distinguish black holes from string stars \cite{Chen:2021dsw}. The marginal case of $d=7$ case exhibits some features of the HP solutions and some of features of the $d>7$ solutions. In the limit of $T\rightarrow T_{\rm H}$, the size of the $d=7$ solution diverges and the profile of $\chi$ converges to $0$ (just like the HP solutions). However, the free energy of the $d=7$ solutions converge to a non-zero value (like the $d>7$ solutions). The existence of higher-dimensional string stars lends more evidence to the Swampland conjecture proposed in \cite{Bedroya:2024uva} that black holes always undergo a thermodynamic phase transition in perturbative corners of quantum gravity. It would be very interesting to further investigate the one-parameter family of non-normalizable solutions that interpolate between the thermal Minkowski saddle and the string star configurations in \( d > 7 \). Particularly, in the context of AdS spacetime, where the large-\( r \) behavior of these solutions can be interpreted as a boundary deformation.

\subsubsection*{Acknowledgements}

We would like to thank Jinwei Chu, Daniel Jafferis, David Kutasov, Juan Maldacena, Erez Urbach, Ethan Sussman, Cumrun Vafa, and Yoav Zigdon for valuable discussions and helpful comments. AB is supported in part by
the Simons Foundation grant number 654561 and by the Princeton Gravity Initiative at Princeton University. DW is supported in part by a grant from the Simons Foundation (602883,CV), the DellaPietra Foundation, and by the NSF grant PHY-2013858.

\appendix
\section{Convergence study}
\label{sec:stability}

Now, we will demonstrate that our results are convergent with respect to different parameters that are artifacts of numerical computations. As the analysis is very much similar across dimensions, in this present appendix, we will only perform the convergence test for numerical results in $d=8$. While we can directly numerically integrate \eqref{eq:EoM}, it is sometimes more convenient to integrate the logarithmic version of this PDE. Namely, let us consider $\chi=\exp(\widetilde{\chi})$. Now, with this, our equation of motion takes on the following explicit form
\[
\partial_r^2\widetilde{\chi}+(\partial_r\widetilde{\chi})^2+\frac{d-2}{r}\partial_r\widetilde{\chi}=\frac{\kappa}{\alpha'}\left(-\frac{1}{\sqrt{2}}\exp(\widetilde{\chi})+\tilde{\kappa}\exp(2\widetilde{\chi})\right)\,,
\]
where, especially for small positive values of $\chi$, the logarithmic version of the equation of motion becomes more numerically stable. However, for analyzing the scaling behavior of solutions and computing the thermodynamic properties, we will revert back to integrating the linear version of the equation of motion, i.e., \eqref{eq:EoM} as the diverging profile of the winding mode plays a dominating role in the integrals when $\chi(0)>\chi_t$. The integration scheme that we have used throughout this present work is \verb+DOP853+ which is an adaptive step-size explicit Runge--Kutta integrator of the 8$^{\rm th}$ order and is available within the \verb+scipy+ library. In particular, in executing each integration to numerically find the solution $\widetilde{\chi}(r)$, conventionally, we will need to specify the following data
\[
\text{initial position: }r_0\,,\qquad \text{final position: }r_f\,,\qquad \text{resolution: }\delta r\,.
\]
Here, the step sizes are not uniform and depending on the behavior of the local solutions, $\delta r$ may become smaller for higher resolution, e.g., if the local physics changes rapidly. What can control $\delta r$ now becomes, instead, local error tolerance parameters. Therefore, the more accurate results will prevail at smaller local error tolerance.

\subsection*{$\chi_t$}

We have presented multiple ways of finding $\chi_t$ in the main text of this paper. To summarize the non-trivial consistency across different methods, given a spacetime dimension, we can find $\chi_t$ through
\begin{itemize}
    \item computing the point of divergence as we perturb away from the perturbative solutions;
    \item computing the exponent at large $r$ (the solution with $b=d-3$ marks $\chi_t$);
    \item finding the solution that has constant, with respect to $r_{\rm max}$, finite $L^2$-norm;
    \item finding the solution that satisfies the scaling constraint;
\end{itemize}
where all such solutions point towards the unique value of $\chi_t$. As much of our results rely on numerical computations, it is worth showing that the value of $\chi_t$ is stable with respect to the chosen values of $\delta r$, $r_0$, and $r_{\rm max}$ for the main text. From the above list, the most straightforward method of doing such a convergence test is via computing the point of divergence for these solutions by perturbing away from the perturbative solutions. From the perspective of numerical accuracy, we have the most reliable solution occurs when $r_0\to 0\,, r_{\rm max}\to \infty\,, \delta r\ll 1$. To this end, the most efficient computational method is to choose the largest $r_0$, smallest $r_{\rm max}$, and largest $\delta r$ such that it is within reasonable deviation away from the smallest $r_0$, largest $r_{\rm max}$, and smallest $\delta r$ simulation results. However, as elaborated previously, with the present integration scheme, we can indirectly tune $\delta r$ by adjusting the local error tolerance parameters at each integration step. To this end, \verb+DOP853+ has two parameters that precisely do so, namely the absolute error tolerance \verb+atol+ and the relative error tolerance \verb+rtol+. In particular, at each integration step, \verb+DOP853+ performs two integrations which are an 8$^{\rm th}$-order and a lower-order Runge--Kutta integration. Then, by comparing the relative difference between these two integration results, e.g., $\chi_{i+1}^{(7)}$ and $\chi_{i+1}^{(8)}$ at each integration step, we can construct an error estimate
\[
\mathrm{error}=|\chi_{i+1}^{(8)}-\chi_{i+1}^{(7)}|\,.
\]
Then the previously mentioned two parameters enforce the integration to optimize the integration step size by ensuring that
\[
\mathrm{error}\leq \verb+atol++\verb+rtol+\cdot |\chi_{i+1}^{(8)}|\,.
\]
In particular, to ensure precision and accuracy, $\verb+atol+<\mathrm{min}\left\{\verb+rtol+\cdot |\chi|\right\}$. Hence, in what follows, we will precisely examine how the value of $\chi_t$ can change as we tune these two error tolerance parameters. 

\begin{figure}
\centering
\hspace*{-.25in}
\begin{subfigure}{.48\textwidth}
        \centering
        \includegraphics[width=\linewidth]{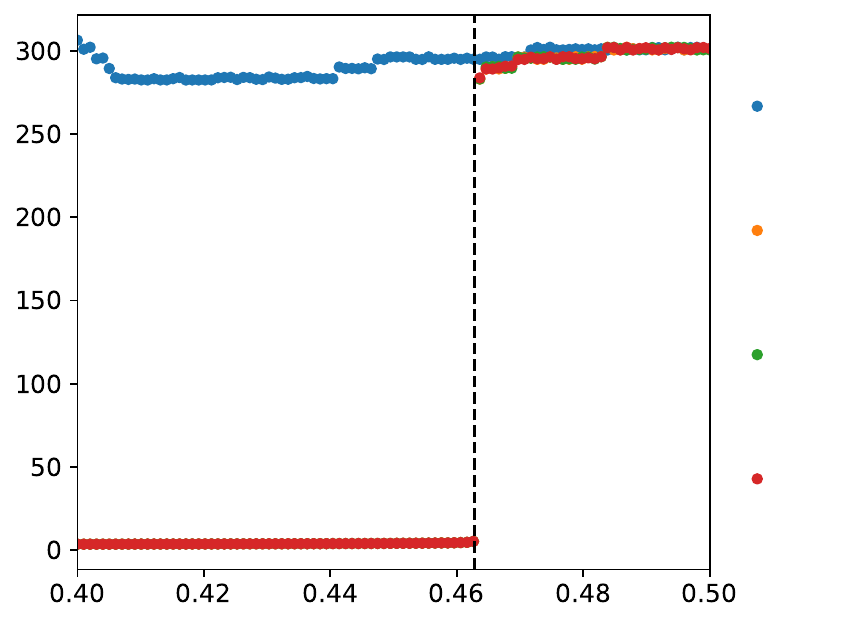}
        \begin{picture}(0,0)\vspace*{-1.2cm}
        \put(100,180){\footnotesize atol}
        \put(105,153){\footnotesize $10^0$}
        \put(105,118){\footnotesize $10^{-6}$}
        \put(105,85){\footnotesize $10^{-12}$}
        \put(105,51){\footnotesize $10^{-22}$}
        \put(-150,105){\footnotesize $\ln(\mathrm{Amp})$}
        \put(-30,5){\footnotesize $\chi(0)$}
        \put(10,5){\footnotesize $\chi_t$}
        \end{picture}\vspace*{-0.6cm}
        \caption{Fixed $\mathrm{rtol}=10^{-9}$.}
        \label{fig:atol}
    \end{subfigure}
    \hfill
    \begin{subfigure}{.48\textwidth}
        \centering
        \includegraphics[width=\linewidth]{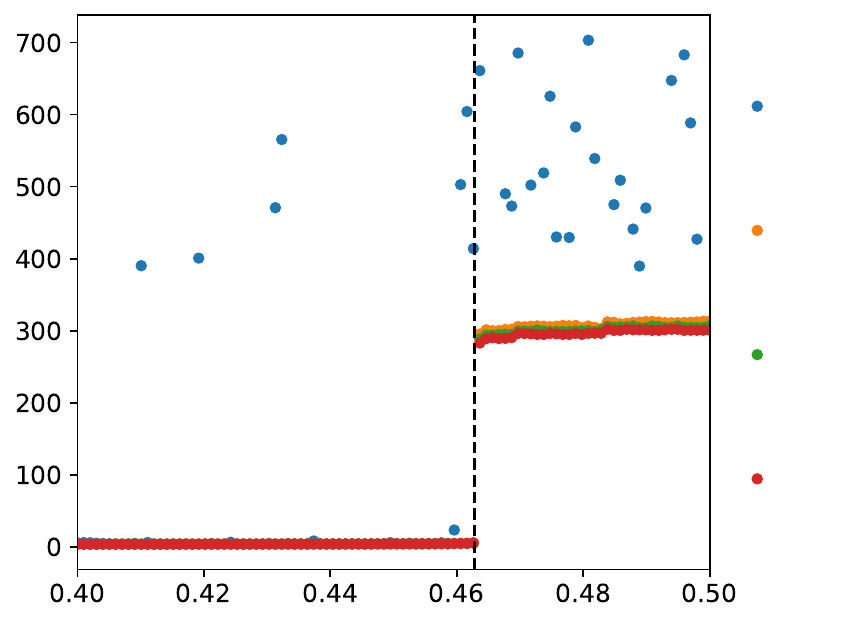}
        \begin{picture}(0,0)\vspace*{-1.2cm}
        \put(100,180){\footnotesize rtol}
        \put(105,153){\footnotesize $10^0$}
        \put(105,118){\footnotesize $10^{-3}$}
        \put(105,85){\footnotesize $10^{-6}$}
        \put(105,51){\footnotesize $10^{-9}$}
        \put(-150,105){\footnotesize $\ln(\mathrm{Amp})$}
        \put(-30,5){\footnotesize $\chi(0)$}
        \put(10,5){\footnotesize $\chi_t$}
        \end{picture}\vspace*{-0.6cm}
        \caption{Fixed $\mathrm{atol}=10^{-6}$.}
        \label{fig:rtol}
    \end{subfigure}
    \caption{Convergence test for $\chi_t$ in $d=8$.}
    \label{fig:convergence for chi0 wrt delta r}
\end{figure}

The results for the convergence test are shown in fig.~\ref{fig:convergence for chi0 wrt delta r} where we examine how computing \eqref{eq:subleading}, and specifically the point of divergence where $\chi_t$ is defined, can depend on \verb+atol+ and \verb+rtol+. These simulations are done with $r_0=e^{-10}\, \sqrt{\alpha'/\kappa}$ and $r_{\rm max}=10^4\, \sqrt{\alpha'/\kappa}$ fixed. In particular, with a small local error tolerance set to $\verb+atol+=10^{-22}$ and $=\verb+rtol+=10^{-9}$, i.e., high resolution, we have the smallest value of the bounded solutions are $\chi_{\rm bounded}(r_{\rm max})\approx 10^{-7}$. ($\verb+atol+$ here is chosen such that $(r_{\rm max})^{3-d}>\verb+atol+$.) Starting from \ref{fig:atol}, with $\verb+rtol+=10^{-9}$ fixed, we observe that the point of divergence between the smaller two magnitudes of order of \verb+atol+ simulations are consistent with each other. Hence, a sensible simulation should be done with absolute error tolerance $\verb+atol+\lesssim 10^{-6}$. Now, in fig.~\ref{fig:rtol}, we observe that for fixed $\verb+atol+=10^{-6}$, \verb+rtol+ is consistent across the latter three smaller order of magnitudes. Hence, for a converging result of $\chi_t$, we need to have $\verb+rtol+\lesssim 10^{-3}$.
Hence, in the main text, due to the low cost of small \verb+rtol+ simulations, we have used the following numerical choice of local error tolerance $\verb+rtol+=10^{-9}$ and $\verb+atol+=10^{-6}$.

\subsection*{Free energy}

\begin{figure}
    \centering
    \includegraphics[width=0.65\linewidth]{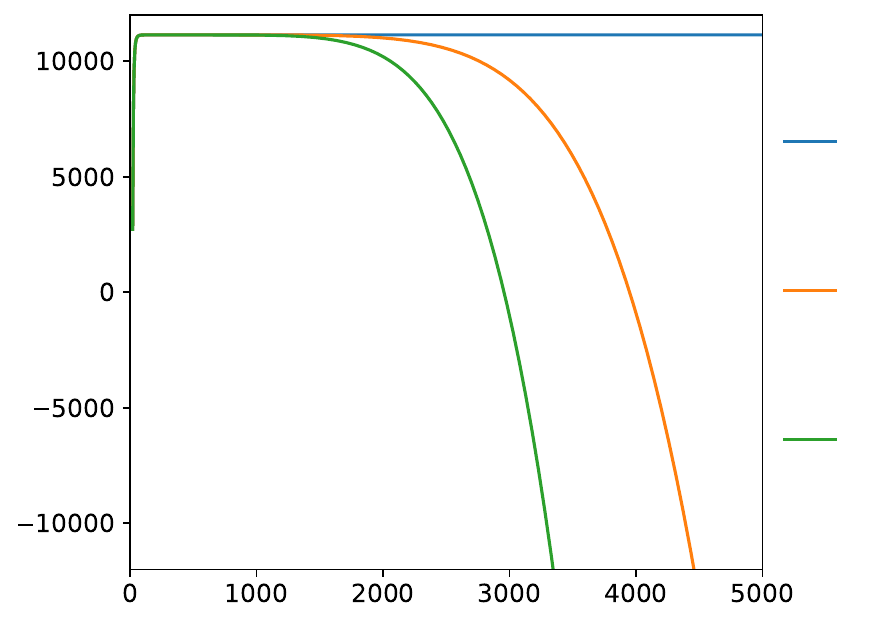}
    \begin{picture}(0,0)\vspace*{-1.2cm}
        \put(-350,110){\footnotesize $F^{(\rm 8d)}$}
        \put(-160,-10){\footnotesize $r_{\rm max}$}
        \put(-20,200){$\chi(0)$}
        \put(0,170){$\chi_t$}
        \put(0,117){$\chi_t-\varepsilon$}
        \put(0,63){$\chi_t-2\varepsilon$}
        \end{picture}\vspace*{0.5cm}
    \caption{The free energy of bounded normalizable solutions and small deviations in $d=8$ as a function of $r_{\rm max}$ with $r_0=e^{-10}\, \sqrt{\alpha'/\kappa}$ with $\mathrm{atol}=10^{-6}$ and $\mathrm{rtol}=10^{-9}$.}
    \label{fig:d8-free-energy-convergence}
\end{figure}

Now, with the above solution that is numerically stable, we integrate $\chi(r)$ up to $r_{\rm max}=5000\, \sqrt{\alpha'/\kappa}$ with different boundary values at $\chi(0)$ and compute the free energy to see if we indeed end up with a convergent free energy. To this end, we have, as can be seen in fig.~\ref{fig:d8-free-energy-convergence}, the bounded normalizable solution has free energy that plateaus for all $r_{\rm max}\gtrsim 10^2\, \sqrt{\alpha'/\kappa}$. However, for solutions that deviate slightly from $\chi_t$, one does not end up with a constant solution, but rather a diverging solution that depends on $r_{\rm max}$. 

The analyses for $d=9,10$ follow an identical pattern and leads to a similar conclusion as $d=8$. Hence, we will omit the analyses here.

\section{Unbounded solutions}
\label{sec:unbounded}

In this section, we show that a large class of higher-order interactions encoded in an effective potential $V_{\rm eff}(\chi)$ will have unbounded solutions for sufficiently large $\chi(0)$. We remind the readers that by unbounded we refer to solutions where $\chi$ vanishes at finite radius. Let us start by potentials that have a local maximum for a positive value of $\chi$. For such potentials, we show that for sufficiently large values of $\chi(0)$, the solution will be unbounded. To show that, it is helpful to think of the equation of motion \eqref{eq:EoM} for $\chi(r)$ in terms of the motion of a one-dimensional particle in the potential $V_{\rm eff}$ with a time-dependent friction term $(d-2)\partial_r\chi /r$. In this equivalence, $r$ and $\chi$ respectively play the roles of time and position of the particle. Suppose $\chi_c$ is the smallest real positive value of $\chi$ that is a local maximum for $V_{\rm eff}$. The constant profile $\chi(r)=\chi_c$ is a solution to the equation of motion. In this solution, the derivative $\partial_r\chi$ is identically zero everywhere. Now, consider the initial condition $\chi(0)=\chi_c-\varepsilon$. For every radius cutoff $L$, we can choose a sufficiently small $\varepsilon$ that would keep the solution arbitrarily close to $\chi(0)=\chi_c$ for $r<L$. Therefore, the derivative $\partial_r \chi$ will remain small for $r<L$. By choosing larger values of $L$, we can make the friction term $(d-2)\partial_r\chi /r$ negligible as far as the evolution of $\chi$ from $\chi_c-\varepsilon$ to $\chi=0$ is concerned. However, after removing the friction term, the motion of the particle will satisfy conservation of energy and is guaranteed to not stop at $\chi=0$ where $V_{\rm eff}$ vanishes. Therefore, for initial conditions that are sufficiently close to the local maximum, the solution will runaway in the negative $\chi$ direction. Our argument proves that in the presence of a local maximum, some initial conditions will lead to solutions that do not vanish at infinity. 

In addition to a potential with a local positive maximum, we can also show the existence of the unbounded solutions for a large class of monotonic potentials. Specifically, when the potential is monotonically increasing for $\chi>0$ and as $\chi\rightarrow\infty$, the potential goes like $\chi^n$ where $n\neq 3$. Note that as shown in \eqref{CIO}, the cubic interactions alone will always be bounded. However, as we know from the equation \eqref{QTP}, there are higher-order corrections including a non-zero quartic term. Therefore, if we keep any finite number of higher-order corrections beyond the cubic term, the potential will satisfy our assumption. 

We want to show that for a sufficiently large $\chi(0)$ the solution cannot be bounded and will cross the $\chi=0$ at finite $r$. If we ignore the quartic and higher-order corrections and only keep the cubic term, the rewriting of the equation of motion as \eqref{EOM2} shows us that the total energy $\partial_x \mathcal{X}^2+\mathcal{V}_{\rm eff}(\mathcal{X})$, where $x=\ln(r\sqrt{\kappa/\alpha'})$, is monotonically decreasing due to the friction term. Since this energy is zero at $x=-\infty$, we have $\mathcal{V}_{\rm eff}(\mathcal{X})\leq 0$ which imposes an upper bound $\mathcal{X}_c=3\sqrt{2}(d-5)$ on $\mathcal{X}$. Suppose $\epsilon\ll 1$ is small enough that if $\chi<\epsilon$, the higher-order corrections are negligible and therefore, $\chi r^2$ has to be bounded from above by $\mathcal{X}_c+\delta$ for some $\delta\ll1$. This implies that the radius $r_\epsilon$ at which $\chi(r_\epsilon)=\epsilon$ must satisfy $r_\epsilon<\sqrt{(\mathcal{X}_c+\delta)/\epsilon}$. In the following, we show that if $n\neq3$ this inequality can be violated by choosing a sufficiently large $\chi(0)$. Let us first consider the case of $n>3$. If we ignore the friction term in \eqref{eq:EoM}, integrating the equation for $V_{\rm eff}\sim \chi^n$ leads to 
\begin{align}
    \int_{\chi(r)} ^{\chi(0)}\frac{\dd\chi}{\sqrt{\chi(0)^n-\chi^n}}\sim r\,.
\end{align}
The value $\chi_f$ at where the friction term becomes non-negligible and begins to compete with the potential is $\chi_f\sim \chi(0)^\frac{n-1}{n}$ and it happens at radius $r_f\sim \chi(0)^{1-\frac{n}{2}}$. Therefore, by choosing a large enough $\chi(0)$, $\chi_f$ can also be large enough such that the potential $V_{\rm eff}$ can be approximated by $\sim \chi^n$. After this point, the friction term is of the same order as the potential.
\begin{align}
-(d-2)\frac{\partial_r\chi}{r}\sim \chi^n\rightarrow \chi(r) \sim r^{-\frac{2}{n-1}}\,.
\end{align}
Suppose the $V_{\rm eff}\propto \chi^n$ is a good approximation for $\chi>\chi_1$. With the above scaling behavior, we can estimate the radius at which $\chi=\chi_1$ to be
\begin{align}
r_f+ \left(\frac{\chi(0)^{(n-1)/n}}{\chi_1}\right)^{(n-1)/2}\,,
\end{align}
Since this radius is smaller than the radius where only cubic terms are significant, we have
\begin{align}
    r_\epsilon>r_f+ \left(\frac{\chi(0)^{(n-1)/n}}{\chi_1}\right)^{(n-1)/2}\,.
\end{align}
However, the right hand side can be much larger than the $\sqrt{(\mathcal{X}_c+\delta)/\epsilon}$ bound for $r_\epsilon$ for a sufficiently large $\chi(0)$. Therefore, we conclude that for $n>3$, a solution with sufficiently large $\chi(0)$ cannot be bounded. On the other hand, for $n<3$, even if we ignore the friction term, the descent of $\chi(r)$ down the potential toward $0$ is not fast enough to satisfy $r_\epsilon<\sqrt{(\mathcal{X}_c+\delta)/\epsilon}$. This is because,
\begin{align}
r_\epsilon \gtrsim\int_{\epsilon} ^{\chi(0)}\frac{\dd\chi}{\sqrt{\chi(0)^n-\chi^n}}>\frac{\chi(0)-\epsilon}{\chi(0)^\frac{n}{2}}\,,
\end{align}
 which for sufficiently large $\chi(0)$, violates the upper bound on $r_\epsilon$. Note that $n=2$ saturates the inequalities. We conclude that any potential $V_{\rm eff}$ that has a local positive maximum, or has a polynomial behavior $\sim \chi^n$ at large $\chi$ where $n\neq 2$ will have unbounded solutions. In particular, keeping the higher-order corrections up to some finite order higher than the cubic term leaves us with a $V_{\rm eff}$ that has unbounded solutions. Our argument is not a proof for a generic potential $V_{\rm eff}$ which is important especially given that the higher-order corrections form an infinite series. It would be interesting to investigate the generality of the existence of the unbounded solutions for a broader class of potentials $V_{\rm eff}$. 

\bibliography{papers}
\bibliographystyle{JHEP}

\end{document}